\documentclass[runningheads]{llncs}
\usepackage[utf8]{inputenc}
\usepackage{amsmath,amssymb,amsfonts}
\usepackage{enumerate,xspace,paralist,tabularx}
\usepackage{subfig,wrapfig,caption}
\usepackage[raiselinks=true,
			bookmarks=true,
			bookmarksopen=true,
			bookmarksopenlevel=2,
			bookmarksnumbered=true,
			hyperindex=true]{hyperref} 
\usepackage[basic]{complexity}
\usepackage[english]{babel}
\usepackage{cite}
\usepackage[capitalise,nameinlink]{cleveref}
\Crefname{figure}{Fig.}{Figs.}
\Crefname{section}{Sec.}{Secs.}
\usepackage[export]{adjustbox}

\usepackage{float,graphicx}
\usepackage[usenames,dvipsnames]{xcolor}
\usepackage{lipsum}
\usepackage{enumitem}
\DeclareGraphicsExtensions{.pdf,.png,.eps}
\graphicspath{{figs/}}
   
\let\doendproof\endproof\renewcommand\endproof{~\hfill$\qed$\doendproof}
\setlist[description]{leftmargin=*}
 
\newcommand{\etal}{{et~al.}}

\usepackage{mathtools}
\def\Oh{\ensuremath{\mathcal{O}}}

\def\RDFS{\texttt{randDFS}\xspace}

\def\BFStree{\texttt{treeBFS}\xspace}
\def\treeBFS{\texttt{treeBFS}\xspace}
\def\SDDFS{\texttt{smlDgrDFS}\xspace}
\def\CONCRO{\texttt{conCro}\xspace}
\def\ConCro{\texttt{conCro}\xspace}
\def\CONGREEDY{\texttt{conGreedy}\xspace}
\def\ConGreedy{\texttt{conGreedy}\xspace}

\def\FCONGREEDY{\texttt{conGreedy+}\xspace}

\def\SLOPE{\texttt{slope}\xspace}
\def\ELEN{\texttt{eLen}\xspace} 
\def\CF{\texttt{ceilFloor}\xspace}
\def\Circular{\texttt{circ}\xspace} 
\def\CIRCULAR{\texttt{circ}\xspace} 
\def\EarDecomposition{\texttt{earDecomp}\xspace}

\def\EAR{\texttt{earDecomp}\xspace}

\def\GreedyAlt{\texttt{greedyAlt}\xspace}
\def\GreedyCombi{\texttt{greedy+}\xspace}
\def\SA{\texttt{simAnn}\xspace}

\usepackage{printlen}

\makeatother

\title{Experimental Evaluation of\\  Book Drawing Algorithms\thanks{Preliminary results of
this paper were presented in a poster at Graph Drawing 2016.}} %

\author{Jonathan Klawitter\inst{1} \and Tamara~Mchedlidze\inst{2} \and Martin~Nöllenburg\inst{3}}
\authorrunning{J. Klawitter et al.}
\institute{
	University of Auckland, New Zealand, \email{jo.klawitter@gmail.com} \and
	Karlsruhe Institute of Technology, Germany, \email{mched@iti.uka.de} \and
	TU Wien, Vienna, Austria, \email{noellenburg@ac.tuwien.ac.at}
}

\begin{document} 

\maketitle
%\linenumbers

\begin{abstract}
	A $k$-page book drawing of a graph $G=(V,E)$ consists of a linear ordering of its vertices along a
	\emph{spine} and an assignment of each edge to one of the $k$ \emph{pages}, which are
	half-planes bounded by the spine.
	In a book drawing, two edges cross if and only if they are assigned to the same page and their
	vertices alternate along the spine. 
	% If the book drawing is crossing-free, it is also called a
	% book embedding and it is a well-known \NP-hard problem to determine the book thickness, i.e., the
	% minimum number of pages for a book embedding to exist.
	Crossing minimization in a $k$-page book drawing is \NP-hard, yet book
	drawings have multiple applications in visualization and beyond. Therefore several heuristic book
	drawing algorithms exist, but there is no broader comparative study on their relative % and absolute
	performance. In this paper, we propose a comprehensive benchmark set of challenging graph classes
	for book drawing algorithms and provide an extensive experimental study of the performance of
	existing book drawing algorithms.
\end{abstract}

\section{Introduction}\label{se:introduction}
\pdfbookmark[0]{Introduction}{Introduction}

\begin{wrapfigure}[13]{r}{.25\textwidth}
	\vspace{-1.2cm}
	\centering
	\includegraphics[width=.22\textwidth]{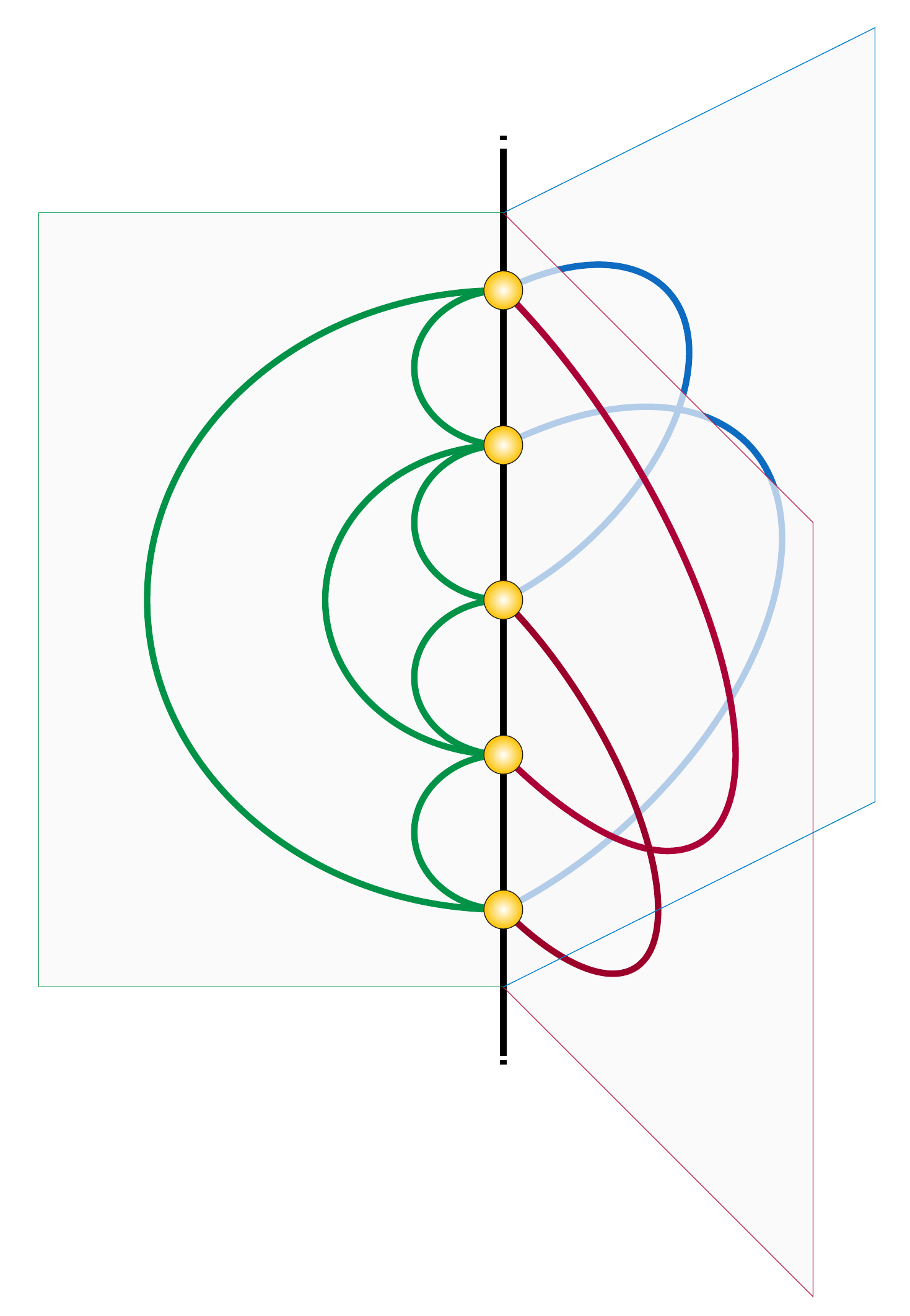}
	\caption{3-page book drawing of $K_5$ with two crossings.}
\end{wrapfigure}
Book embeddings and book drawings are a fundamental and well-studied topic in graph theory and graph
drawing. Combinatorially, a \emph{$k$-page book drawing} of a graph $G = (V, E)$ consists of a
cyclic linear ordering of its vertices along a \emph{spine} and an assignment of each
edge to one of the $k$ \emph{pages}, which are half-planes bounded by the spine.
The spine and the $k$ pages form a \emph{book}.
Clearly, two edges $\{u, v \}$ and $\{w, z\}$ in a book drawing cross if and only if they
are assigned to the same page and the four vertices alternate on the spine.

A book drawing is called a \emph{book embedding} if it is crossing-free.
The \emph{book thickness} (or \emph{pagenumber}) of a graph $G$ is the smallest $k$ such that $G$
admits a $k$-page book embedding~\cite{BK79}.
Deciding whether a graph can be embedded on $k$ pages is an \NP-complete
problem even for $k=2$~\cite{MNKF90,CLR87} and there are many results about lower and upper bounds on the book
thickness of specific graph classes.
A long-standing open question~\cite{DW07} is to determine whether the book thickness of planar
graphs is three or four.
Yannakakis~\cite{Yan89} showed that any planar graph can be embedded on four pages and there are
planar graphs that cannot be embedded on two pages.
Likewise, the book thickness of $k$-planar graphs is open. 
Alam et al.~\cite{ABK15} showed that there are 1-planar graphs that need four pages and that 16 is
an upper bound. 

If the number $k$ of pages is given, a $k$-page book embedding may not exist.
In this case, crossing minimization becomes the primary optimization goal.
It reduces to two basic and interdependent combinatorial problems: the \emph{vertex ordering} (VO)
along the spine and the \emph{page assignment} (PA)  for the edges.

Again, computing the $k$-page \emph{book crossing number}, i.e.,
the minimum number of crossings over all $k$-page book drawings of a graph, is an \NP-hard
problem~\cite{MNKF90,SSSV96} and fixed-parameter tractable algorithms for 1- and 2-page crossing
minimization are known~\cite{BE14}.
Book drawings are motivated by several applications, e.g., 
network visualization~\cite{w-dvss-02,GK06,BB05,HeSyk04,GiacomoDLW06}, VLSI design~\cite{y-lbeg-86}, RNA
folding~\cite{CDKU12}, and knot theory~\cite{Dynnikov1999}.
Various heuristic algorithms have been proposed in the literature. In addition, crossing
minimization in book drawings has been the challenge problem of the Graph Drawing Contests %\footnote{\url{www.graphdrawing.de}} 
in 2015 and 2016.
Yet there are no broader comparative studies of these algorithms and no established set of challenging benchmark graph classes.
In this paper, we introduce a comprehensive benchmark set for book drawing algorithms and provide
the first extensive experimental study of the performance of state-of-the-art book drawing algorithms for
multiple numbers of pages.

There are several heuristics for 2-page crossing minimization~\cite{Cim02,Cim06,CiM07} with a fixed
linear vertex ordering,  as well as algorithms for the general 2-page crossing minimization
problem~\cite{HSV05,HSSV05}.
Genetic and evolutionary crossing minimization algorithms have been proposed for one
page~\cite{HeSyk04}, two pages~\cite{HSM07,PMH07,BSSV08}, and any number of pages~\cite{SSS13}.
Further, neural networks have been used for 2-page crossing minimization~\cite{HSM06,Wan08} and for
$k$-page crossing minimization~\cite{RCLG07}.

Experimental evaluations have been performed by Satsangi et al.~\cite{SSS13}, who, however, excluded
previously best performing algorithms by Baur and Brandes~\cite{BB05} and tested
algorithms for VO and PA problems only independently from each other, not in combination.
He \etal~\cite{HSMV15} performed an experimental study of several heuristics, but only for the
2-page crossing minimization problem. 

\paragraph{Contributions and outline.}
In this paper, we determine the strengths and weaknesses as well as the relative performance of
 heuristic algorithms for the book drawing problem by means of a detailed
quantitative experimental study.
To this end, we present a list of state-of-the-art heuristic algorithms
from the literature as well as some newly proposed heuristics in Section~\ref{se:algo}.
Section~\ref{se:design} presents a collection of different graph classes together with suitable
random graph generators to be used for creating benchmark graphs for our evaluation.
Finally, Section~\ref{se:evaluation} contains our comparative experimental evaluation. The main
focus of our study is the relative performance of the heuristics in terms of crossing minimization
depending on the properties of the benchmark instances, such as book thickness, graph size, edge
density, and graph structure. 
Since out implementations are not optimized for a fast performance, we refrained from a detailed
running time analysis. Some preliminary indications of the running times can be found in the
appendix (\cref{fig:appendix:time:heuristics}).
The code of our benchmark graph generators and of the book drawing algorithms can be found
online\footnote{Graph generators: github.com/joklawitter/GraphGenerators, book drawing
algorithms: github.com/joklawitter/BookDrawingAlgorithms}.

%!TEX root = /Users/noelle/Documents/work/00-aktuell/bookdraw/BookEmbedding/sources/booksPaper.tex
\section{Algorithms}\label{se:algo}
\pdfbookmark[0]{Algorithms}{Algorithms}

We distinguished between constructive heuristics that have the
common property that they consider each vertex  and edge once, and local search heuristics that
make several rounds re-considering the same vertices and edges iteratively. 
We evaluate these algorithms separately, as the latter can be seen as local search heuristics, which
also use much more computation time.
The constructive heuristics themselves can be characterized as VO heuristics, PA heuristics, and
\emph{combined} heuristics, that construct both VO and PA simultaneously.
 
\subsection{Constructive Heuristics}\label{se:construct}
Four of the  heuristics presented in this section have not appeared in the literature earlier,
namely \BFStree, \ConGreedy, \FCONGREEDY, and \EarDecomposition (see~\cite{Klawitter16} for more
details).
Several additional heuristics are referenced, but not included in our study because they were always
outperformed by the other presented heuristics in previous experimentation.

\subsubsection{VO Heuristics.}
A VO heuristic considers vertices in some particular order and places them on the spine based on
some criteria. An edge $\{u,v\}$ where only one of $u$ and $v$ (resp. both) has been placed on
the spine is called \emph{open} (resp. \emph{closed}).

\begin{description}
\item[smallest degree DFS (\SDDFS)~{\normalfont\cite{HeSyk04}}.] DFS-based heuristics set the VO to
be the order in which the vertices are visited by a depth-first traversal of the graph. The \SDDFS
heuristic starts with a smallest degree vertex and chooses a neighbor with smallest degree to
proceed.

\item[random DFS (\RDFS)~{\normalfont\cite{BSSV08}}.]  In contrast to \SDDFS, \RDFS starts with
a random vertex and proceeds with a random neighbor. 

\item[\textbf{tree-based BFS} (\BFStree).] This heuristic generates a breath-first spanning tree
of the graph and embeds it crossing-free in a 1-page book yielding the VO.
All three search based heuristics have a running time of $\Oh(m + n)$.

\item[connectivity based (\ConCro)~{\normalfont\cite{BB05}}.] This heuristics builds the VO step by
step.
At each step it selects the vertex with the most neighbors already placed and breaks ties in favor of vertices
with fewest unplaced neighbors (connectivity $\rightarrow$ \texttt{con}). It places the vertex on
that end of the already computed spine, where it introduces fewer crossings with open edges (crossings
$\rightarrow$ \texttt{Cro}). The intuition behind this heuristic is that the chosen vertex closes
most open edges and opens fewest at ties. Its running time is $\Oh((m + n)\log n)$.

\item[greedy connectivity based (\ConGreedy).] Like \ConCro it selects the next vertex to place based on
connectivity, however, it places it on any position (not just one of the end points) of the current spine where it
introduces fewer crossings with closed edges. With $\Oh(m^2 n)$ it has the highest running time. 
\end{description}
Heuristics excluded, due to relatively poor performance, are among others a maximum neighborhood
heuristic, a vertex-cover heuristic, a simple BFS heuristic~\cite{SSS13},
and variations of \ConCro~\cite{BB05}.

\subsubsection{PA Heuristics.}
% greedy
The following first three heuristics share a general framework. They first compute an
edge order according to some strategy and then place the edges one by one on the
page where the increase in crossings is minimal.

\begin{description}
\item[ceil-floor (\CF)~{\normalfont\cite{KRSZ02}}.] In this strategy the edges are ordered non-in\-crea\-sing\-ly by
their length in a circular drawing. 

\item[length (\ELEN)~{\normalfont\cite{Cim02, SSS13}}.] Here the edges are ordered non-increasingly by the
distance of their end vertices on a spine. Thus edge $\{1,n\}$ is
listed first and any edge $\{i, i+1\}$ last. Like \CF, it has a $\Oh(m^2)$ running time. 

\item[circular (\Circular)~{\normalfont\cite{SSS13}}.]
The edges are enumerated in the order they are visited by the paths 
$P_1\dots P_{\lceil \frac n 2 \rceil}$, where path $P_i$ starts at vertex $i$ and
visits vertices $i+1, i-1, i+2, \ldots, i + \lceil \frac n 2 \rceil$. This heuristic is inspired by
the fact that it achieves zero crossings for complete graphs on $\lceil \frac n 2 \rceil$ pages by
placing edges of each path on a distinct page. It has a running time of $\Oh(n^4)$.

\item[ear decomposition (\EarDecomposition).]
Consider a circular drawing $\Gamma_C$ of a graph $G = (V, E)$.  The edge intersection graph is
defined as $G_C=(E, \{\{e,e'\} \mid e,e' \in E  \land e,e' \textrm{ cross in }\Gamma_C\})$.
The heuristic \EarDecomposition considers the circular drawing for the given VO,
constructs its intersection graph $G_C$, and an ear decomposition of $G_C$, and then assigns the
vertices of each ear (i.e., the edges of $G$) alternatingly to different pages.
The intuition behind the heuristic is that it tries to put the conflicting edges to different
pages. \EarDecomposition can be implemented to run in $\Oh(m^2)$ time.

\item[slope (\SLOPE)~{\normalfont\cite{HSV05}}.] 
Consider the circular drawing with equally distributed vertices for a graph and VO. Then, the more
the geometric slopes of two non-incident edges in this drawing differ, the more likely they cross.
\SLOPE groups the edges based on their slopes and assigns each group to a page. It has a linear
running time $\Oh(m)$.

\end{description}
Again, due to relatively poor performance, we excluded several other greedy
variations~\cite{Cim02,HSSV05,SSS13}, a dynamic programming and a divide and conquer approach~\cite{Cim02}. 

\subsubsection{Combined Heuristics.} 
Almost all existing constructive heuristics compute a VO and a PA
independently.
He et al.~\cite{HSSV05} first combined the two problems. They extended \SDDFS such that whenever an
edge is closed it is assigned to the page where it introduces the smallest number of crossings. 
We experimented with such extensions for \SDDFS and \RDFS heuristics and concluded that
they performed worse than using them in combination with another PA heuristic~\cite{Klawitter16}.
The following heuristic utilizes this idea for \ConGreedy.

\begin{description}
\item[combined greedy connectivity based (\FCONGREEDY).] 
While constructing \linebreak the VO like \CONGREEDY, this heuristic considers the PA of already
placed edges. More precisely, the best position for a new vertex is the position where this
vertex's incident and now closed edges induce fewest new crossings. The PA is then accordingly
extended to these newly closed edges. The heuristic's overall asymptotic running time is $\Oh(m^2 n)$.
\end{description}

We note that the immediate page assignment done by  \FCONGREEDY also affects the computed VO. 
Hence, \FCONGREEDY can also be used as VO heuristic by discarding the produced PA.

\subsection{Local Search Heuristics} \label{se:localsearch}

Local search heuristics take a given book drawing and try to reduce its number of crossings by
performing local changes. Heuristics \GreedyAlt and \GreedyCombi are newly proposed, while \SA has
been proposed by Cibulka~\cite{Cib17}, who won the Automated Graph Drawing Challenge in
2015~\cite{Kindermann2015}.

\begin{description}
\item[alternating greedy search (\GreedyAlt).] A single \emph{vertex round} of this heuristic 
considers vertices in a random order, takes each of them in this order and places it on the
position on the spine where it produces the least number of crossings. Here edges stay on the pages
they are. A single \emph{edge round} does the same with the edges: it considers edges in a random
order and places them on the page where it produces the least number of crossings. 
\GreedyAlt alternates between vertex and edge rounds until it converges to a local minimum.
\item[combined greedy search (\GreedyCombi).] A single round of this heuristic is similar
to \FCONGREEDY, but the vertices are considered in a random order. Several rounds are performed
until a local minimum is found. 
\item[simulated annealing (\SA)~{\normalfont\cite{Cib17}}.]  This algorithm, depending on a
temperature that decreases with each iteration, makes local changes to the book drawing and accepts
them if they either improve the drawing, or with a certain probability 
depending on the temperature and how many crossings the move introduces. The moves in each
iteration are (1) $m$ times moving an edge to a random page, (2) $n \sqrt{n}$ times swapping a
random vertex with its neighbor, (3) $n$ times moving a vertex to a random position and greedily
improving the assignment of its incident edges, and (4) $\frac{4}{n}$ times searches, in the
fashion of \GreedyCombi, for the best position of a vertex. It runs 1000 iterations.
\end{description}

The literature contains similar greedy optimization algorithms for the VO~\cite{HSSV05,ST06}, 
another simulated annealing approach~\cite{SSS13}, evolutionary~\cite{KRSZ02,GSS11},
and neural network algorithms~\cite{RCLG07,Wan08}. However, they all were either restricted to only
two pages or were outperformed by other heuristics in previous experiments~\cite{Klawitter16}.

%!TEX root = /Users/noelle/Documents/work/00-aktuell/bookdraw/BookEmbedding/sources/booksPaper.tex
\section{Benchmark graphs}\label{se:design}
\pdfbookmark[0]{Design}{Design}

Our previous experiments have shown that there is no fixed ranking for the performance of the
construction heuristics in terms of crossing minimisation~\cite{Klawitter16}. On the
contrary, rankings depend on the number of pages, the edge densities of the graphs and their
structuredness. We therefore selected nine different benchmark graph classes that vary in terms of
density and structuredness and that are challenging for book drawing algorithms.
This excludes some previously used graph classes such as trees or complete graphs.
With our choices we aim to establish a set of benchmark graphs that will also serve as a basis for
future investigations on book drawings. 
% We refrained from testing our algorithms with real world graphs, as this would not allow for a
% systematic investigation of the performance of the algorithms based on graph structure. Testing the
% algorithms with  graphs stemming from  applications of book embedding (e.g., RNA folding, knot
% theory) is an interesting research direction. \todo{added sentence about real world graphs}

\begin{description}
\item[Random.] We use \emph{random graphs} (Erd\H{o}s-R\'enyi model) with linear density
$a$, i.e., $n$-vertex graphs with $an$ edges for $a = 2, \ldots, 10$, and with quadratic density in
terms of $n$, i.e., edge probabilities $p$.
\item[Topological planar.] To generate $n$-vertex \emph{triangulated planar graphs}, we used a random edge-flip
walk of length $n^3$ in the space of planar triangulations with a random Apollonian network as starting point.
Known bounds suggest that $n^3$ is a suitable and still practical length \cite{MST99}.
\item[Topological $1$-planar.] We generated 1-planar graphs by augmenting the 4-cycles in our
planar triangulations with diagonals in a random order. This yielded on average $93\%$
of the maximal number of edges in a 1-planar graph. 
\item[Geometric $k$-planar.] 
For $k$ from zero to four, we generated $k$-planar graphs as follows. 
Taking a random set of points in the plane, we sort them lexicographically, and then add an edge
from a vertex (processed in sorted order) to an already processed vertex (in reversed order) only if
the segment connecting them would not create more than $k$ crossings in the current drawing. 
This process achieved on average $85\%$ of the maximal number of edges.
\item[$k$-tree.] A \emph{$k$-tree} is a recursively defined graph that is formed by starting with
a $k$-clique and adding vertices and connecting them to all vertices of a $k$-clique
of the current graph. We used this process to construct $k$-trees.
\item[Hypercube.] We used \emph{hypercubes} $Q_d$ of dimension $d$. They have $n = 2^d$ vertices and
$m = \frac{1}{2}nd = \frac{1}{2}n \log n$ edges. Their book thickness is $d
-1$~\cite{HKT89}.
\item[Cube-connected cycles.] A \emph{cube-connected cycle} $CCC_d$ of dimension $d$ is a hypercube
 $Q_d$ with vertices replaced by cycles of length $d$. They have $n = d2^d$ vertices and $m = 1.5n$
 edges.
\item[Toroidal mesh.] A \emph{toroidal mesh} $C_i \times C_j$ is the product graph of two cycles of
length $i$ and $j$. It has $n = ij$ vertices and $m = 2n$ edges.
\item[3-toroidal mesh.] A \emph{3-toroidal mesh} $C_i \times C_j \times C_k$ is the product graph of
three cycles of length $i$, $j$ and $k$. It has $n = ijk$ vertices and $m = 3n$ edges.
\end{description}
The structuredness of these graph classes varies from very symmetric graphs, such as hypercubes and
toroidal meshes (we call them \emph{homogeneous}) to less homogeneous but still geometrically
structured graphs, such as $k$-trees and $k$-planar graphs (we call them \emph{structured}) to
finally \emph{random}, unstructured graphs.

%!TEX root = /Users/noelle/Documents/work/00-aktuell/bookdraw/BookEmbedding/sources/booksPaper.tex
\section{Evaluation}\label{se:evaluation}
\pdfbookmark[0]{Evaluation}{Evaluation}

In this section, we present the results of our experiments on the performance of the
heuristic algorithms presented in Sect.~\ref{se:algo} on the different benchmark graph classes
introduced in Sect.~\ref{se:design}. Our main focus in the evaluation is to analyze the relative
performance of the book drawing heuristics, based on the density and structuredness of
the graph classes, as well as the specified number of pages.

\subsection{Experimental Setup}
For each experiment, we used specific graphs, like hypercubes, 200 times or 200 graphs of the
same class. For each graph, in the data representation, the order of the vertices and adjacency
lists were randomized. The maximal number of pages considered in an experiment was either determined by
the book thickness of a graph (if known), or limited to the first number where the best
heuristic achieved less than ten crossings, or a 20 pages otherwise.

\subsection{Constructive Heuristics}
We first evaluate the constructive heuristics of \cref{se:construct} by considering
all possible combinations of VO and PA heuristics. In previous experiments we observed that
the right combination of them is crucial~\cite{Klawitter16}. We thus refrained from testing them
independently as done by Satsangi et al.~\cite{SSS13}.
By plotting the number of produced crossings for all the heuristic combinations and various graph
classes, we observed that the parameters density, number of pages, and structure have a
significant impact.   
To analyze how the performance depends on these three factors we consider each graph class
individually, grouping them into homogeneous, structured and random graphs.

\begin{figure}[tbp]
\subfloat[Hypercube $Q_7$.\label{fig:regular:details:hypercube}]{%
     \includegraphics[valign=t,width=0.48\textwidth]{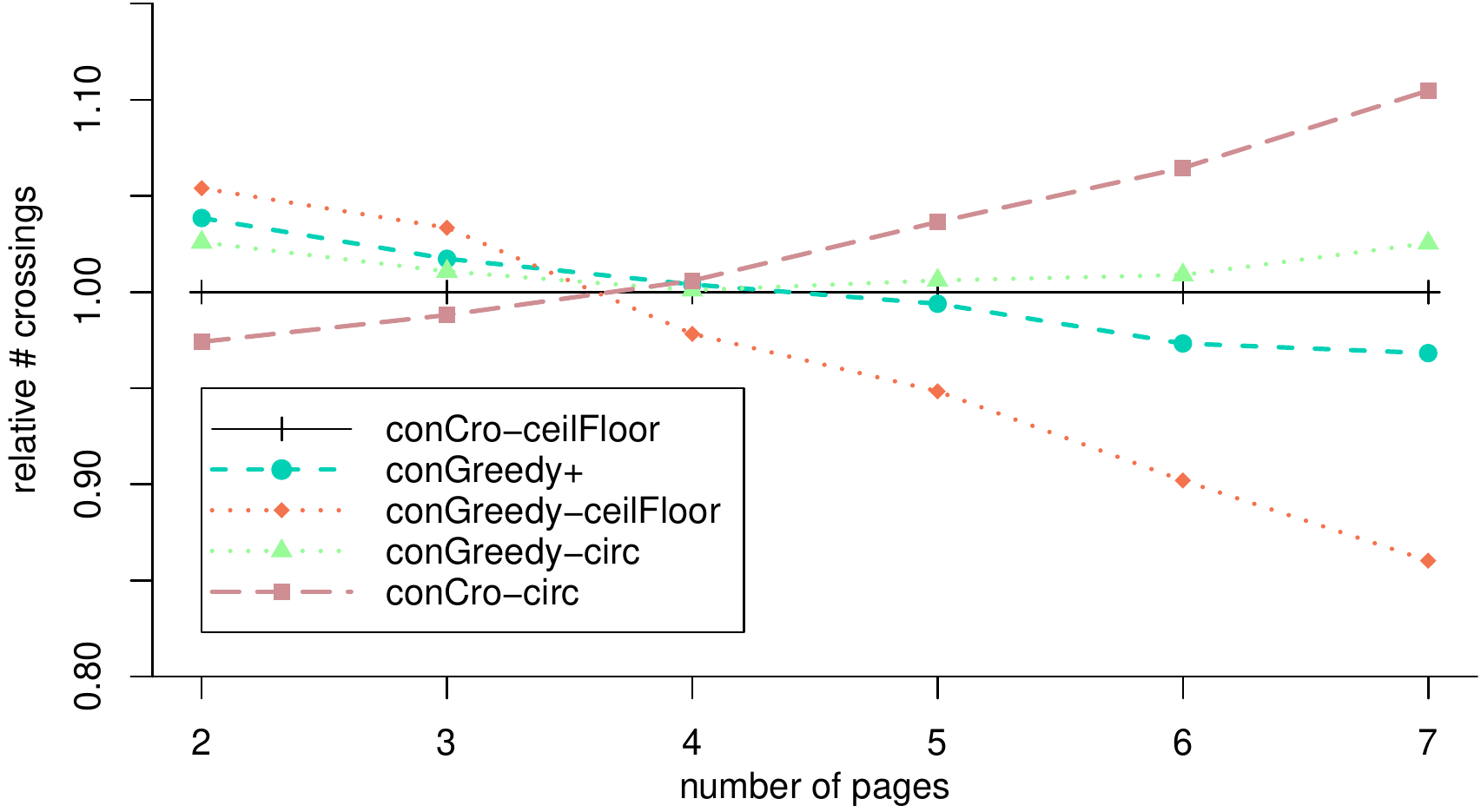}
	\vphantom{\includegraphics[width=0.48\textwidth,valign=t]{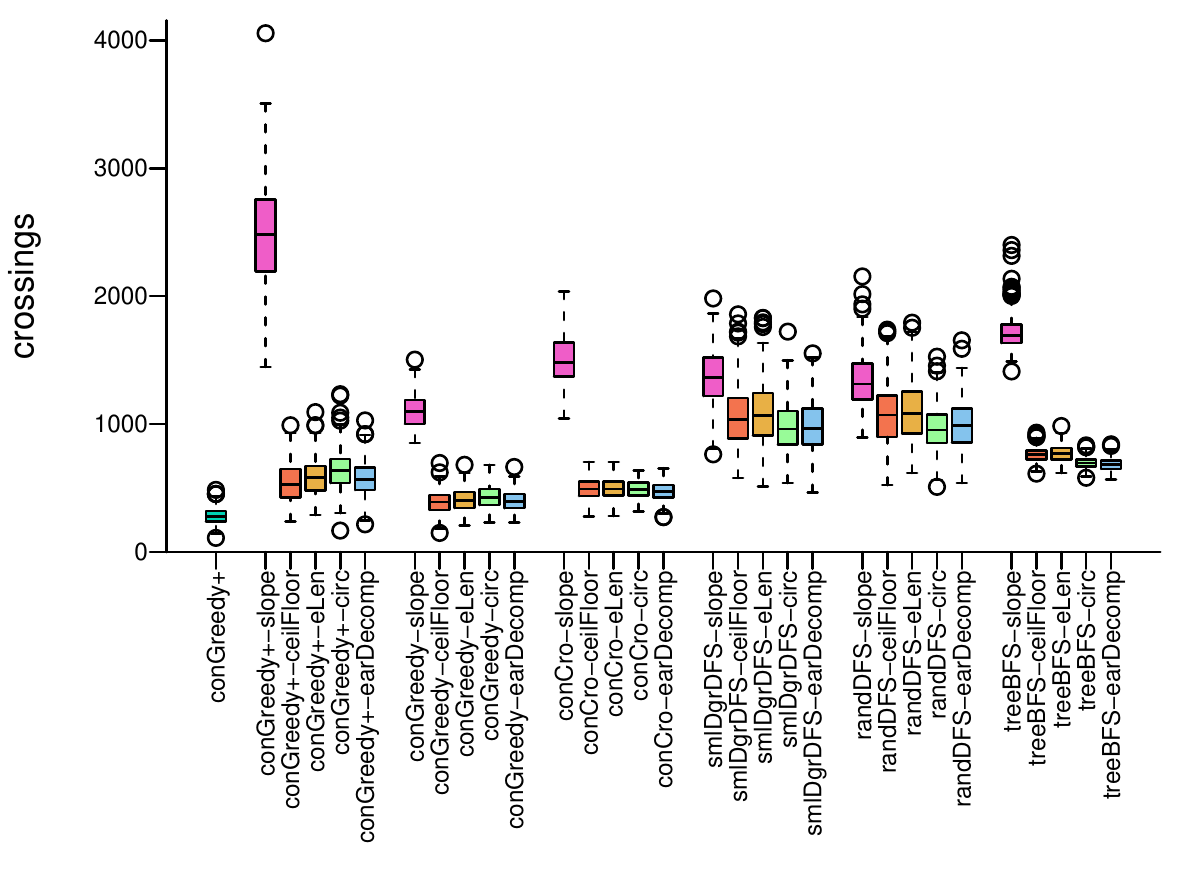}}%
}
\hfill 
\subfloat[$C_{16} C_{16}$, 3 pages.\label{fig:regular:details:toroidal}]{%
      \includegraphics[valign=t,width=0.48\textwidth]{C16C16-ALL-k3-boxplot} 
}
\caption{Number of crossings of the heuristics (a) with respect to \ConCro-\CF (lower means less)
for $Q_7$ depending on the number of pages, and (b) in absolute values for the toroidal mesh $C_{16}
C_{16}$ and three pages.} 
\label{fig:regular:details}
\end{figure}

\paragraph{Homogeneous graphs.}
\begin{wrapfigure}[14]{r}{.65\textwidth}
	\vspace{-.5cm}
	\includegraphics[width=0.65\textwidth]{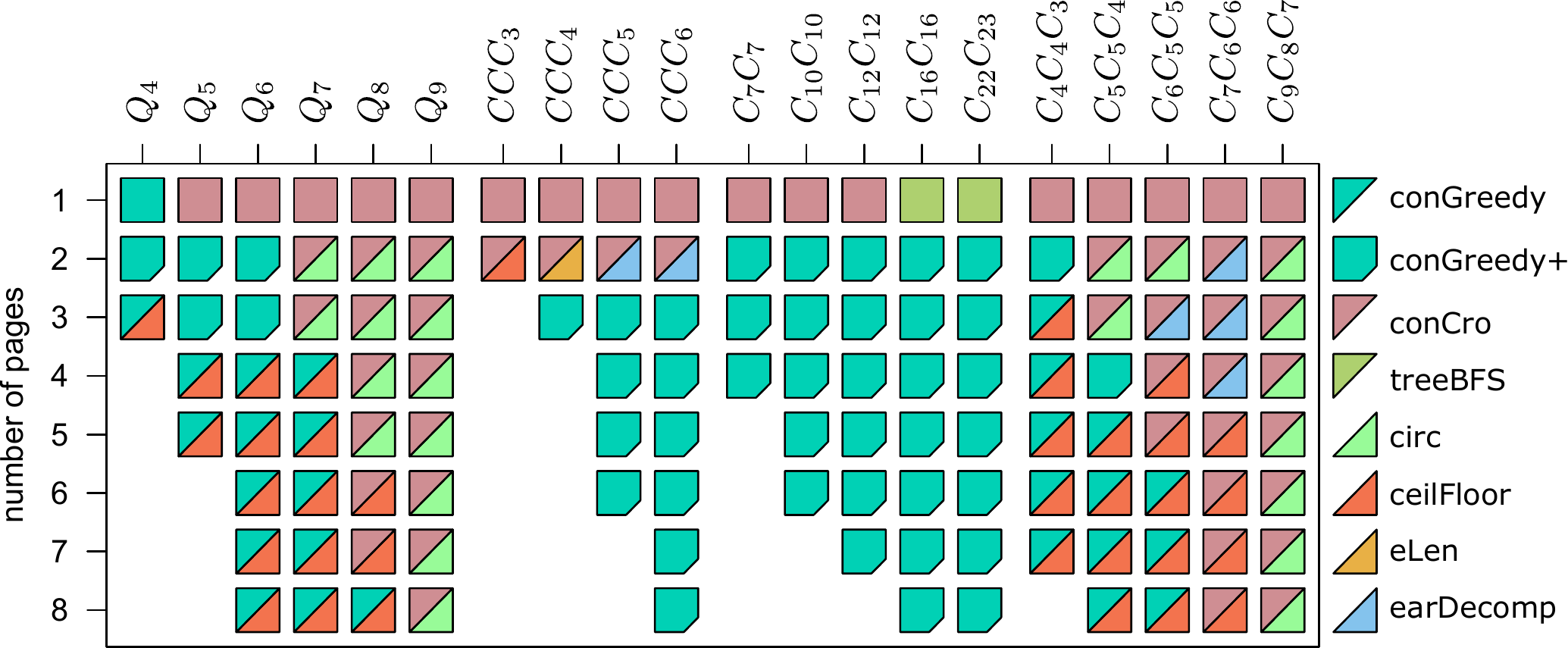}
	\caption{Tile diagram for homogeneous graphs. One tile represents the heuristic or heuristic
	combination that achieved the best mean of crossings for the specific number of pages (row) and
	graph (column).}
	\label{fig:tiles:regular}
\end{wrapfigure}
\mbox{}\newline\Cref{fig:tiles:regular} shows the best heuristics on the hypercubes $Q_4$ to $Q_9$
and on different toroidal meshes.  
For $Q_4$, having book thickness $d- 1 = 3$, \CONGREEDY-\CF\\ could almost achieve its book
embedding, however as the dimension increases the performance of the heuristics gets worse.
The same holds for toroidal meshes and cube-connected cycles, which both have book thickness
three~\cite{Klawitter16,ShTa10}. 
As we can observe by more detailed analysis (\cref{fig:appendix:all:cube} and
\cref{fig:appendix:all:tori}), all heuristics have on average more than a hundred crossings for
hypercubes and toroidal meshes on book thickness many pages.
For example, as shown in \cref{fig:regular:details:toroidal}, the best heuristics have on
average more than 250 crossings for $C_{16} C_{16}$ on three pages. We suspect that the book
thickness of 3-toroidal meshes is constant, and most likely below $8$. If this holds, the
performance of the heuristics is also poor for this graph class.
The best performing heuristics (refer to \cref{fig:tiles:regular}) for hypercubes are \CONGREEDY-\CF
and \ConCro-\CIRCULAR. For 3-toroidal meshes \ConCro-\CF is also often the winner. For
toroidal meshes and cube-connected cycles \FCONGREEDY performs best.

\cref{fig:regular:details:hypercube} illustrates the vertical dimension of the tile diagram for
$Q_7$. It shows the performance of heuristics depending on the number of pages. 
We see that the changes are smooth and that, however, for a high number of pages 
\CONGREEDY-\CF is substantially better than \ConCro-\CIRCULAR and the other heuristic.
\Cref{fig:regular:details:toroidal} shows the results for all heuristics on $C_{16} C_{16}$
and three pages. Here \FCONGREEDY performed best. Overall, we see that the choice of the VO heuristics
has higher significance than PA, except if in combination with \SLOPE. It is also interesting that \treeBFS
performs significantly better than the other search based heuristics. Similar results appear for the
other homogeneous graphs and diagrams can be found in the appendix
(see \cref{fig:appendix:lines,fig:appendix:all:cube,fig:appendix:all:tori}). 

\begin{figure}[tbp]
\centering
\includegraphics[width=0.9\textwidth]{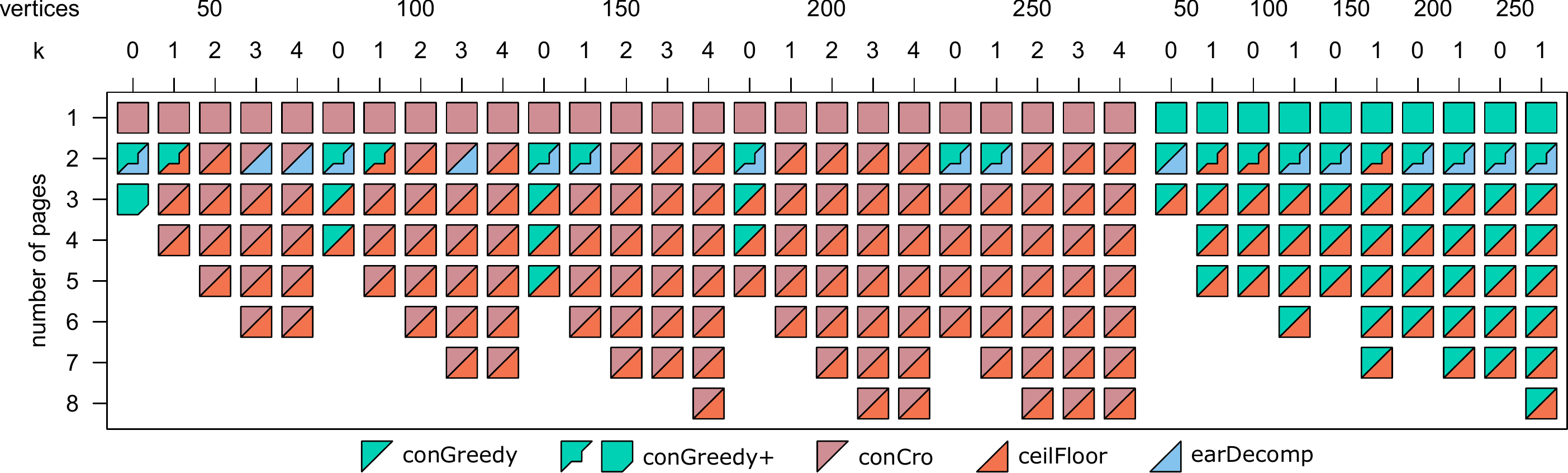}
\caption{Tile diagram for $k$-planar graphs, geometric (left) and topological (right).}
\label{fig:tiles:kplanar} 
\end{figure}

\paragraph{Structured graphs.}
The structured graphs that we investigate are the $k$-planar graphs and $k$-trees. 
\Cref{fig:tiles:kplanar} presents the best performing heuristics for geometric $k$-planar graphs and 
topological planar and $1$-planar graphs. We observe that the difference in the
structure of these graphs is crucial for the performance of the heuristics. For topological
planar graphs \CONGREEDY-\CF dominates, while for geometric $k$-planar
graphs \ConCro-\CF is ahead in the majority of the cases. The PA heuristic \EAR performs well for
two pages.

\begin{figure}[tbp]
\subfloat[Topological planar, 4 pages.\label{fig:planar:lines}]{%
      \includegraphics[width=0.48\textwidth,valign=t]{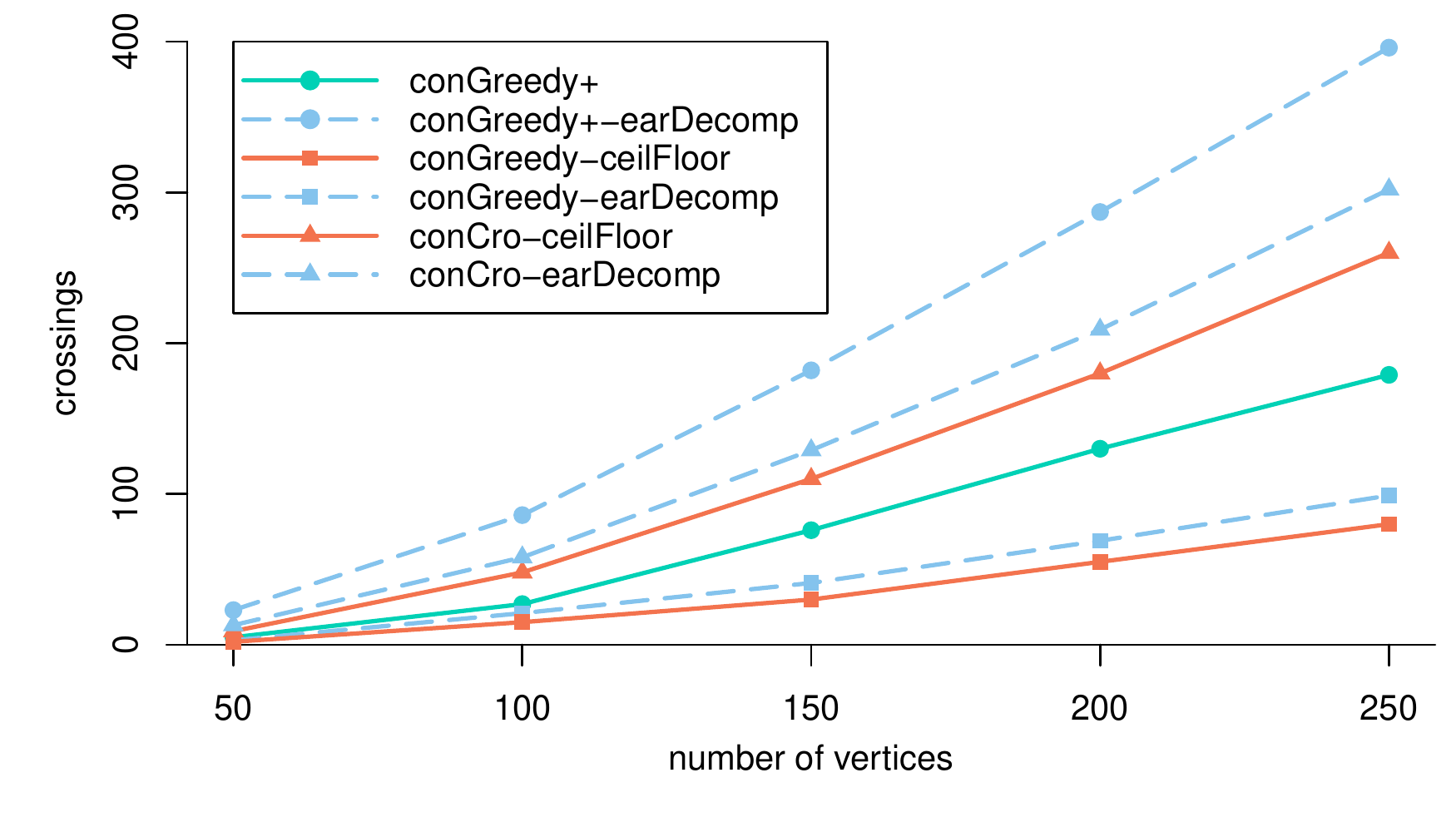}
\vphantom{\includegraphics[width=0.48\textwidth,valign=t]{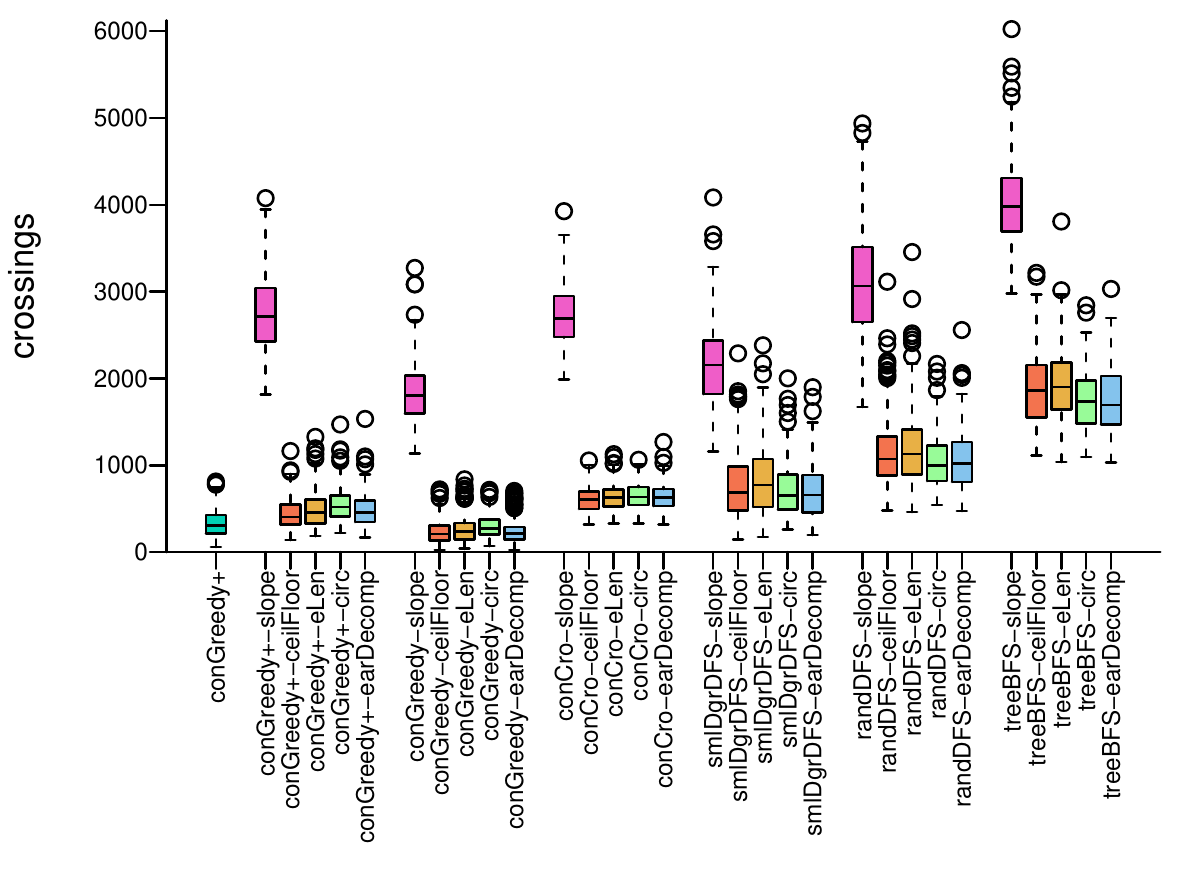}
}}
\hfill
 \subfloat[Topological planar, $n = 250$, 3 pages.\label{fig:planar:boxplot}]{%
 \includegraphics[width=0.48\textwidth,valign=t]{triangulation-ALL-n250-k3-boxplot}
 }
\caption{Performance of the heuristics on topological planar graphs.}
\label{fig:planar}
\end{figure}

\Cref{fig:planar:lines} shows the performance of the heuristics for topological planar graphs
plotted as a function of the number of vertices for four pages, the upper bound for the book
thickness of planar graphs~\cite{Yan89}.
The leading heuristic \CONGREEDY-\CF is not close to the optimum of zero, but 
achieves, for example, for graphs with 250 vertices on average 62 crossings.
In contrast to the homogeneous graphs, we see in \cref{fig:planar:boxplot}, that the two DFS-based heuristics perform
better on topological planar graphs, while \treeBFS performs worst.
Similar observations can be made for $k$-planar graphs (see \cref{fig:appendix:all:planar,fig:appendix:all:onePlanar}).

\begin{figure}[ht]
\centering
\includegraphics[width=1\textwidth]{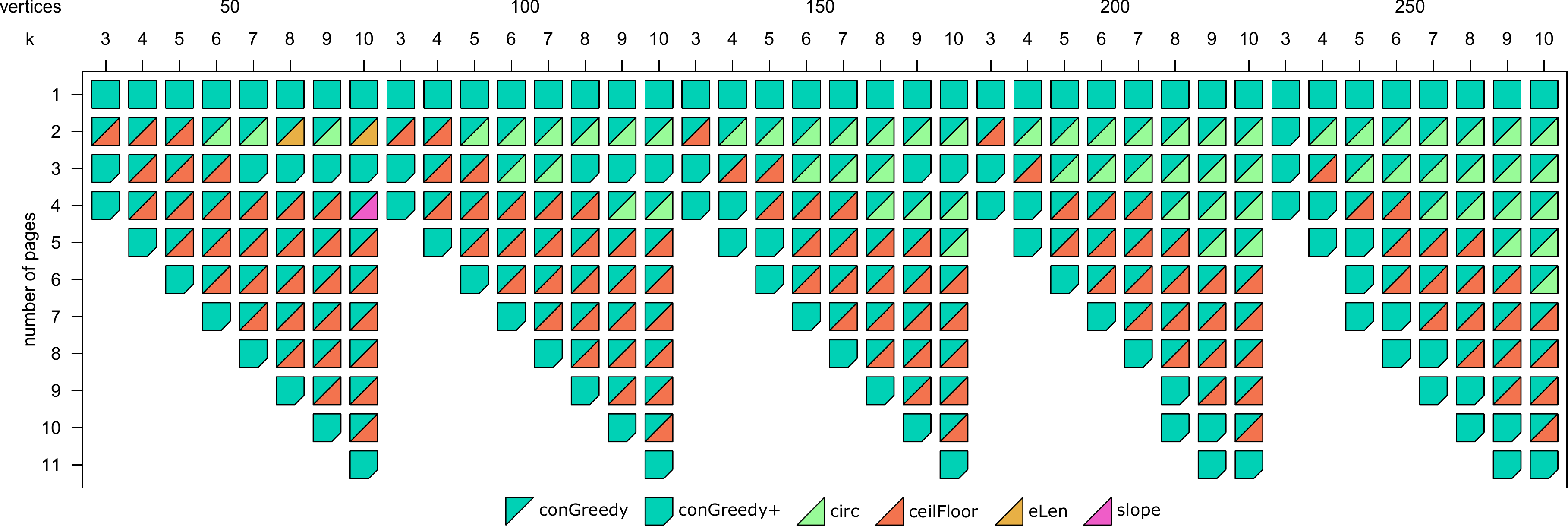} 
\caption{Tile diagram for $k$-trees.} 
\label{fig:tiles:ktree}
\end{figure}

\begin{figure}[tbp] 
\centering
  \includegraphics[width=0.6\linewidth]{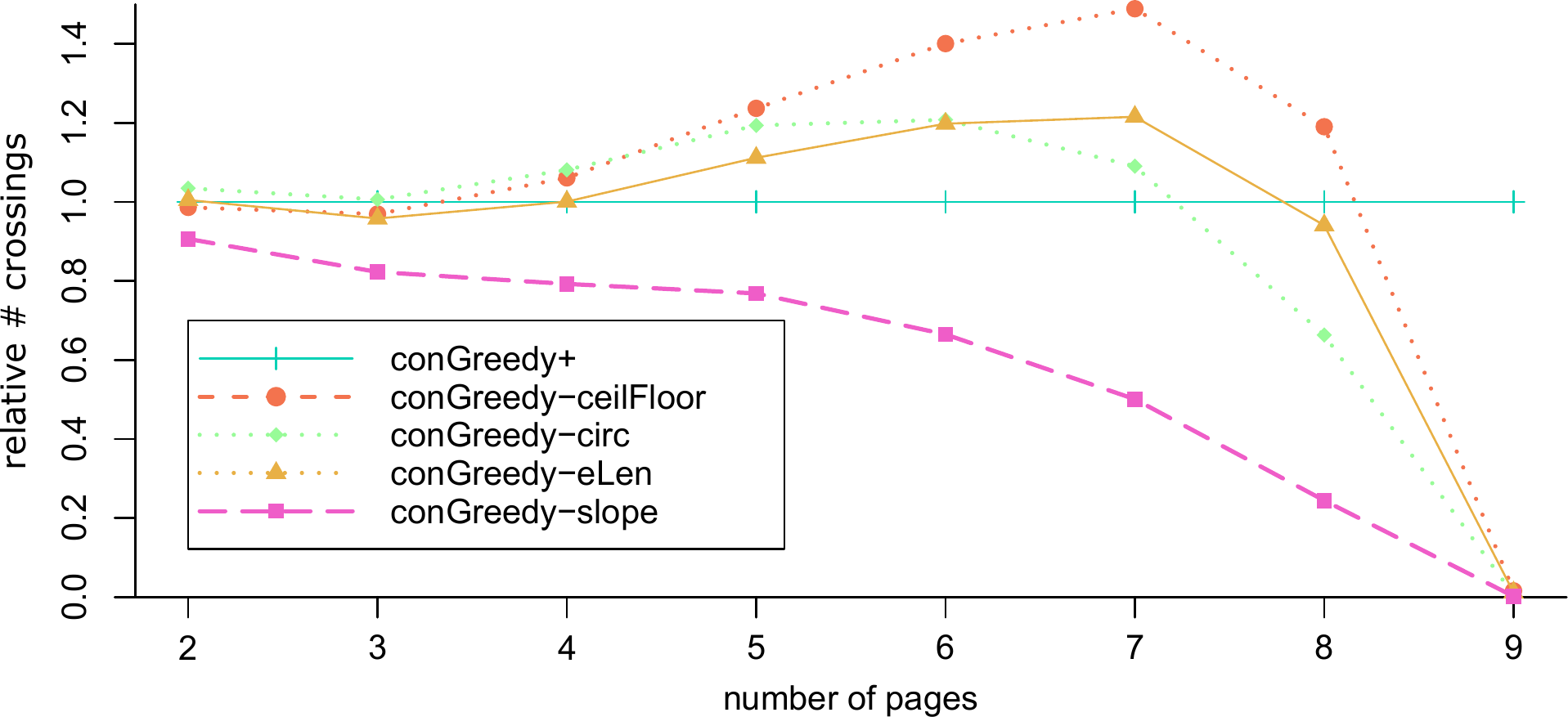}
  \caption{Performance of \FCONGREEDY relative to \CONGREEDY in combination with PA
  heuristics, for 8-trees, $n = 150$ and two to nine pages. A higher value means thus fewer
  crossings compared to \FCONGREEDY.}   
  \label{fig:ktreeTrend}
\end{figure}

\Cref{fig:tiles:ktree} shows the overview of the results for $k$-trees. The diagram is dominated
by \CONGREEDY-\CIRCULAR and \CONGREEDY-\CF. We note that the
book thickness of $k$-trees is at most $k + 1$~\cite{GH99,DW07} and then observe that \FCONGREEDY
achieves less than ten crossings on average for $k + 1$ pages. This becomes more apparent in
\cref{fig:ktreeTrend}, which shows the performance as a function of the 
number of pages for $8$-trees with 250 vertices.
We see that for a small number of pages, where \CONGREEDY-\CIRCULAR dominates, the performance of
all heuristics is comparable, while for more pages \CONGREEDY-\CF performs clearly better and
finally, on nine pages, \FCONGREEDY takes significant
lead.
 
\paragraph{Random graphs.}

\begin{figure}[htb]
\centering
\includegraphics[width=1\textwidth]{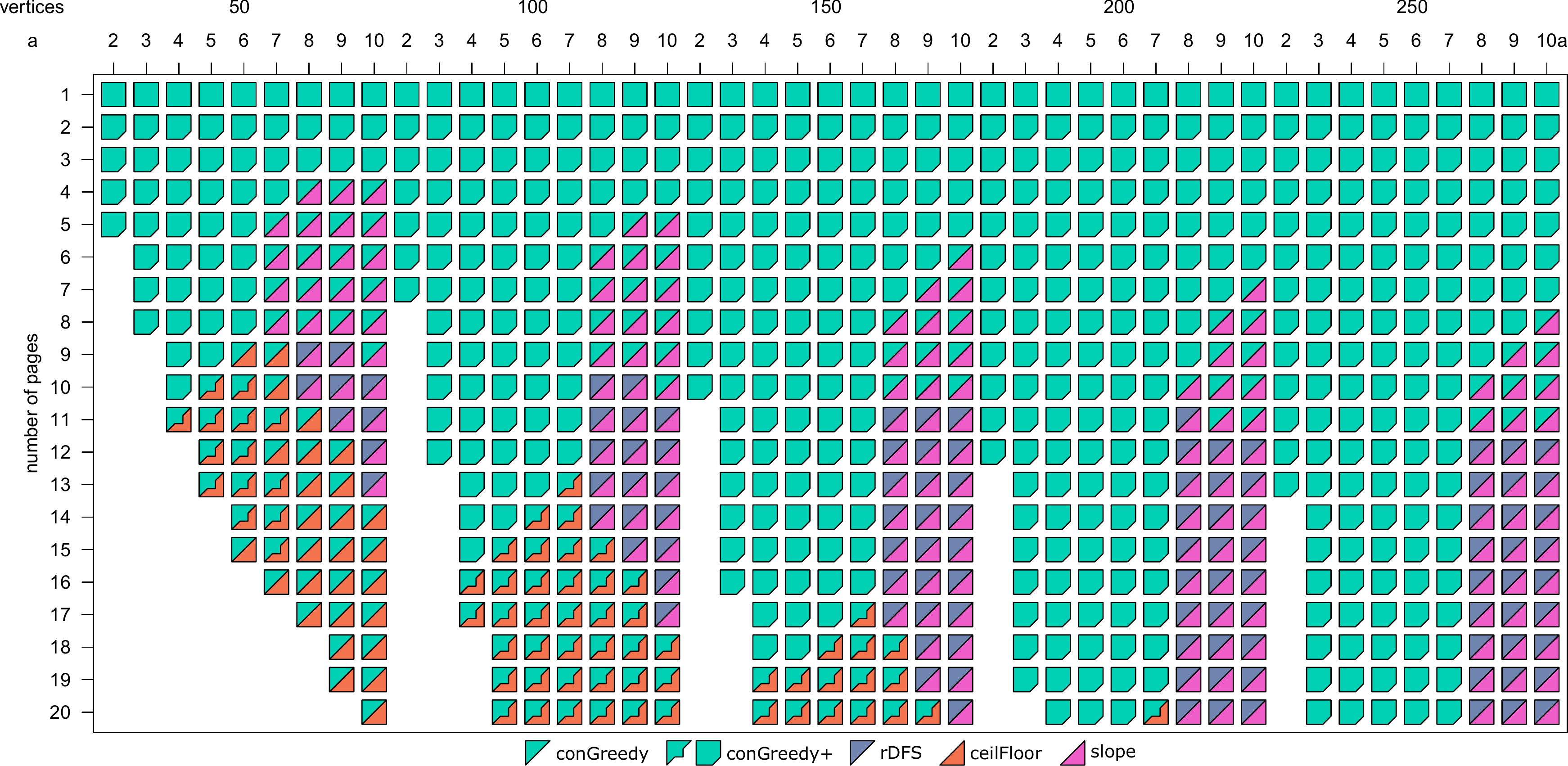}
\caption{Tile diagram for random graphs with linear density, i.e. roughly $an$ edges.}
\label{fig:tiles:random}
\end{figure}

The tile diagram in \cref{fig:tiles:random} for random graphs with linear number of edges shows again a clear pattern. 
Further investigation (see \cref{fig:appendix:all:randomL}) shows that the transition between the
best heuristics in \cref{fig:tiles:random} shifts smoothly along the number of vertices, pages and density. 
The heuristic \FCONGREEDY dominates for small number of pages and not too high density. 
However, if the number of pages is relatively high, we observe that \FCONGREEDY-\CF and \CONGREEDY-\CF perform best.
The differences between PA heuristics becomes more apparent for both
higher density and more pages. The performance of \SLOPE gets significantly better with higher
density, either with \CONGREEDY or \RDFS. The search based VO heuristics perform nearly equally, as
do the greedy VO heuristics with conGreedy however slightly in the lead.
 
The good performance of \SLOPE seems natural, as with high density, and thus more edges per page,
one edge with a slope different from other edges on the same page, is very likely to produce a lot of crossings.
\Cref{fig:tiles:randomQuadratic,fig:appendix:all:randomQuadratic} on random graphs
with quadratic density in the appendix illustrate this even further. In fact, de Klerk
\etal~\cite{KPS13} conjecture that on complete graphs \SLOPE finds an optimal solution for any number of pages.
They proved this for two and $\lfloor \frac n 2 \rfloor$ pages, with the latter being the book
thickness of complete graphs $K_n$~\cite{BK79}.

\subsection{Local Search Heuristics}
In this section, we evaluate the local search heuristics \GreedyAlt,
\GreedyCombi and \SA (the core of the algorithm that won the 2015 Graph
Drawing Contest), described in \cref{se:localsearch}.
Recall that \FCONGREEDY is a combined constructive heuristic that
considers VO and PA simultaneously, and as seen above often outperforms other heuristic combinations. 
With \GreedyCombi, we extended the idea of \FCONGREEDY to a local search heuristic, which does
multiple rounds on all vertices and edges, until a local minimum is found.

We tested the local search heuristics, similarly to the constructive heuristics, on
graphs of different sizes, densities, structure, and with different numbers of pages
(see~\cref{fig:opti:results}, as well as \cref{fig:appendix:opti1} in the appendix).
Here our findings are more clear-cut. In all our experiments \GreedyCombi performed best, followed
by \GreedyAlt.
The heuristic \SA had sometimes difficulties to improve the given book
drawings, and performed worse than \GreedyAlt. 
The good performance of \GreedyCombi comes with a high trade-off in running time compared to
\GreedyAlt. However, our implementation of \SA was even slower\footnote{We implemented our algorithms in Java and tested on a standard home computer
(Intel\textsuperscript{\textregistered} Core\textsuperscript{TM} i5-6600, 3.3GHz, 8GB RAM and
Windows OS).}.

\begin{figure}[htbp]
% \subfloat[Topological planar, $n = 250$, 2 pages.\label{fig:opti:results:planar}]{%
%       \includegraphics[width=0.3\textwidth]{opti/planar-n250-k2}
% }
% \hfill
 \subfloat[Topological 1-planar, $n = 250$, 3 pages.\label{fig:opti:results:onePlanar}]{%
 \includegraphics[width=0.3\linewidth]{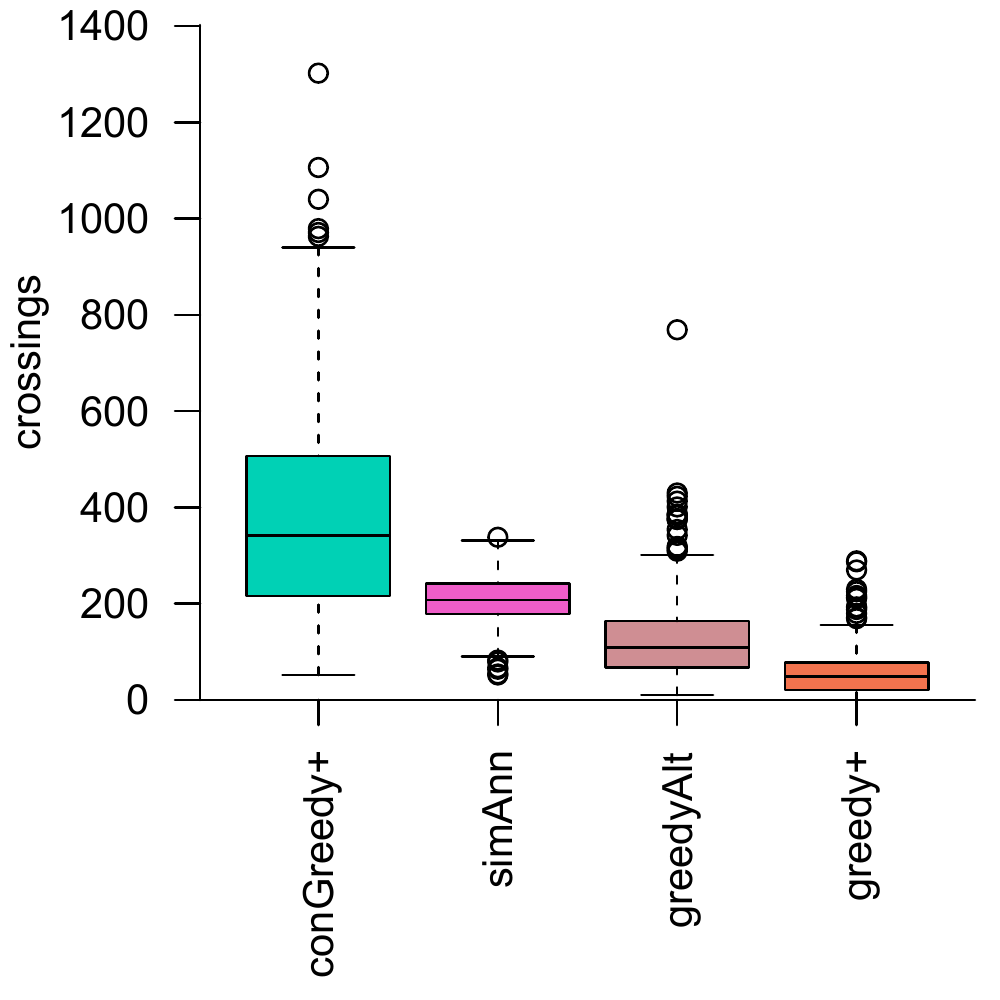}
 }
%  \hfill
%  \subfloat[4-tree, $n = 250$, 3 pages.\label{fig:opti:results:fourTree}]{%
%  \includegraphics[width=0.3\linewidth]{opti/4Tree-n250-k3}
%  }
%  \hfill 
%  \subfloat[Random (linear density 4), $n = 250$, 3
%  pages.\label{fig:opti:results:planar}]{%
%       \includegraphics[width=0.3\textwidth]{opti/randomL4-n250-k3}
% }
\hfill
 \subfloat[Hypercube $Q_7$, 6~pages.\label{fig:opti:results:onePlanar}]{%
 \includegraphics[width=0.3\linewidth]{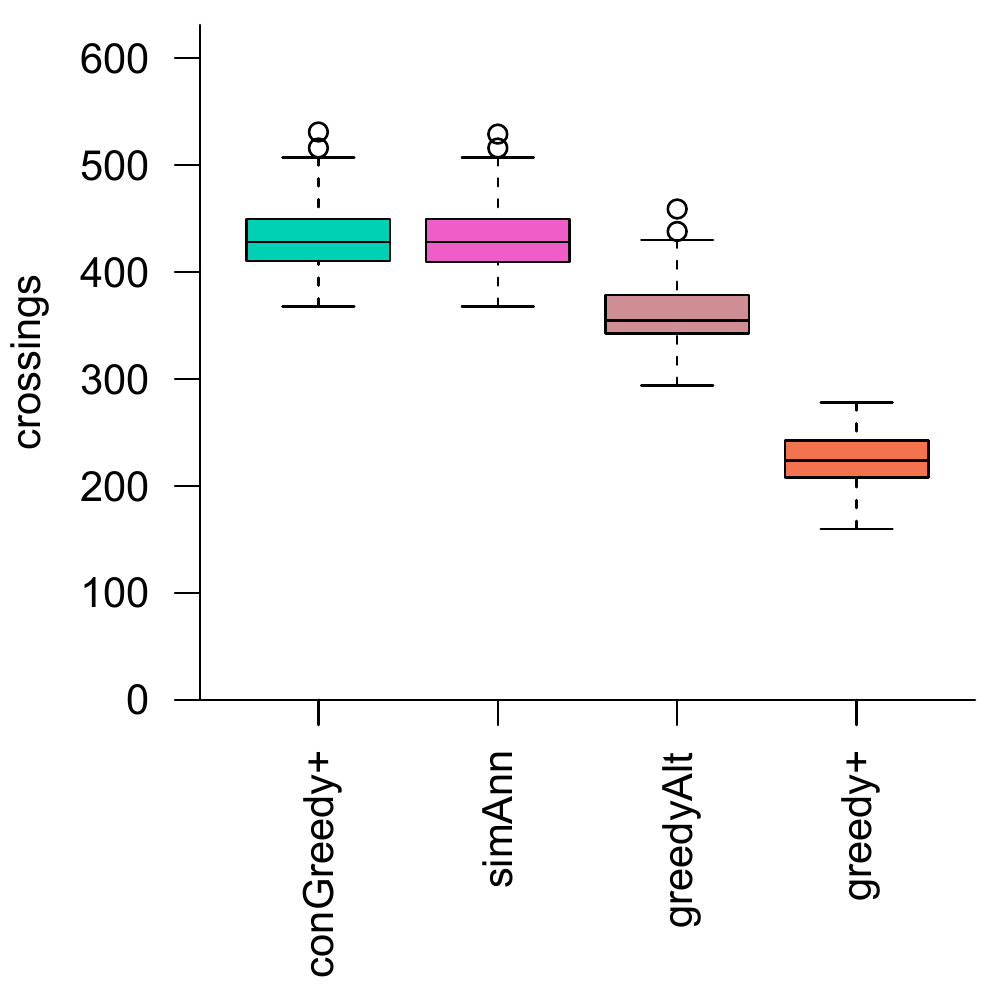}
 }
 \hfill
 \subfloat[Random (quadratic density 0.5), $n = 100$, 6~pages.\label{fig:opti:results:fourTree}]{%
 \includegraphics[width=0.3\linewidth]{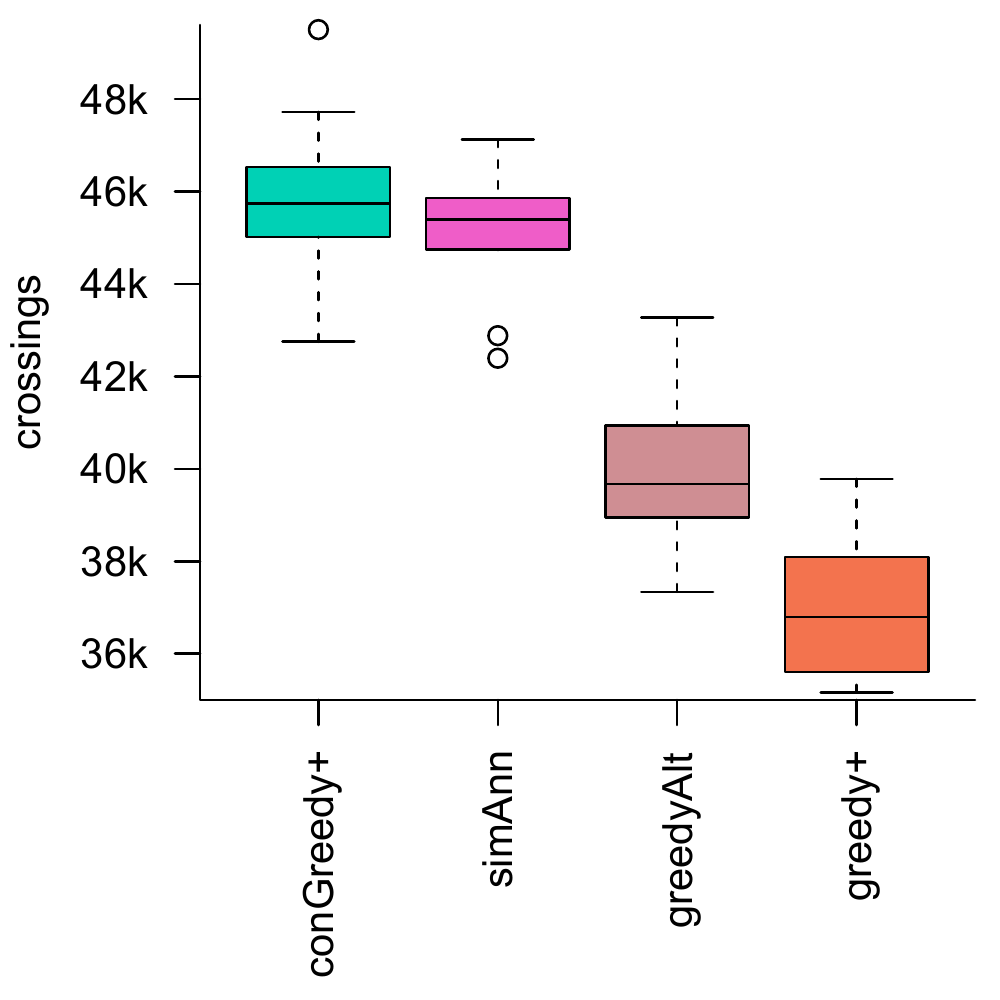}
 }
\caption{Performance of the local search heuristics on graphs of various classes.}
\label{fig:opti:results}
\end{figure}

%!TEX root = /Users/noelle/Documents/work/00-aktuell/bookdraw/BookEmbedding/sources/booksPaper.tex

\section{Discussion and Conclusions}\label{se:concl}
\pdfbookmark[0]{Conclusions}{Conclusions}

% Dependence on factors
In our experiment, we investigated the relative performance of the heuristics presented in
\cref{se:algo}.
We saw that the choice of the best constructive heuristic depends on several factors: 
density of the graphs, their structural properties and the number of pages. 
We could also saw that the performance strongly depends on the selected combination of VO and PA
heuristics.

% quality in general
We observed that constructive heuristics are mostly unable to achieve optimal results. 
For graph classes with known book thickness, the constructive heuristics could achieve very
low crossing numbers only on $k$-trees.
For homogeneous and structured graphs the results were far from optimal.
We also observed that whenever the constructive
heuristics performed poorly, the results of local search heuristics were also far from optimal.
This fact, however, is not surprising, as even for trees, starting from random
configuration, local search heuristics cannot achieve a book embedding~\cite{Klawitter16}.
Since most crossing numbers for the considered graph classes are unknown, we could not further
investigate the relative performances in these cases.
 
% particular results on heuristics
The constructive heuristics  \CONGREEDY  and its combined version \FCONGREEDY performed best most
of the times, but at a cost of higher running time. In several cases
the VO heuristic \CONCRO  performed better than \CONGREEDY, even though the former is just a
restricted version of the latter.
We observed that the \SLOPE heuristic performed well on dense graphs or in the case of a high ratio
of the number of edges to the number of pages, which complies with the earlier conjecture about the
power of \SLOPE to achieve optimal results for complete graphs.
We also saw that our new VO heuristic \treeBFS and PA heuristic \EAR perform best or comparably to
the other heuristics for homogeneous graphs and few pages.
Regarding the local search heuristics, \GreedyCombi performed significantly better than the other
local search heuristics on all graph classes, densities and number of pages. 
The simulated annealing algorithm performed even worse. 

% 	limitations
% When building our benchmark set we aimed to cover graphs with various properties in order to
% investigate carefully the relative performance of the algorithms.
% We did not aim to cover all previously used graph classes in past book drawing experiments.
% Rather we tried to draw a big picture by considering as many heuristics as possible. 
% Optimizing specific implementations of the heuristics was out of the scope of this work.

Our experiment can be extended into several directions that were beyond the limit of this paper:
other theoretically and practically interesting graphs or graph classes, concentration on particular
graph classes, more sophisticated implementations of the heuristics that would make it possible to test
larger graphs, and finally a more detailed analysis of the results.

With respect to the tested heuristics, closer investigation would be necessary to understand why
\CONCRO performs better than \CONGREEDY in several cases.
It would also be of interest to see whether the performance of \treeBFS on regular graphs could be
further improved by derandomizing the way the BFS tree is constructed.
Concerning local search heuristics, an interesting open question is whether the optimisation of the
VO or the PA is more influential on the overall performance of a heuristics.

\newpage
\bibliographystyle{abbrvurl}
\bibliography{bibliography}

\newpage
\appendix
\section{Appendix}\label{se:appendix}

This appendix contains additional figures to give more details of the performance of the heuristics. They
show that our reported finding were not specific to single number of pages, graph classes, or
graph sizes. The captions of the figures contain comments recalling some of these findings.

% lines
\begin{figure}[htb]
\centering
\subfloat[Topological 1-planar, 4 pages.\label{fig:appendix:lines:kPlanar:topOnePlanar}]{% 
     \includegraphics[width=0.45\textwidth]{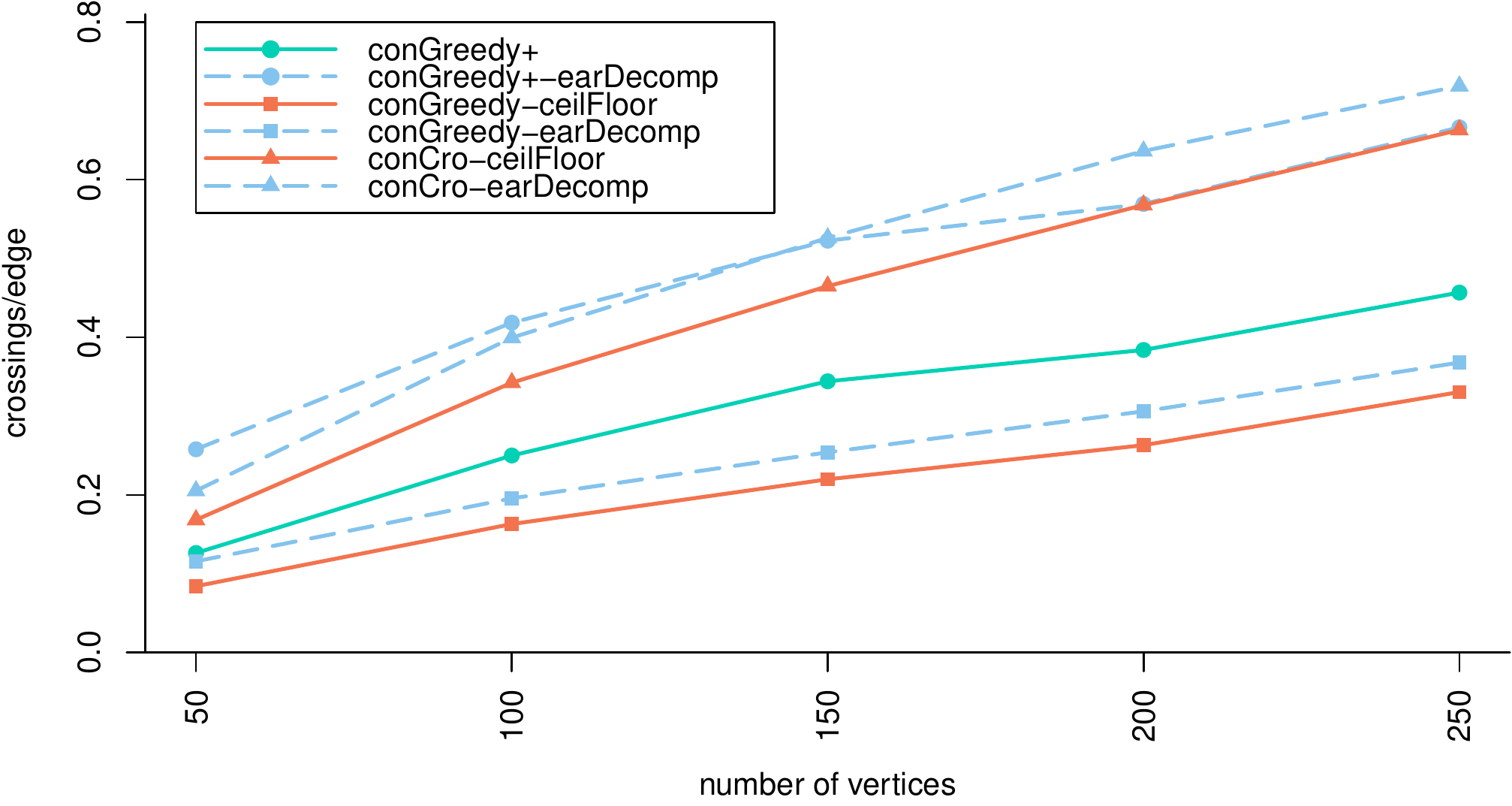} 
}\hfill
\subfloat[$k$-planar graphs, 2 pages.\label{fig:appendix:lines:kPlanar:twoPages}]{%
      \includegraphics[width=0.45\textwidth]{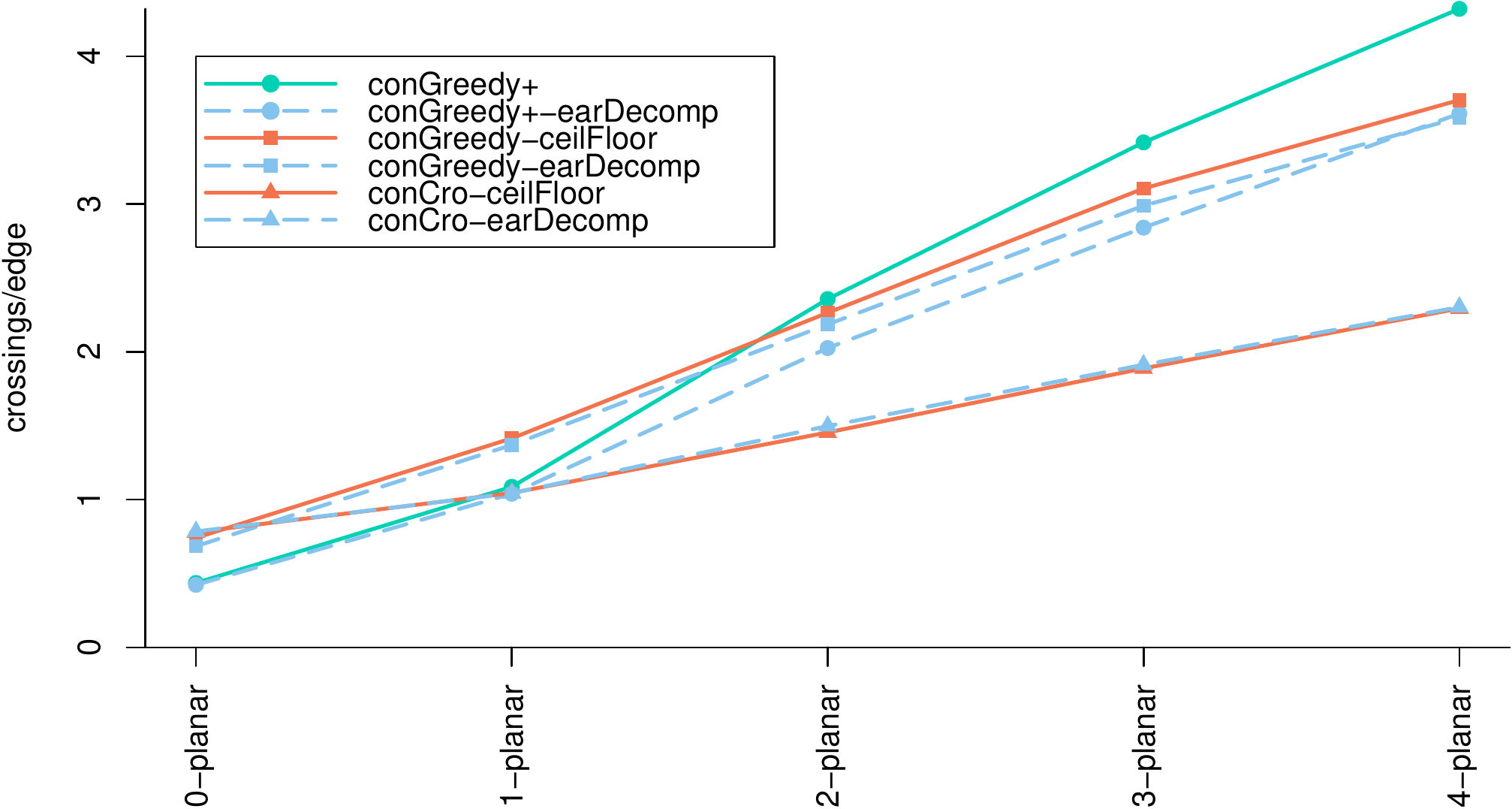}  
}\\
\subfloat[$k$-planar graphs, 150 vertices, 3 pages.\label{fig:planar:lines:kPlanar}]{% 
     \includegraphics[width=0.45\textwidth]{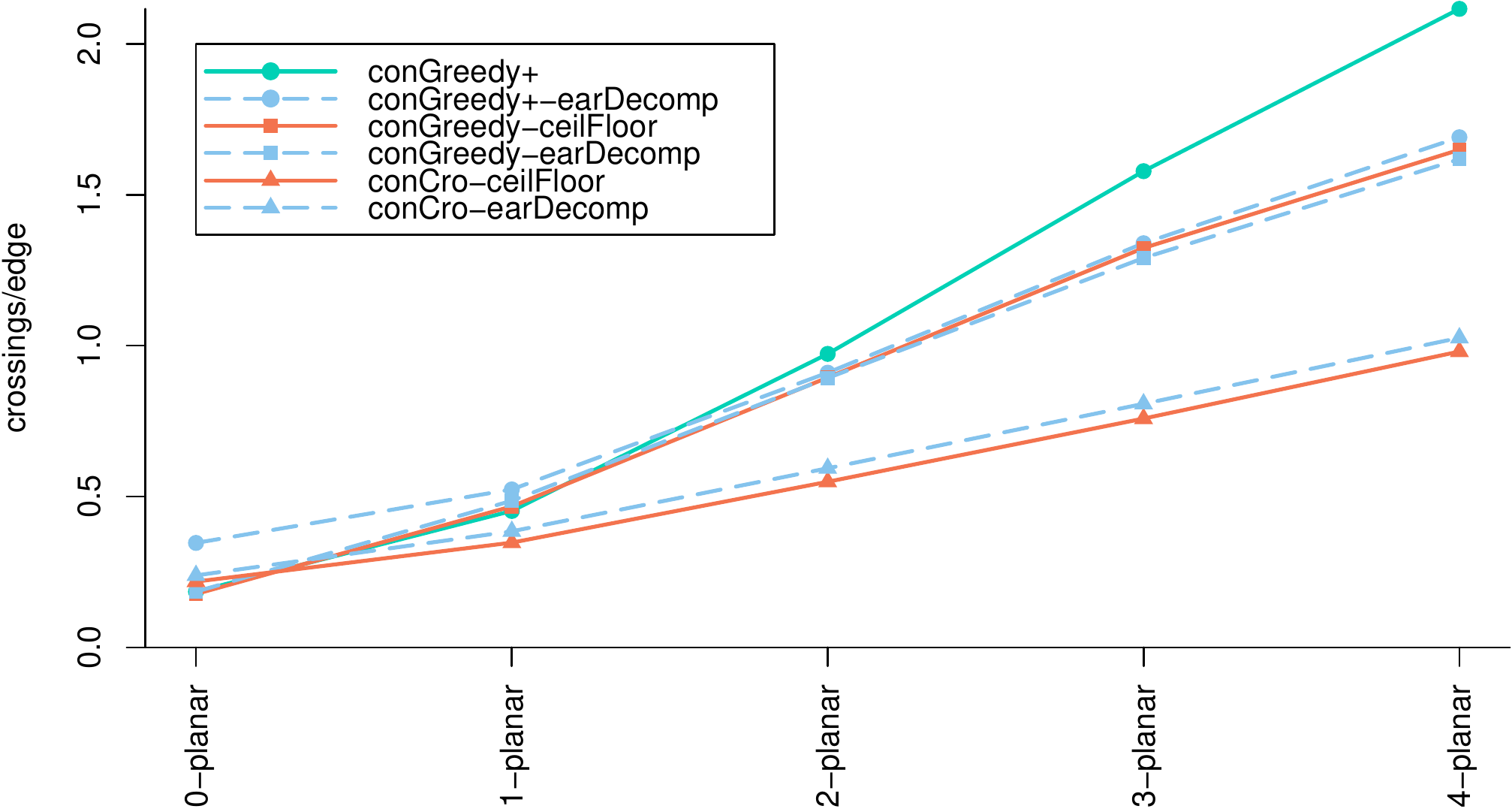}  
}
\hfill
\subfloat[3-toroidal meshes, 4 pages.\label{fig:appendix:lines:toroidal}]{%
      \includegraphics[width=0.45\textwidth]{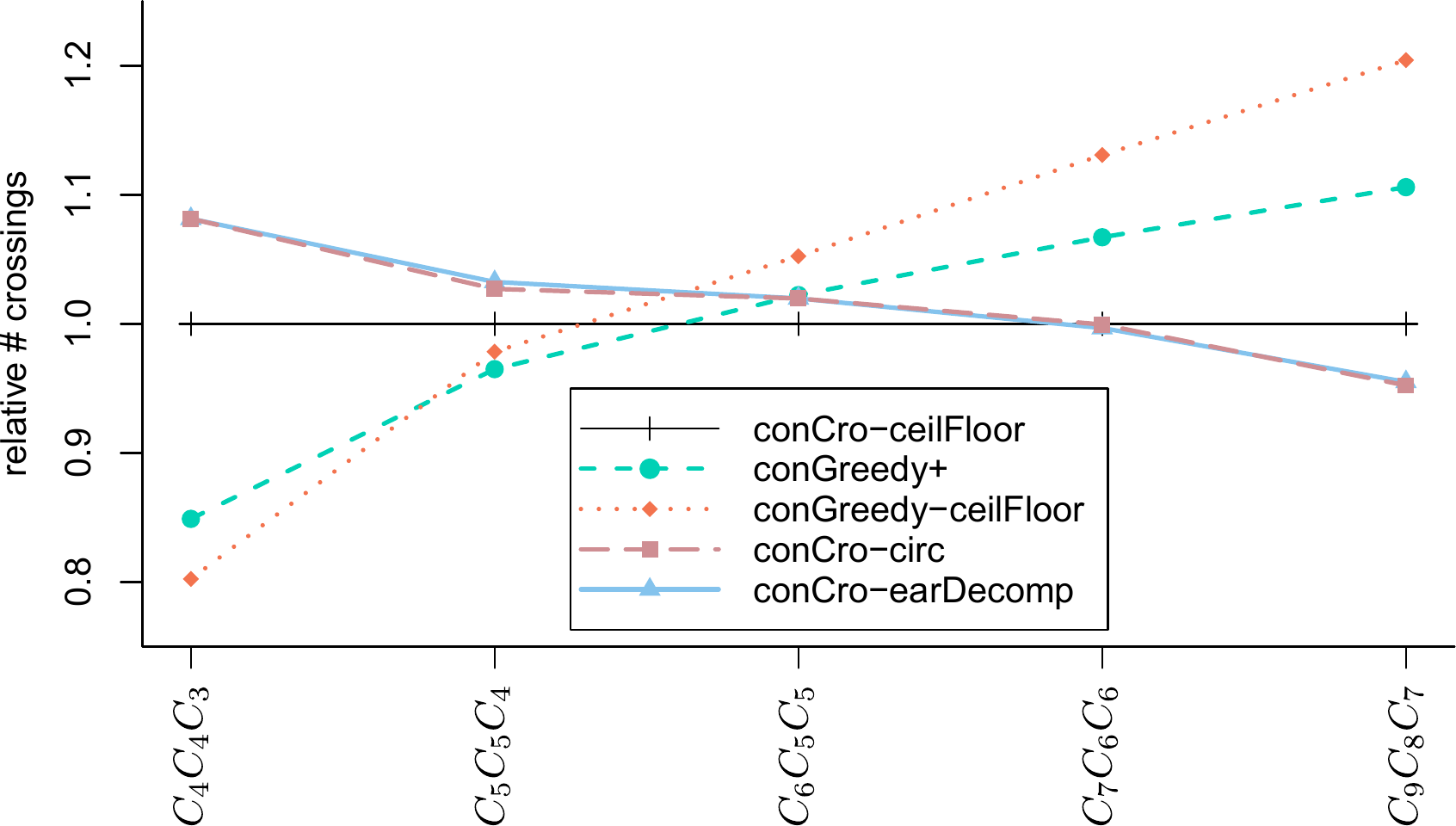} 
}
 \caption{Diagrams showing various changes of performance on different graphs.}
 \label{fig:appendix:lines} 
\end{figure}

% hypercube vs ccc
\begin{figure}[htb]
 \centering
 \subfloat[$Q_6$, 2 pages.\label{fig:appendix:all:cube:Q6k2}]{%
 \includegraphics[width=0.45\linewidth]{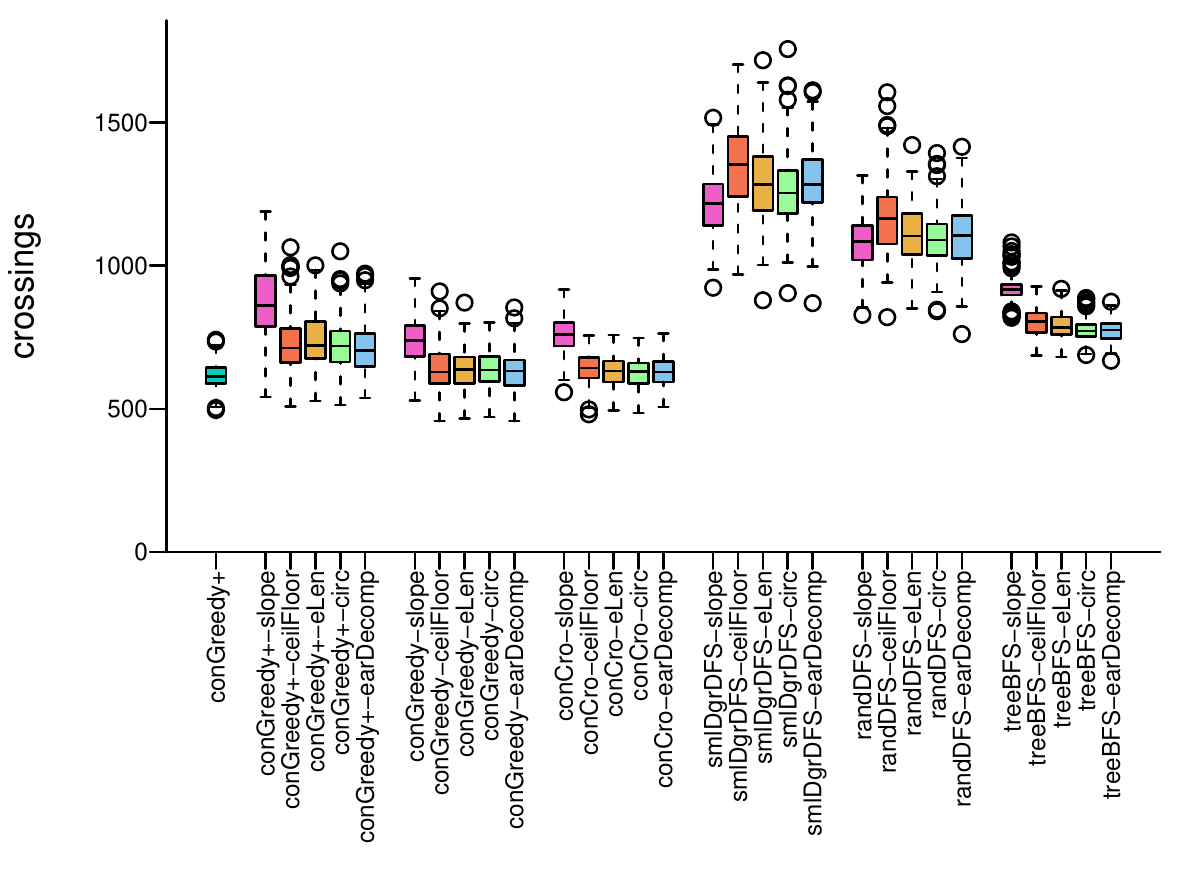}
 }\hfill
 \subfloat[$CCC_6$, 2 pages.\label{fig:appendix:all:cube:CC6k2}]{%
 \includegraphics[width=0.45\linewidth]{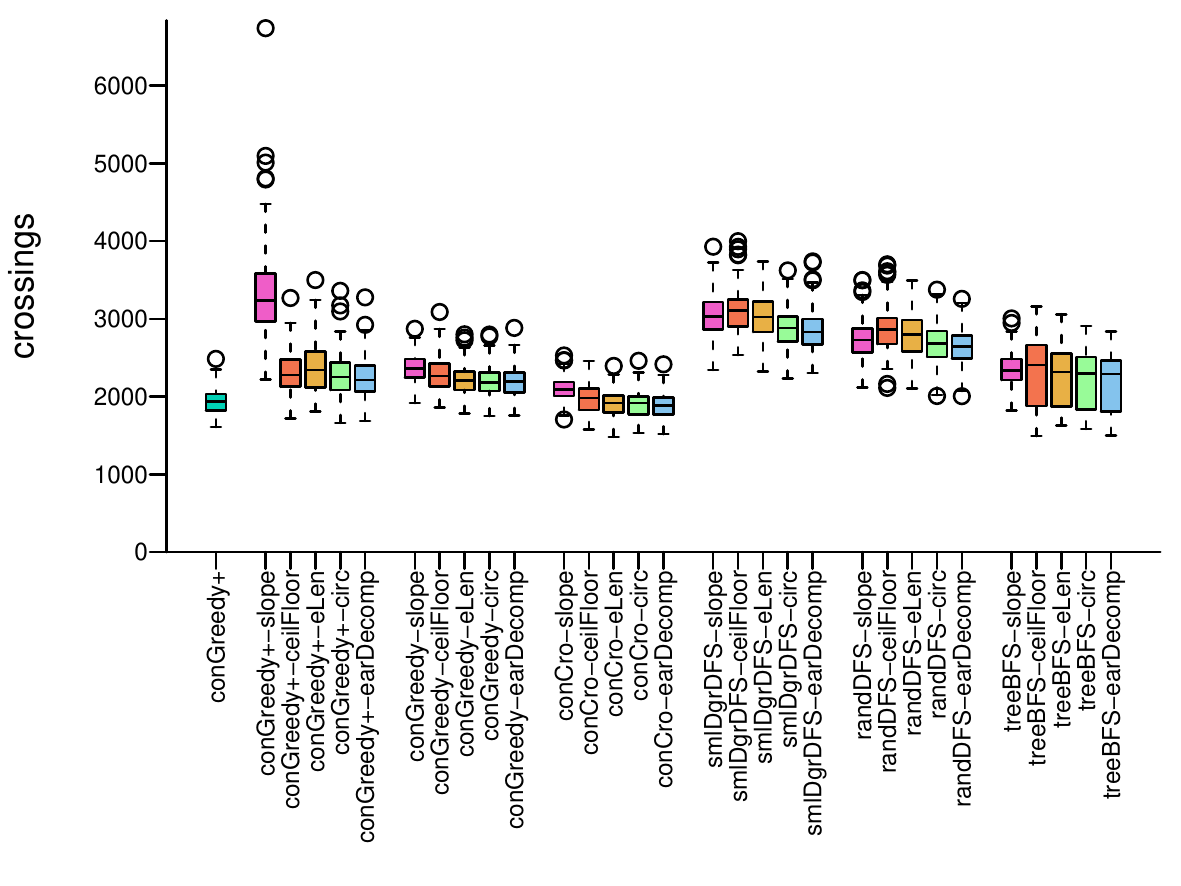}
 }\\
 \subfloat[$Q_6$, 5 pages.\label{fig:appendix:all:cube:Q6k5}]{%
 \includegraphics[width=0.45\linewidth]{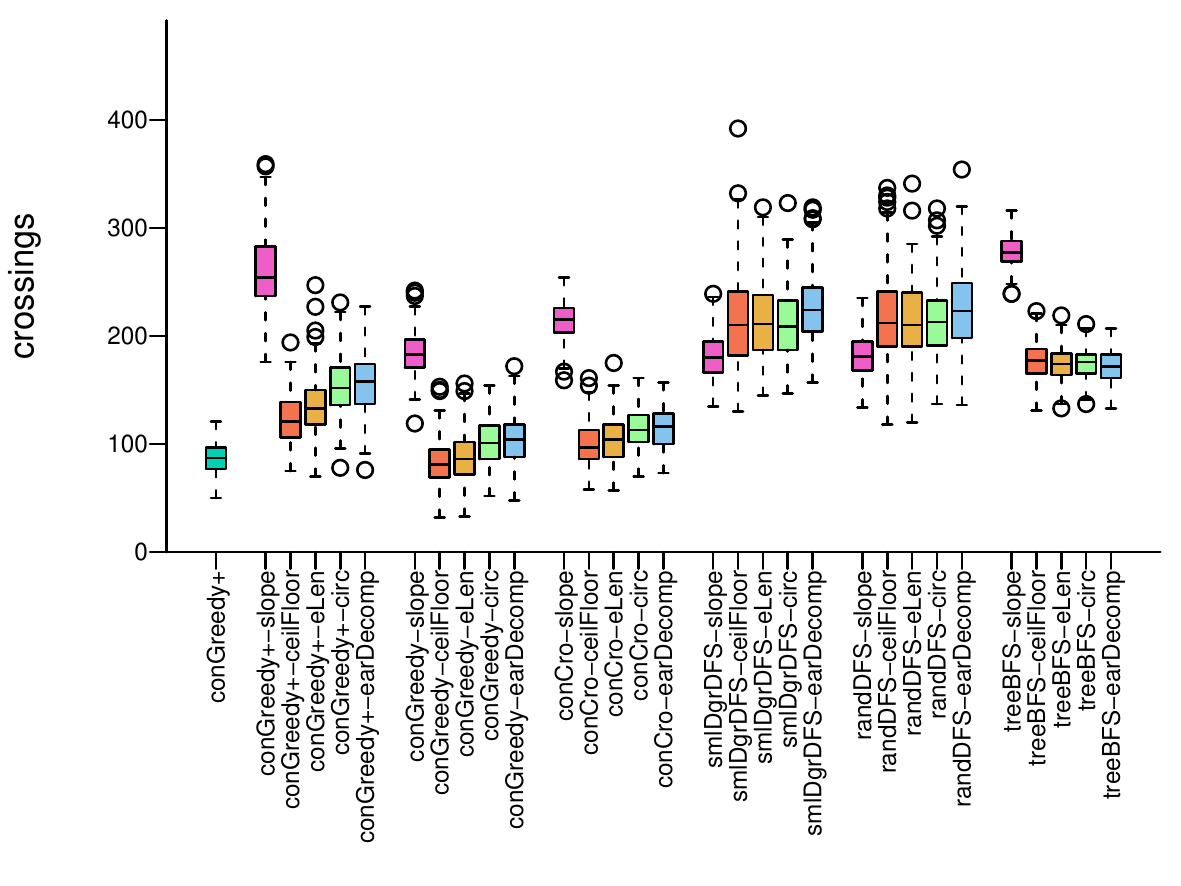}
 }\hfill
 \subfloat[$CCC_6$, 5 pages.\label{fig:appendix:all:cube:CC6k5}]{%
 \includegraphics[width=0.45\linewidth]{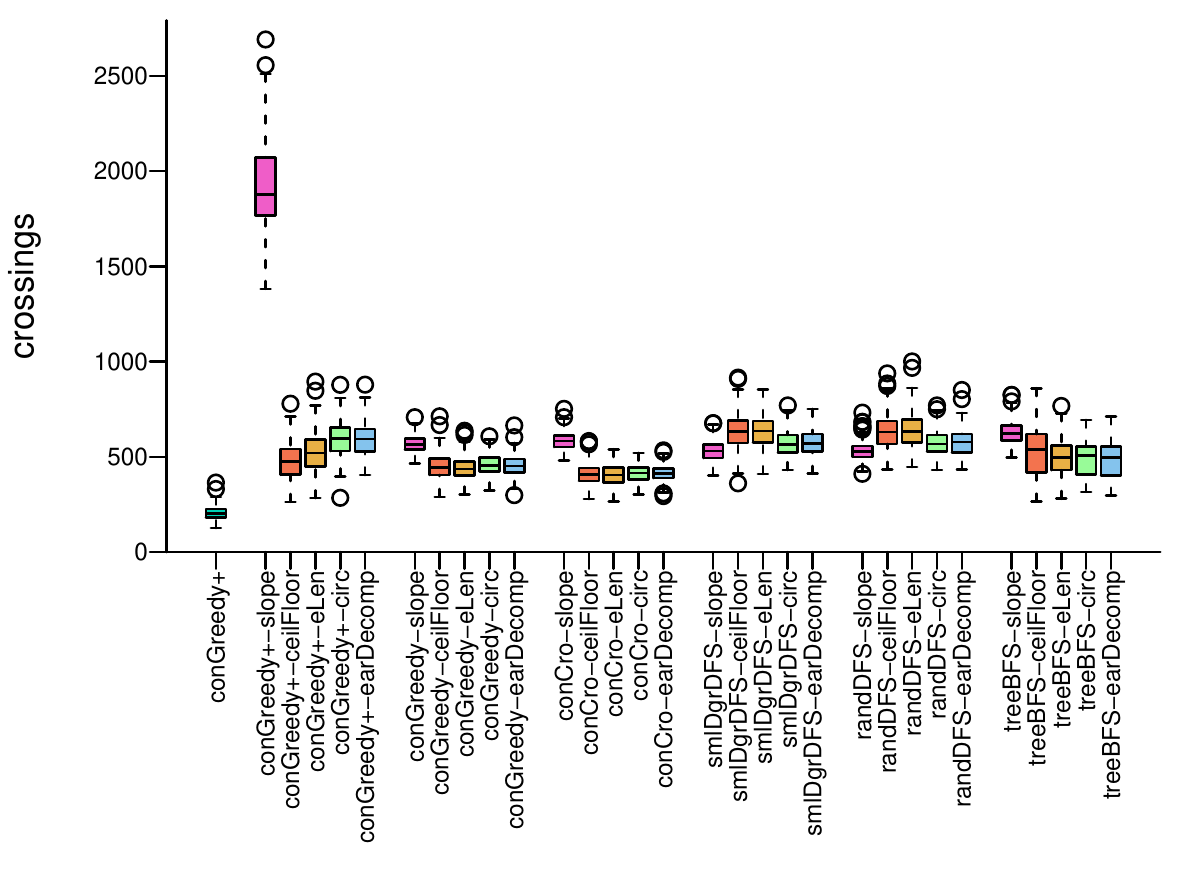}
 }\\
 \subfloat[$Q_6$, 6 pages.\label{fig:appendix:all:cube:Q6k6}]{%
 \includegraphics[width=0.45\linewidth]{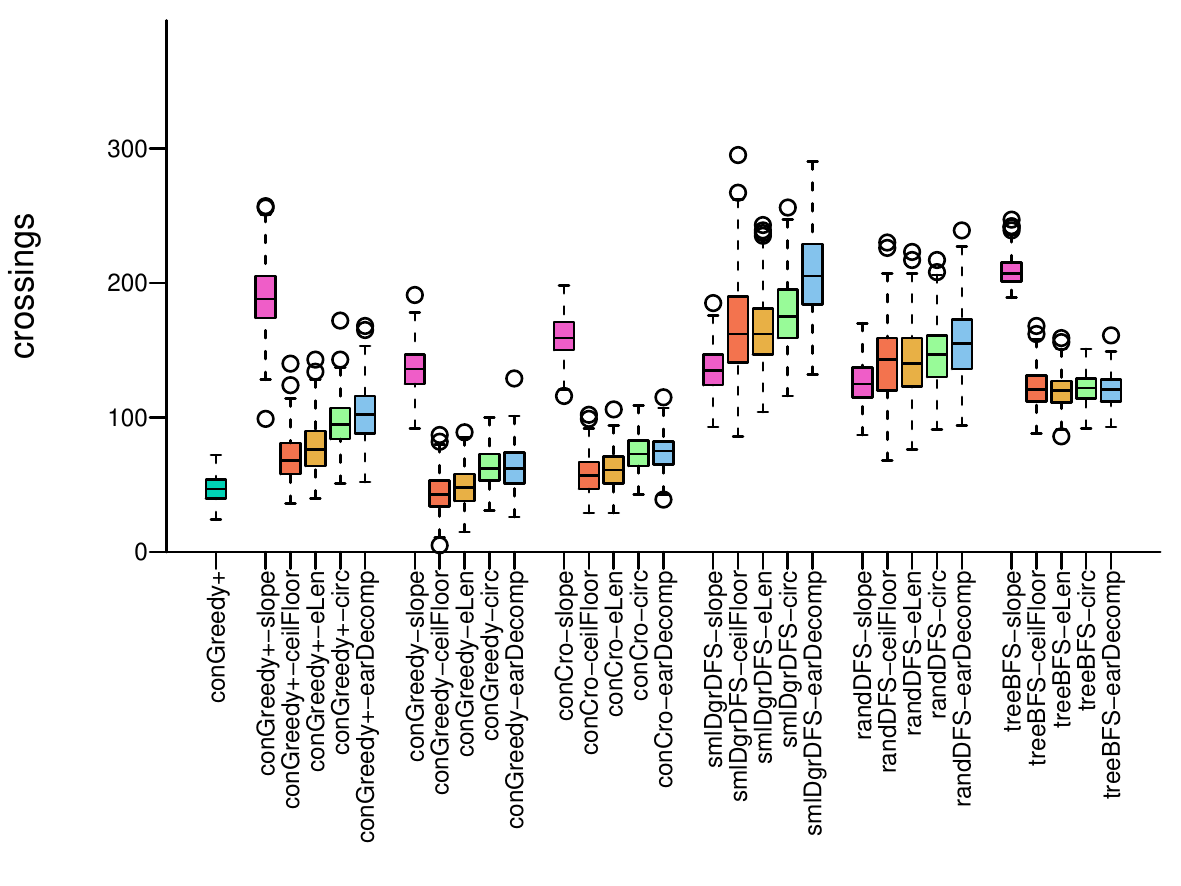}
 }\hfill
 \subfloat[$CCC_6$, 6 pages.\label{fig:appendix:all:cube:CC6k6}]{%
 \includegraphics[width=0.45\linewidth]{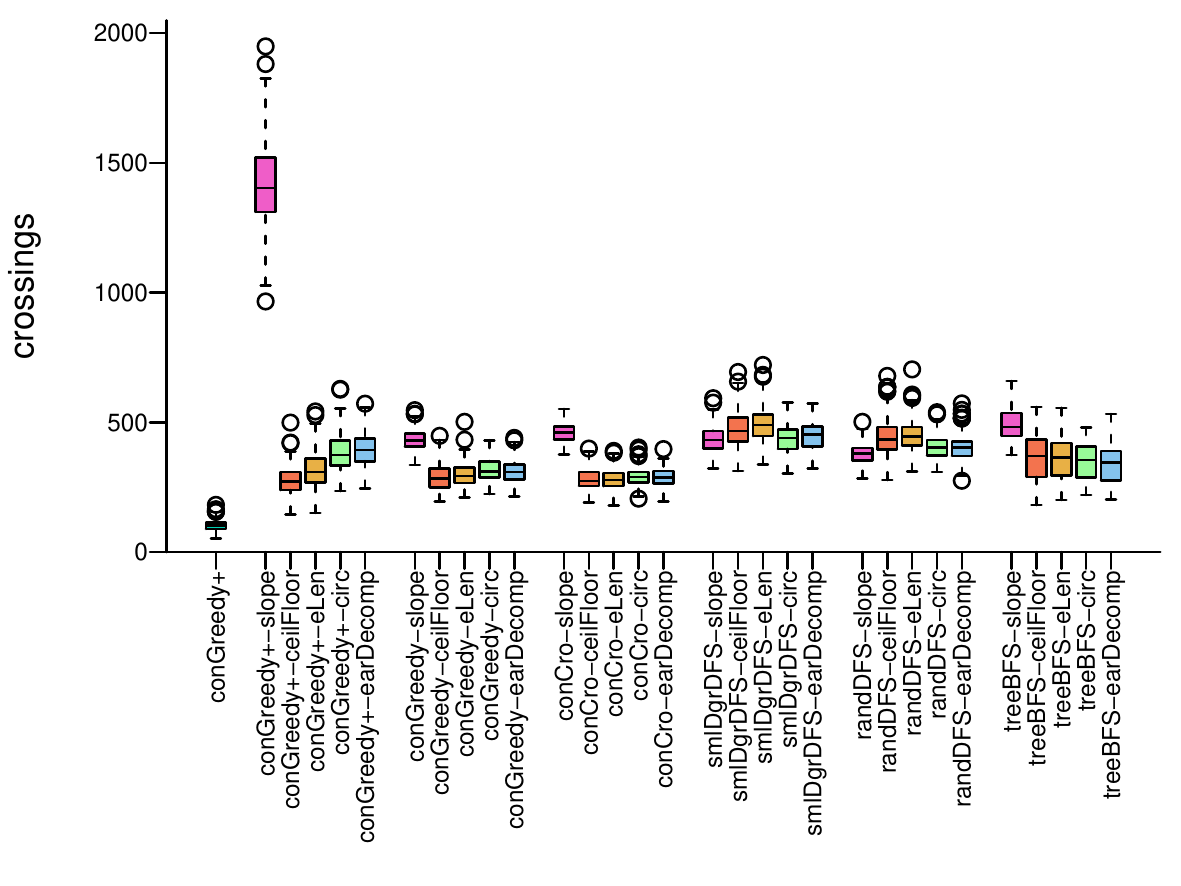}
 }
 \caption{All heuristics on the hypercube $Q_7$ and the cube-connected cycle $CCC_6$. It illustrates
 that the constructive heuristics are not able to achieve nearly zero crossings for five pages,
 which is the book thickness of $Q_6$. The book thickness of cube-connected cycles of dimension $d
 > 3$ is three~\cite{ShTa10}. However, the heuristics are again far off from zero crossings even
 for five and six pages, and perform worse than on their minor $Q_6$.}
 \label{fig:appendix:all:cube}
\end{figure} 

% toroidal 
\begin{figure}[htb]
 \centering
 \subfloat[$C_{16} C_{16}$, 2 pages.\label{fig:appendix:all:tori:cc2}]{%
 \includegraphics[width=0.45\linewidth]{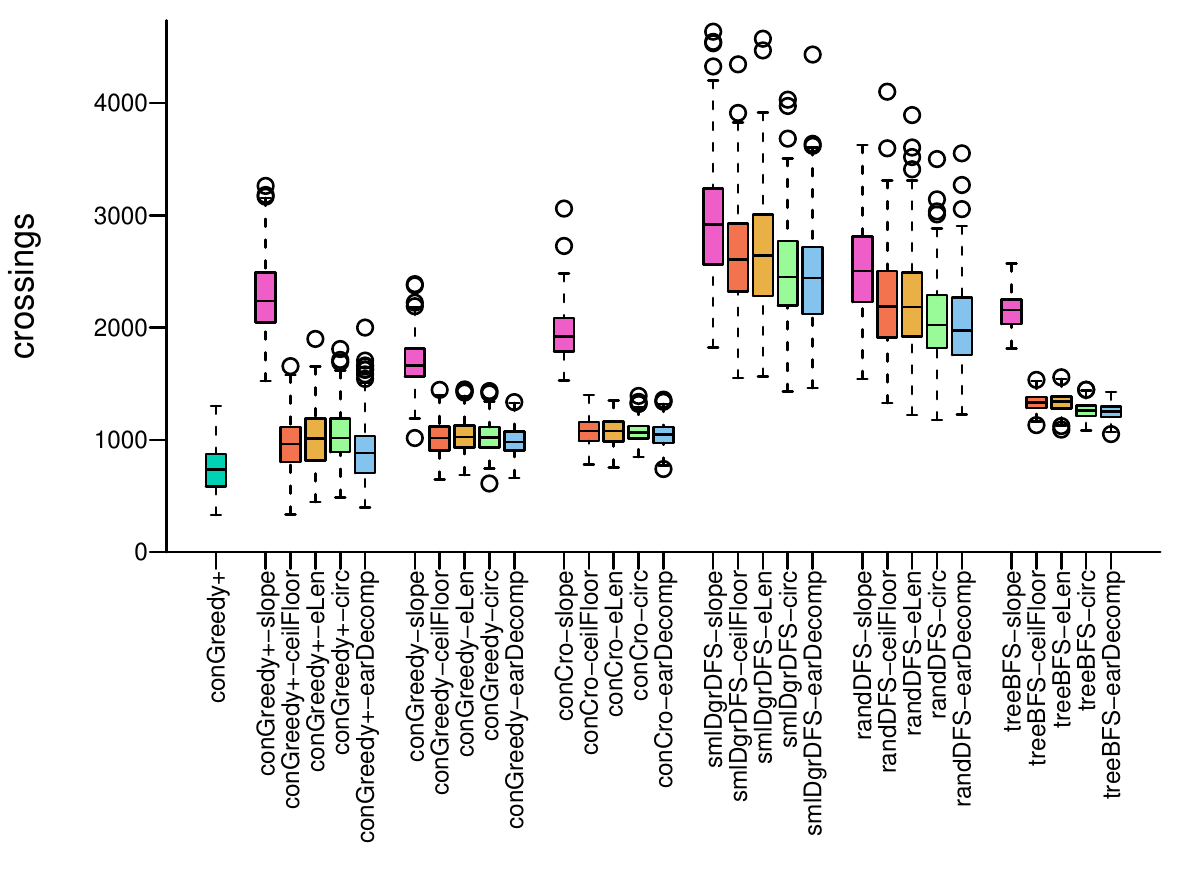}
 }\hfill
 \subfloat[$C_7 C_6 C_6$, 2 pages.\label{fig:appendix:all:tori:ccc2}]{%
 \includegraphics[width=0.45\linewidth]{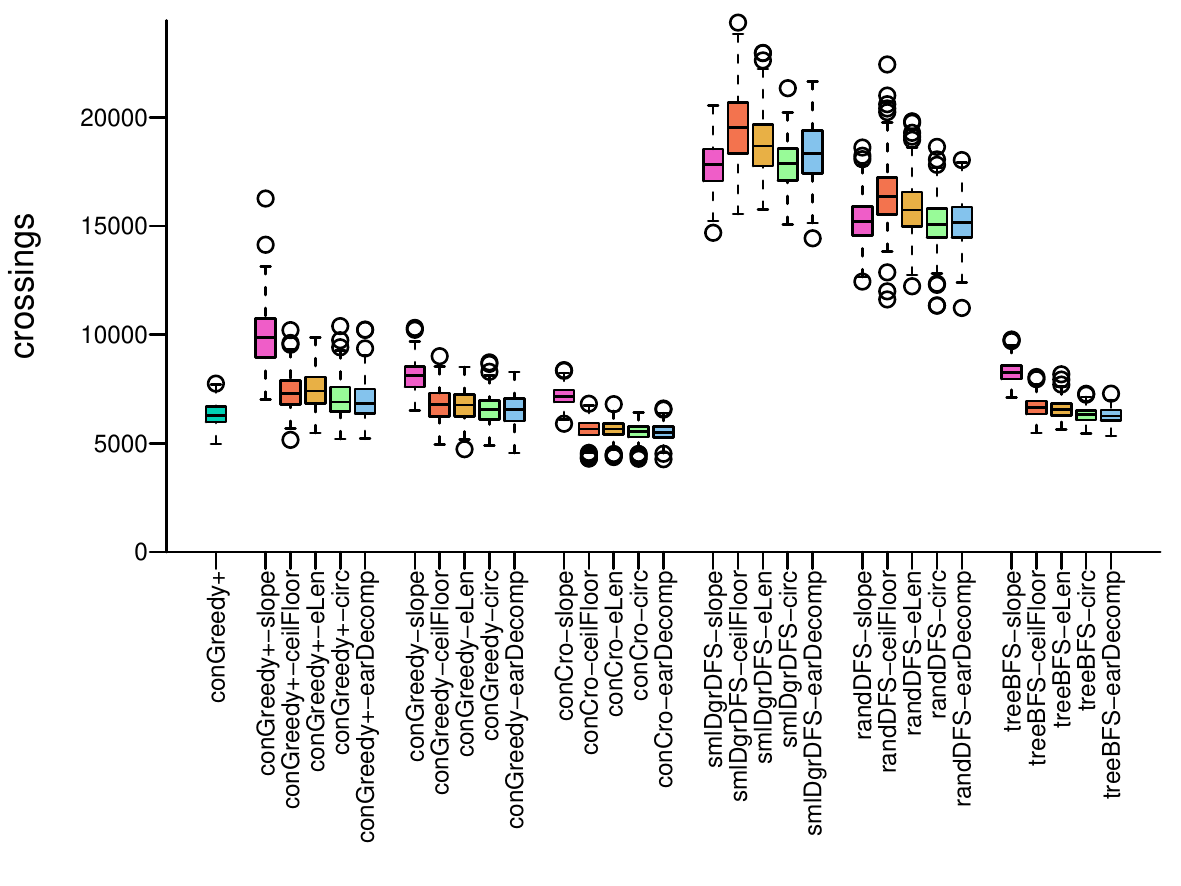}
 }\\
 \subfloat[$C_{16} C_{16}$, 3 pages.\label{fig:appendix:all:tori:cc3}]{%
 \includegraphics[width=0.45\linewidth]{C16C16-ALL-k3-boxplot}
 }\hfill
 \subfloat[$C_7 C_6 C_6$, 3 pages.\label{fig:appendix:all:tori:ccc3}]{%
 \includegraphics[width=0.45\linewidth]{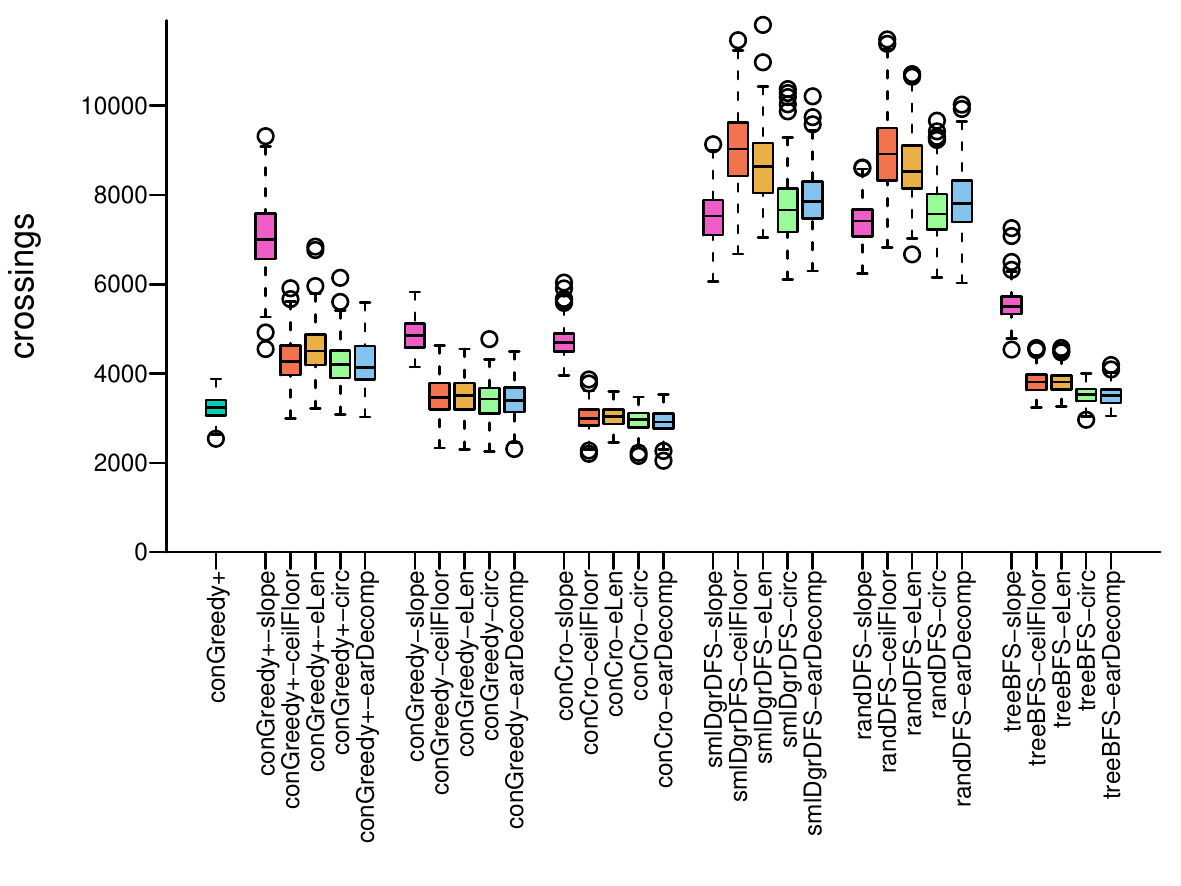}
 }\\
 \subfloat[$C_{16} C_{16}$, 4 pages.\label{fig:appendix:all:tori:cc4}]{%
 \includegraphics[width=0.45\linewidth]{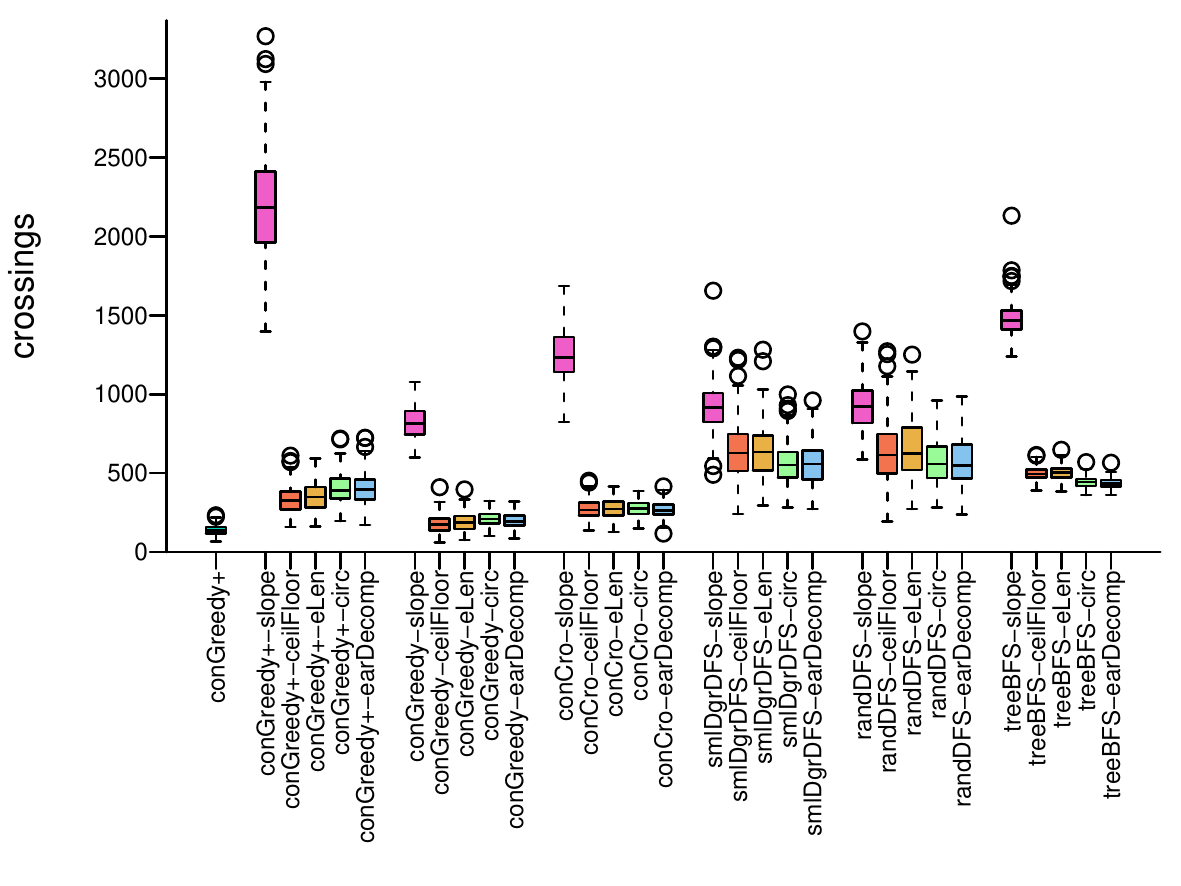}
 }\hfill
 \subfloat[$C_7 C_6 C_6$, 4 pages.\label{fig:appendix:all:tori:ccc4}]{%
 \includegraphics[width=0.45\linewidth]{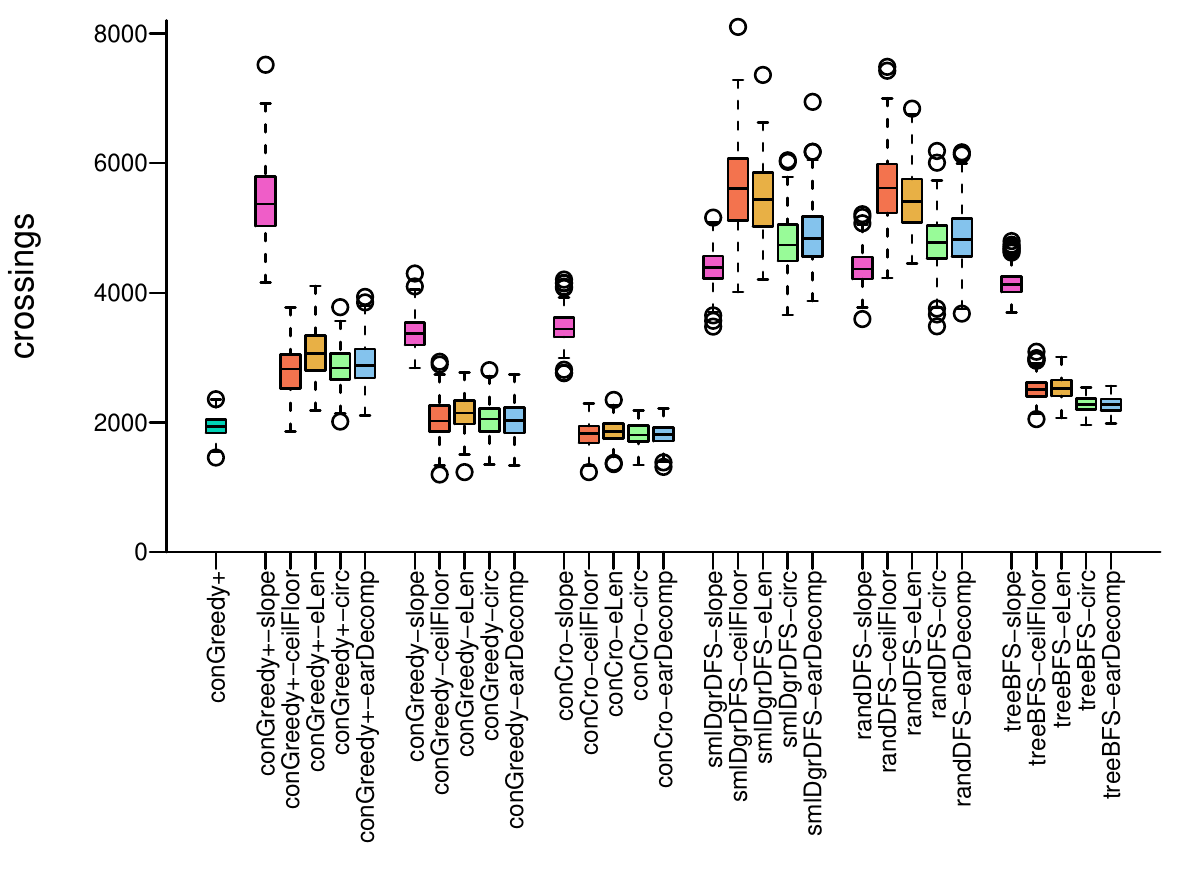}
 }
 \caption{All heuristics on $C_{16} C_{16}$ and $C_7 C_6 C_6$. The book thickness of toroidal meshes is
 three~\cite{Klawitter16}, 
 and unknown for us for 3-toroidal meshes. For both classes we can
 observe again that \treeBFS performs better than the two DFS-based heuristics, and that \CONCRO is
 comparable to \ConGreedy.}
 \label{fig:appendix:all:tori}
\end{figure}

% planar
\begin{figure}[htb]
 \centering
 \subfloat[Topological planar graphs, $n = 250$, 2 pages.\label{fig:...}]{%
 \includegraphics[width=0.45\linewidth]{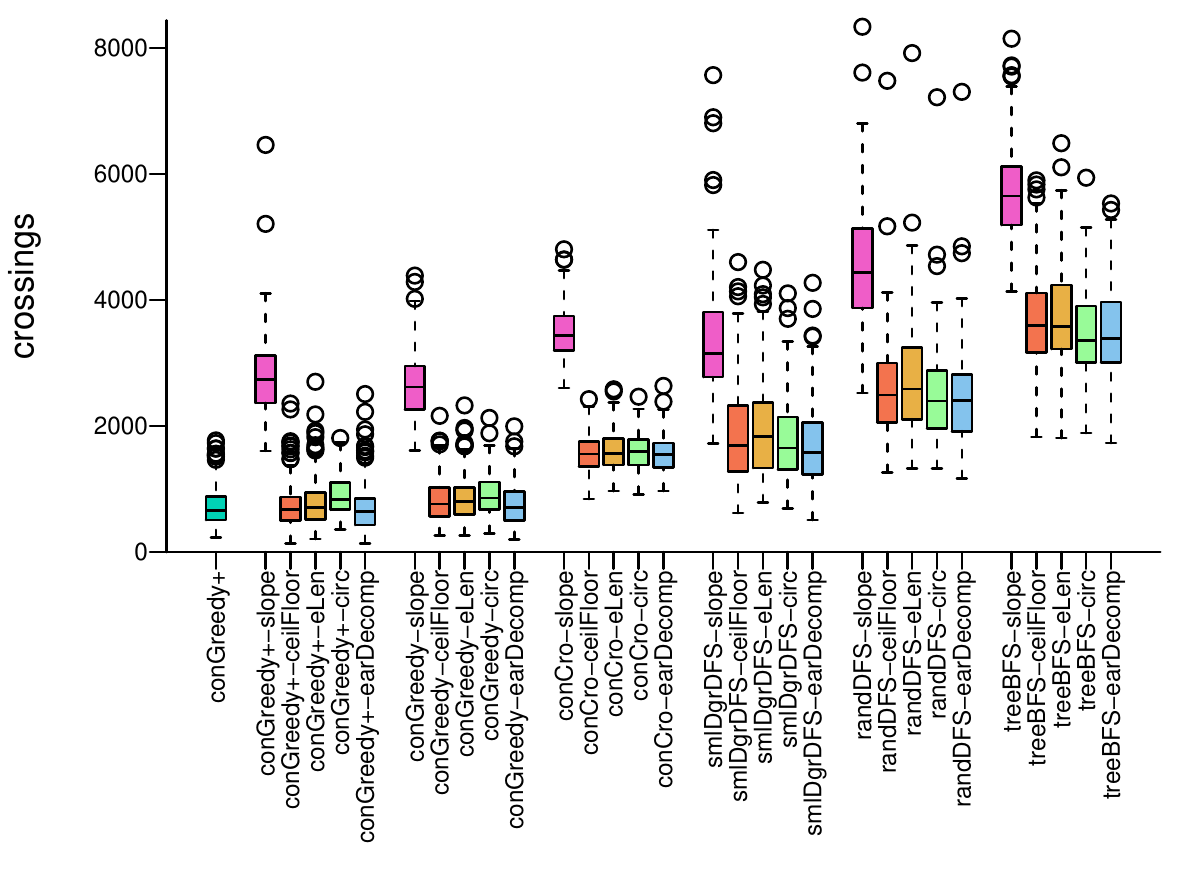}
 }\hfill
 \subfloat[Geometric planar graphs, $n = 250$, 2 pages.\label{fig:...}]{%
 \includegraphics[width=0.45\linewidth]{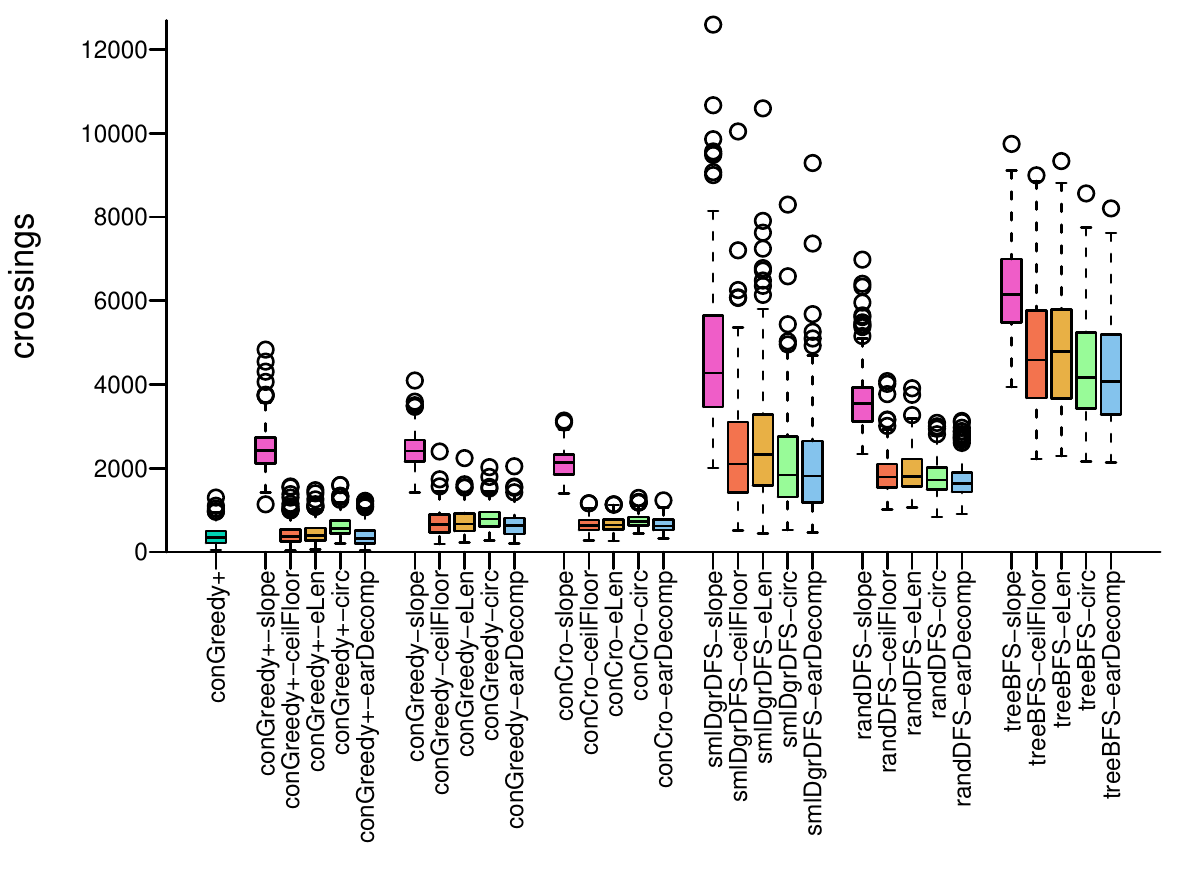}
 }\\
 \subfloat[Topological planar graphs, $n = 250$, 3 pages.\label{fig:...}]{%
 \includegraphics[width=0.45\linewidth]{triangulation-ALL-n250-k3-boxplot}
 }\hfill
 \subfloat[Geometric planar graphs, $n = 250$, 3 pages.\label{fig:...}]{%
 \includegraphics[width=0.45\linewidth]{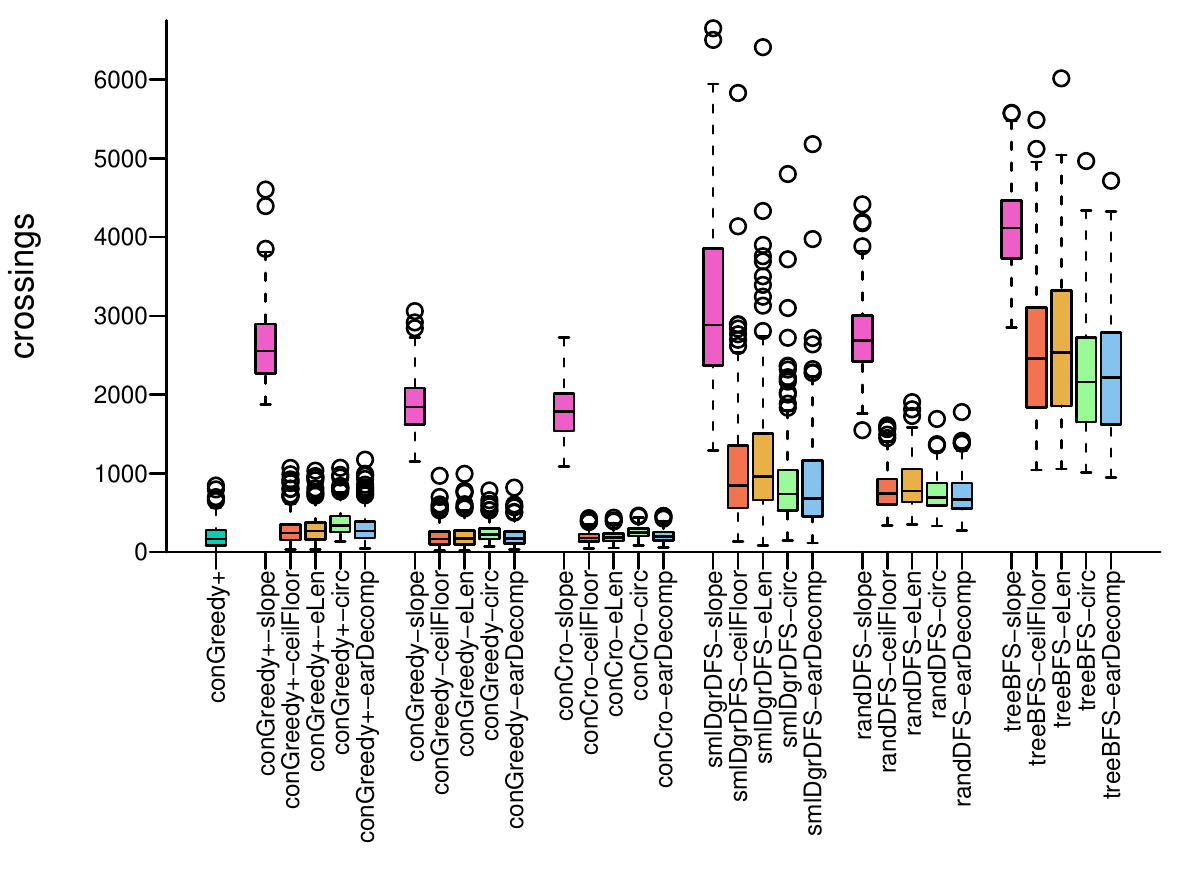}
 }\\
 \subfloat[Topological planar graphs, $n = 250$, 4 pages.\label{fig:...}]{%
 \includegraphics[width=0.45\linewidth]{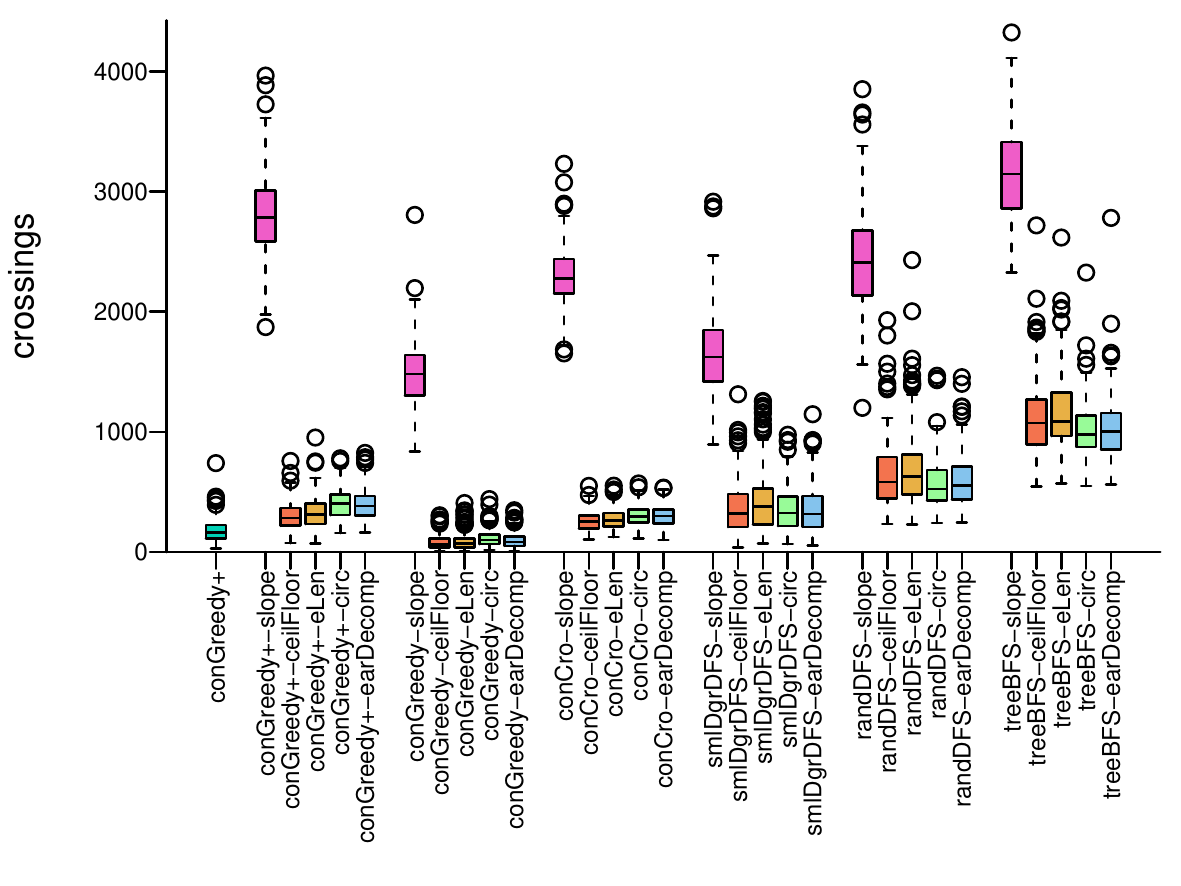}
 } \hfill
 \subfloat[Geometric planar graphs, $n = 250$, 4 pages.\label{fig:...}]{%
 \includegraphics[width=0.45\linewidth]{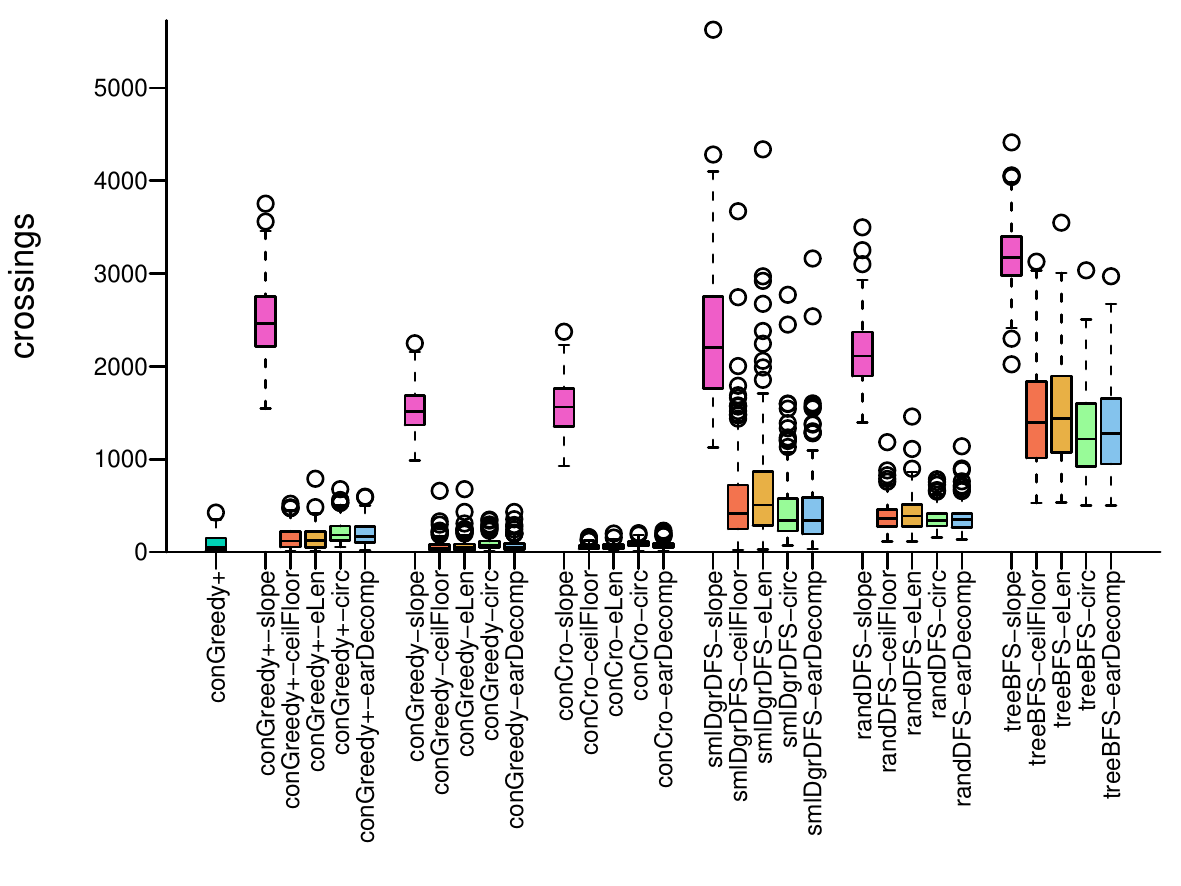}
 }
 \caption{All heuristics on planar graphs with 250 vertices. The choice of the VO heuristic is more
 important than the choice of the PA heuristic, as long as it is not \SLOPE. \CONGREEDY performs
 better on topological planar graphs, while \CONCRO performs better on geometric planar graphs. In
 both cases are the heuristics far from 0 crossings for 3 or even 4 pages.}
 \label{fig:appendix:all:planar}
\end{figure} 

% 1-planar
\begin{figure}[htb]
 \centering
%  \subfloat[1-planar (triang.), $n = 250$, 2 pages.\label{fig:...}]{%
%  \includegraphics[width=0.49\linewidth]{onePlanar-T-ALL-n250-k2-boxplot}
%  }\hfill
%  \subfloat[1-planar (point set), $n = 250$, 2 pages.\label{fig:...}]{%
%  \includegraphics[width=0.49\linewidth]{onePlanar-X-ALL-n250-k2-boxplot}
%  }\\
 \subfloat[Topological 1-planar graphs, $n = 250$, 3 pages.\label{fig:...}]{%
 \includegraphics[width=0.45\linewidth]{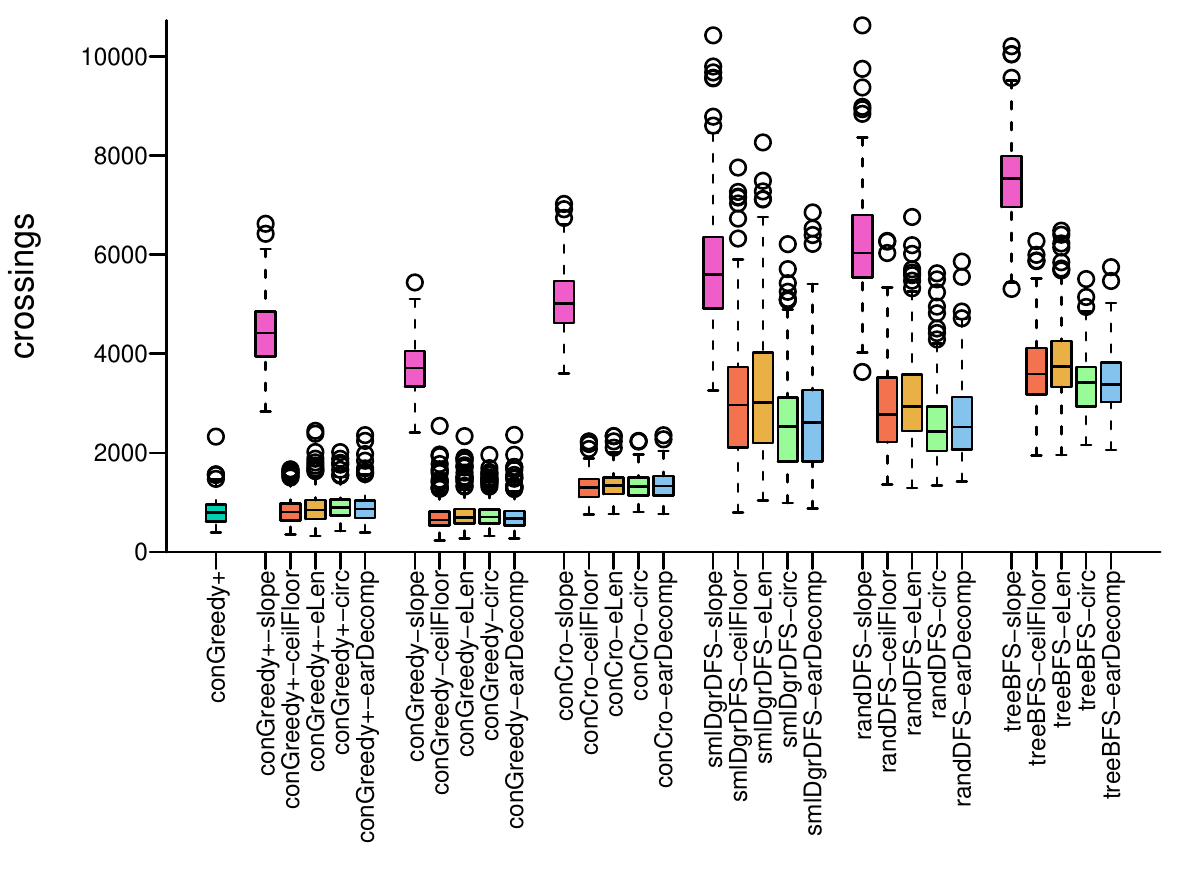}
 }\hfill
 \subfloat[Geometric 1-planar graphs, $n = 250$, 3 pages.\label{fig:...}]{%
 \includegraphics[width=0.45\linewidth]{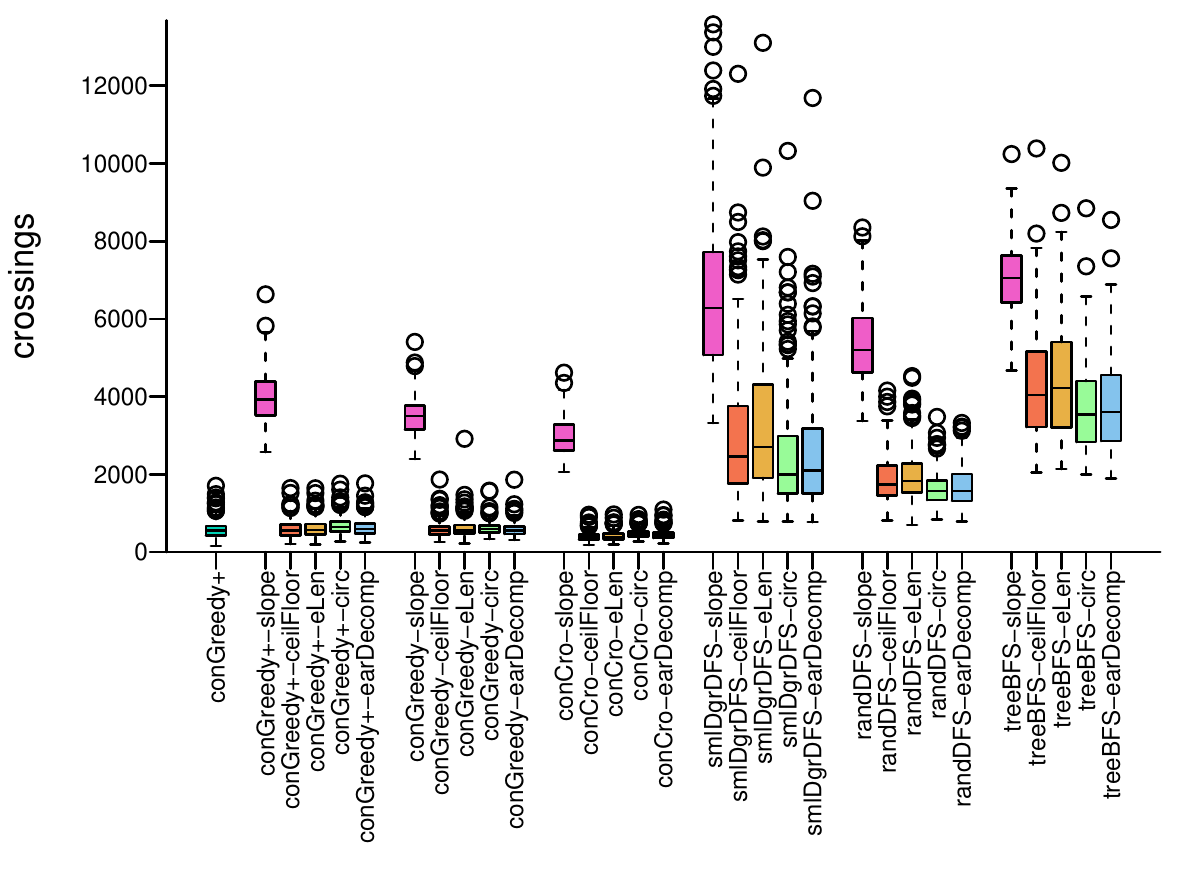}
 }\\
 \subfloat[Topological 1-planar graphs, $n = 250$, 4 pages.\label{fig:...}]{%
 \includegraphics[width=0.45\linewidth]{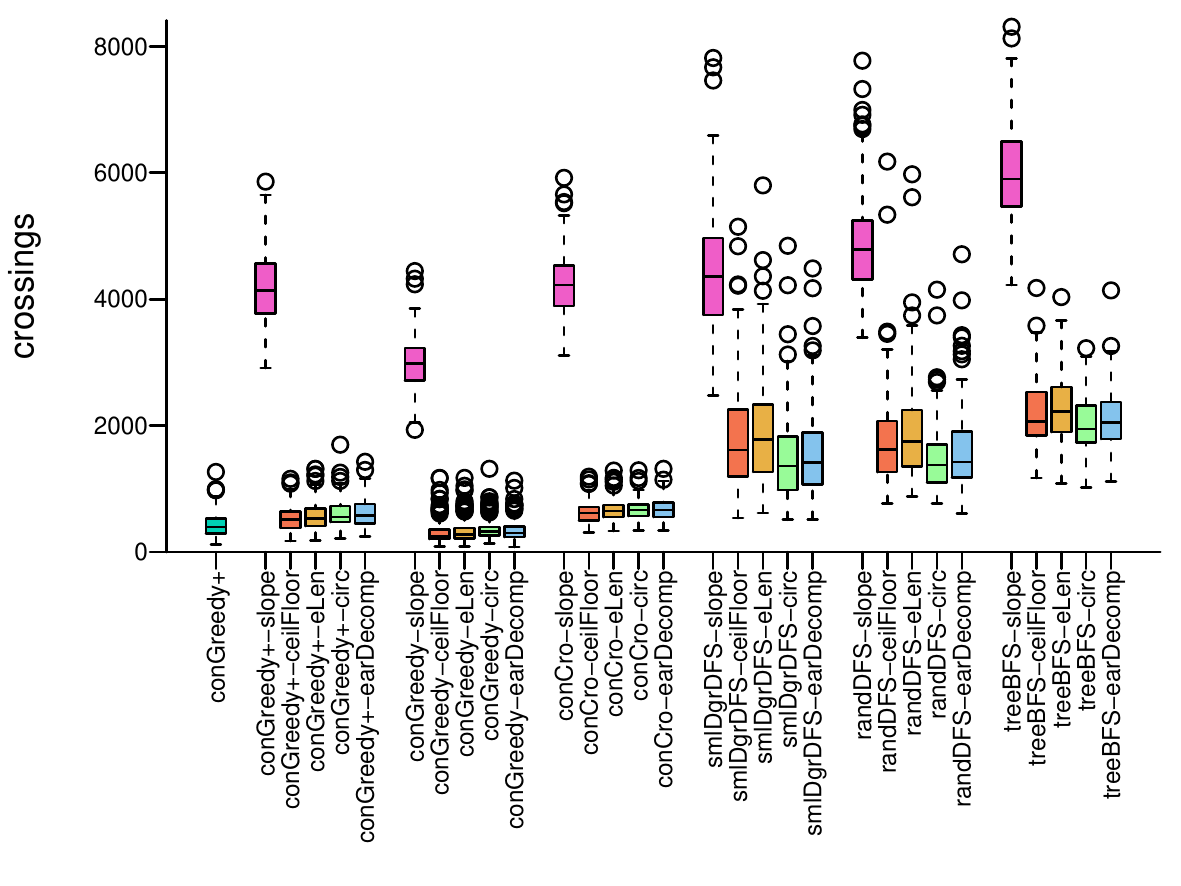}
 } \hfill
 \subfloat[Geometric 1-planar graphs, $n = 250$, 4 pages.\label{fig:...}]{%
 \includegraphics[width=0.45\linewidth]{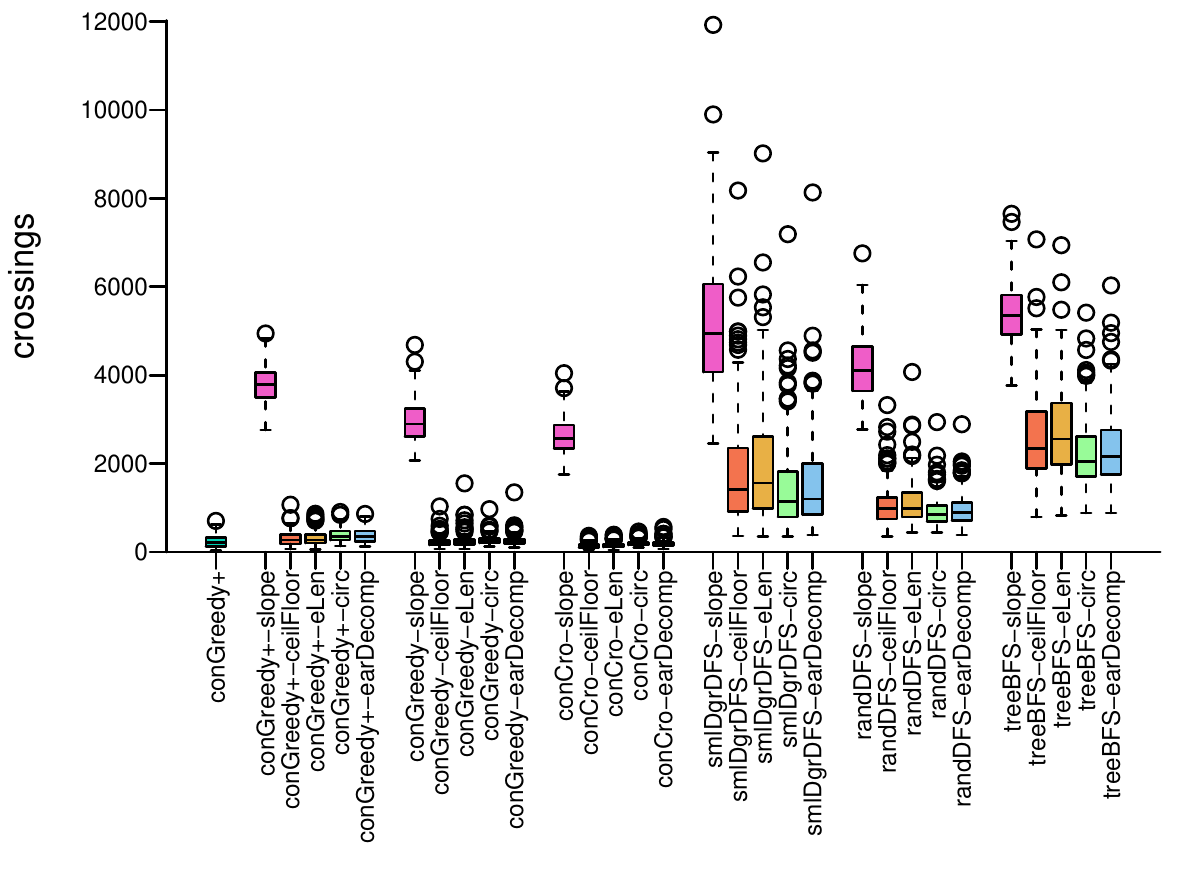}
 }\\
 \subfloat[Topological 1-planar graphs, $n = 250$, 5 pages.\label{fig:...}]{%
 \includegraphics[width=0.45\linewidth]{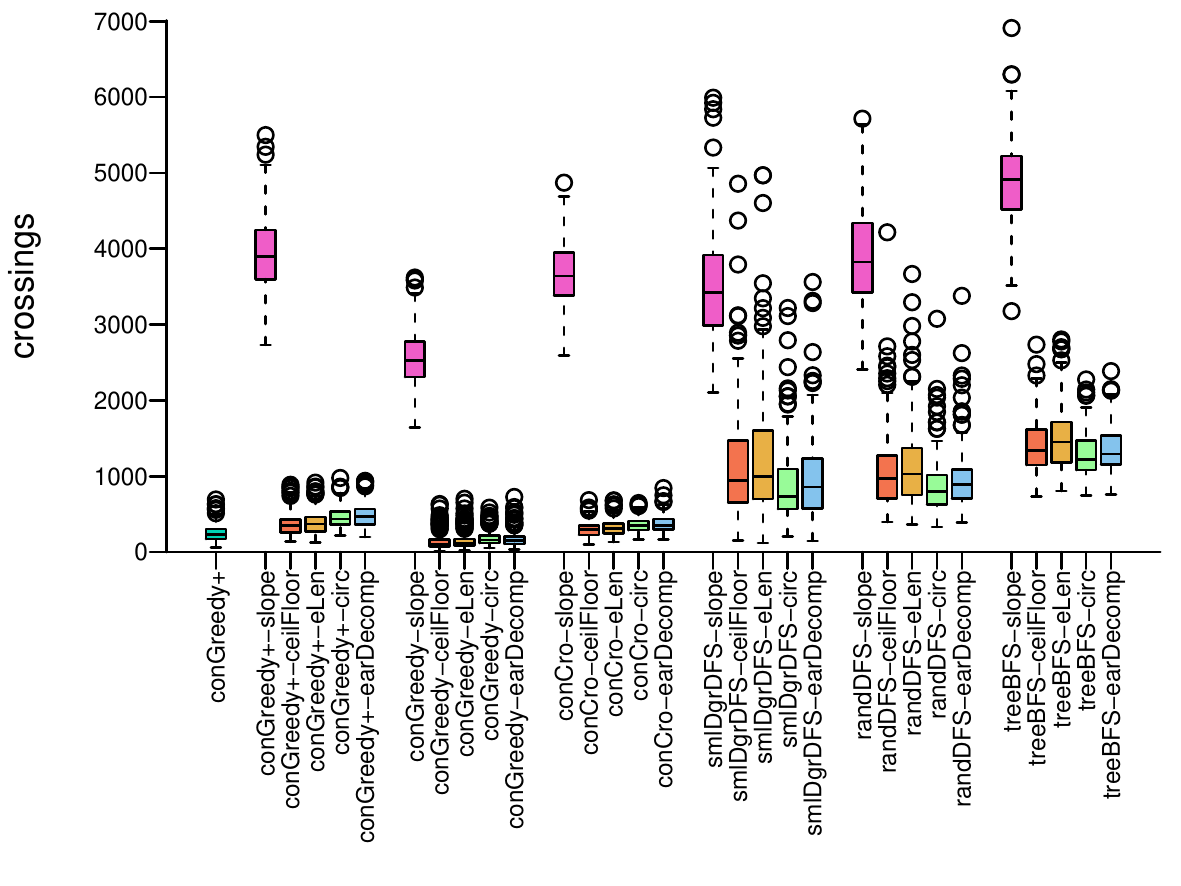}
 } \hfill
 \subfloat[Geometric 1-planar graphs, $n = 250$, 5 pages.\label{fig:...}]{%
 \includegraphics[width=0.45\linewidth]{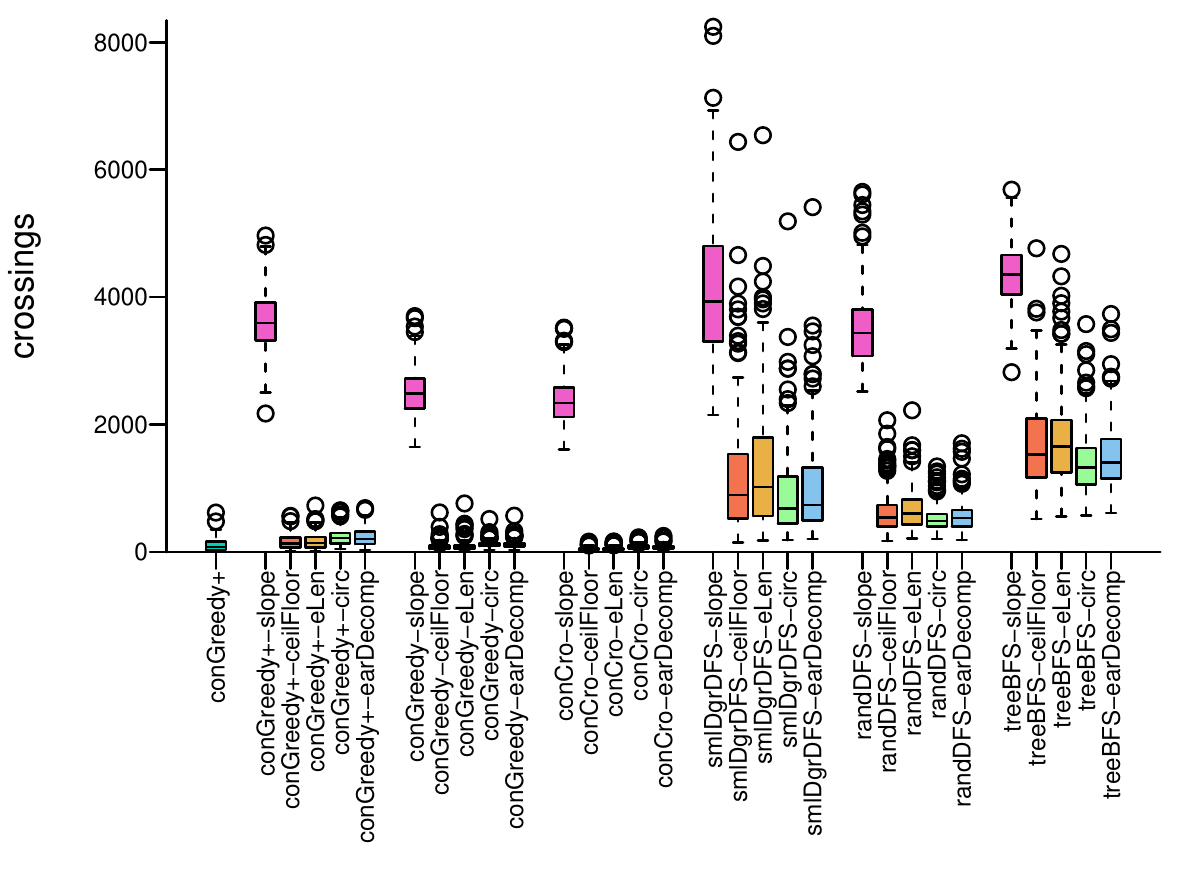}
 }
 \caption{All heuristics on topological and geometric 1-planar graphs with 250 vertices.
 The conjectured book thickness is 4. For 5 pages \CONCRO achieves near zero crossings for
 geometric 1-planar graphs.}
 \label{fig:appendix:all:onePlanar} 
\end{figure} 

% k-tree
\begin{figure}[htb]
 \centering
 \subfloat[6-tree, $n = 250$, 2 pages.]{%
 \includegraphics[width=0.45\linewidth]{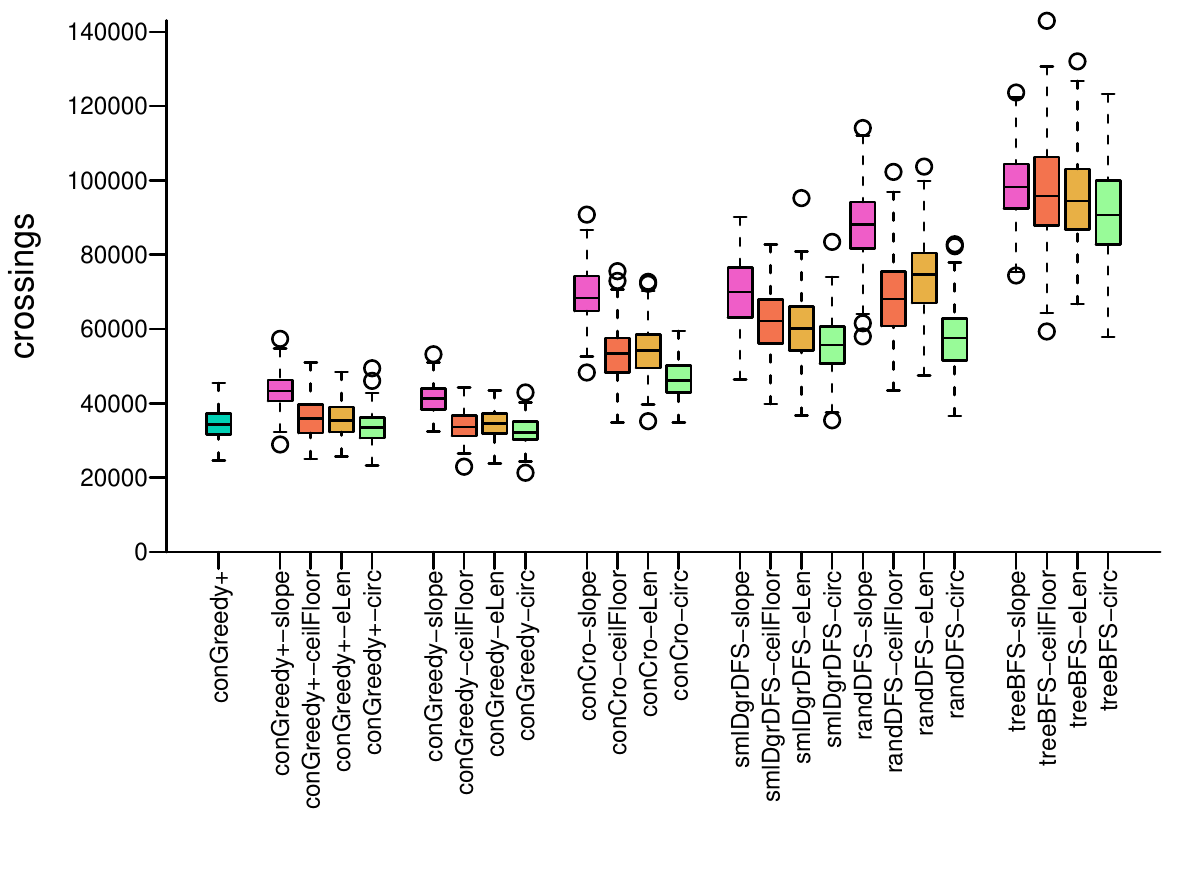}
 }\hfill
 \subfloat[6-tree, $n = 250$, 4 pages.]{%
 \includegraphics[width=0.45\linewidth]{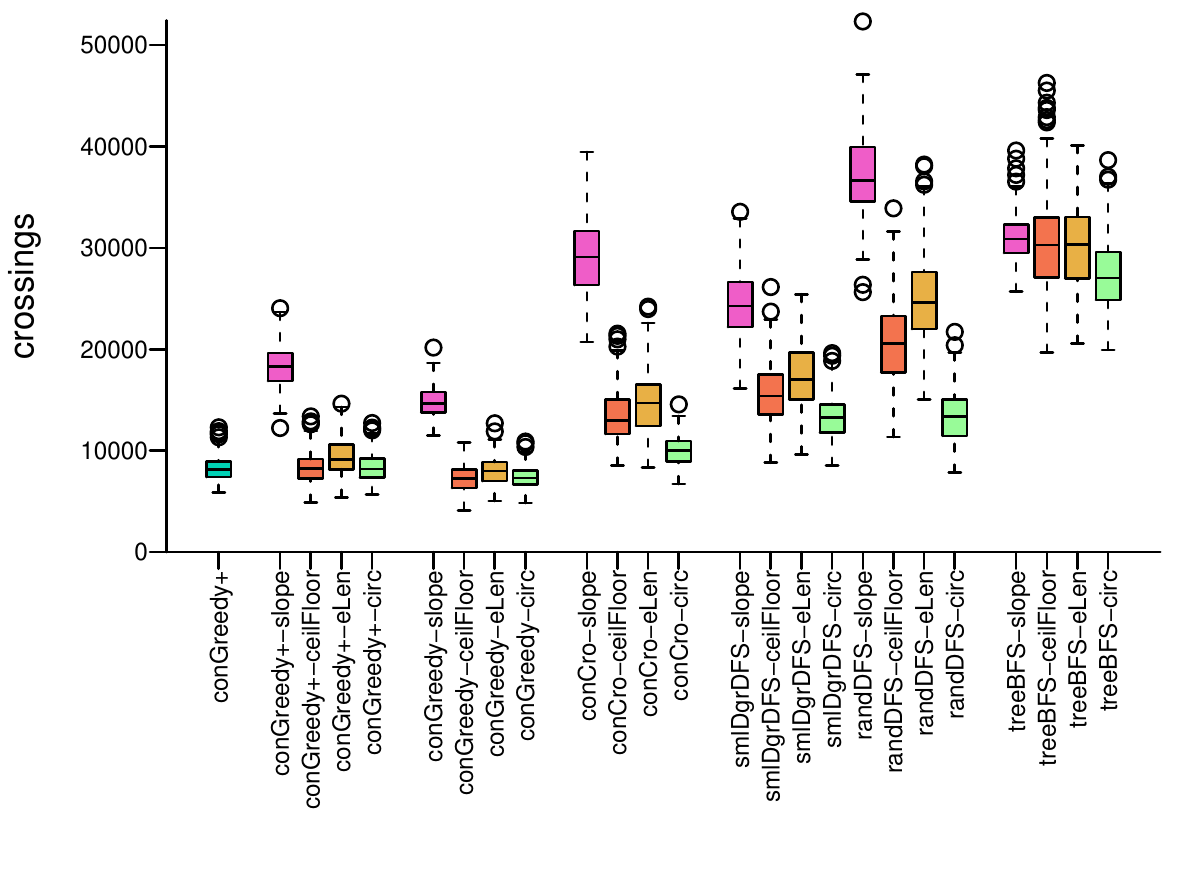}
 }\\
 \subfloat[6-tree, $n = 250$, 6 pages.]{%
 \includegraphics[width=0.45\linewidth]{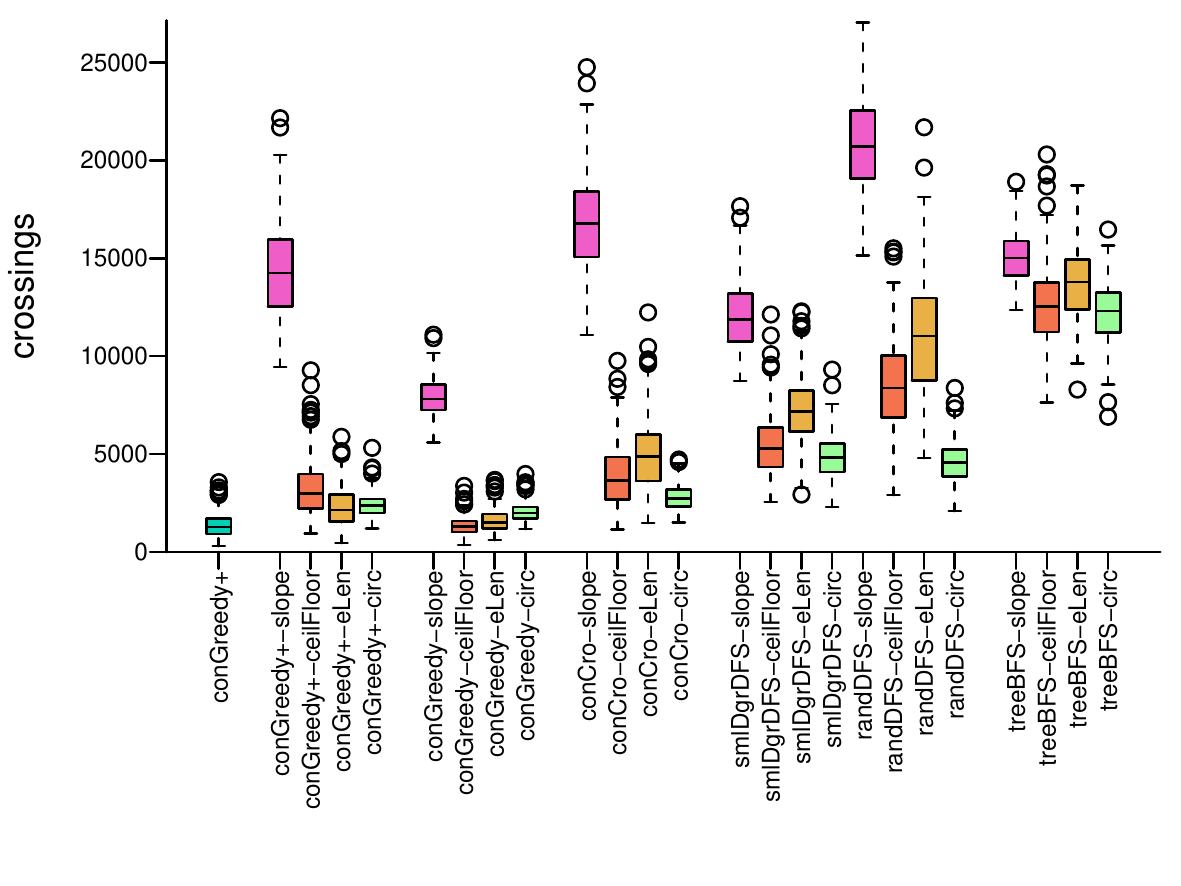}
 } \hfill
 \subfloat[6-tree, $n = 250$, 7 pages.]{%
 \includegraphics[width=0.45\linewidth]{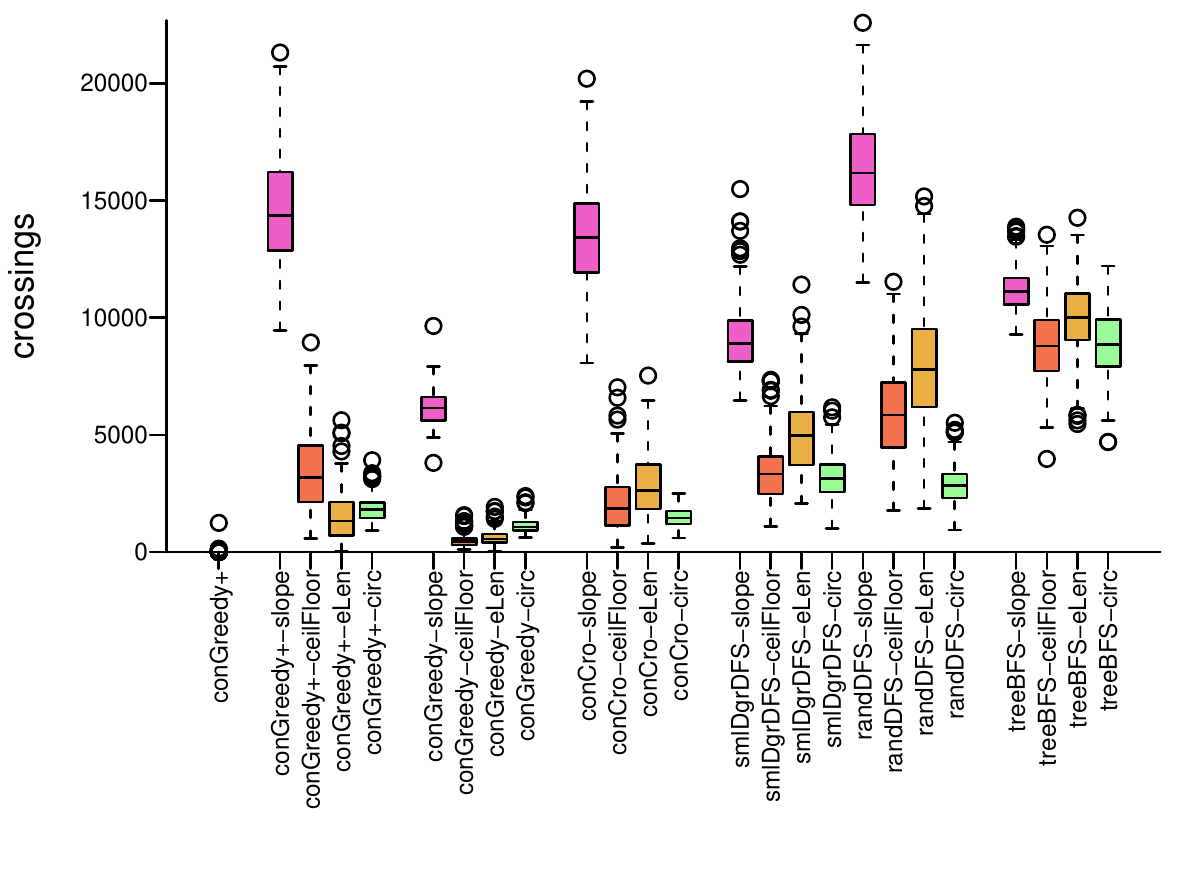}
 }
 \caption{Heuristics on 6-trees with 250 vertices, which have book thickness at most 7. For 7 pages
 we see that \FCONGREEDY achieves mostly near 0 or 0 crossings. However, \CONGREEDY performs better
 than \FCONGREEDY when used in combination with a PA heuristic. The higher the number of pages get,
 the worse is the performance of \SLOPE.}
 \label{fig:appendix:all:kTree} 
\end{figure} 

% random linear
\begin{figure}[htb]
 \centering
 \subfloat[Random (linear 3), $n = 250$, 5 pages.\label{fig:appendix:all:randomL:3k5}]{%
 \includegraphics[width=0.45\linewidth]{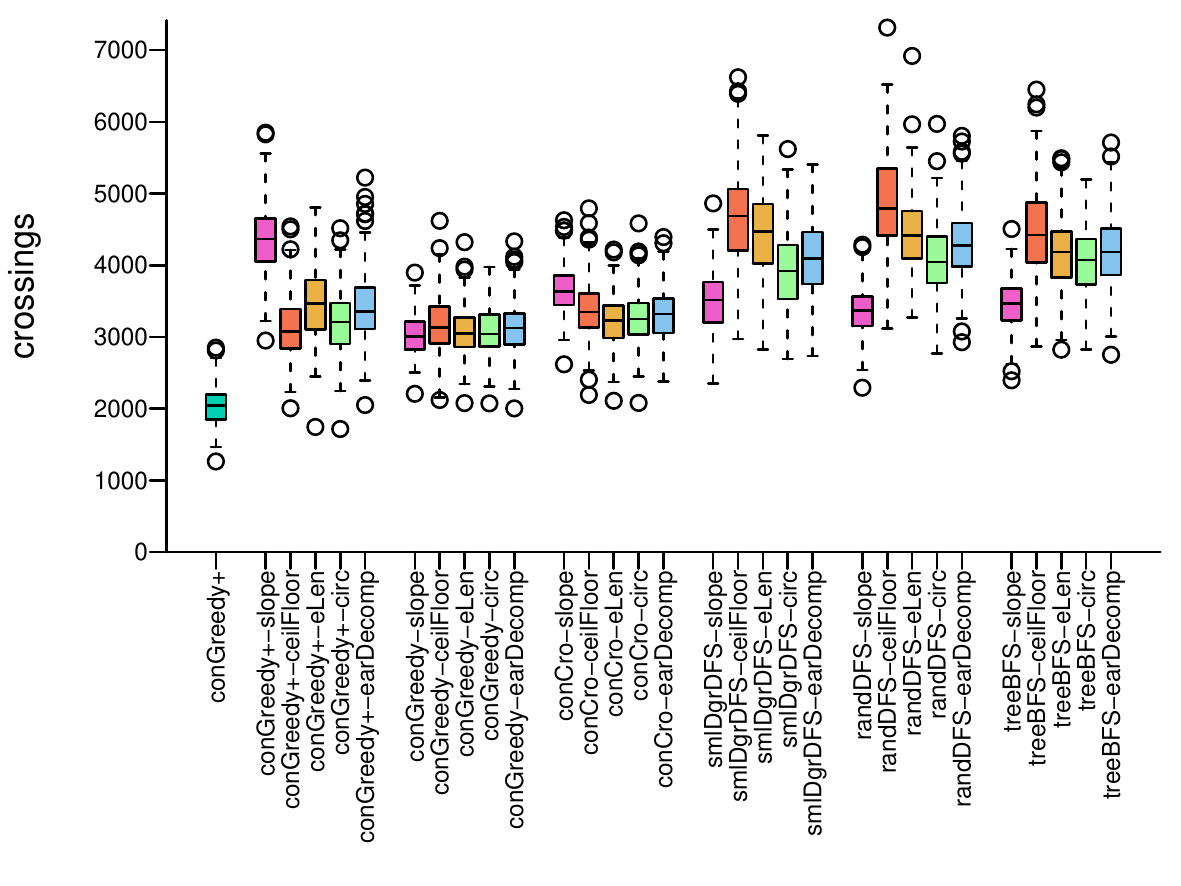}
 }\hfill
 \subfloat[Random (linear 6), $n = 250$, 5 pages.\label{fig:appendix:all:randomL:6k5}]{%
 \includegraphics[width=0.45\linewidth]{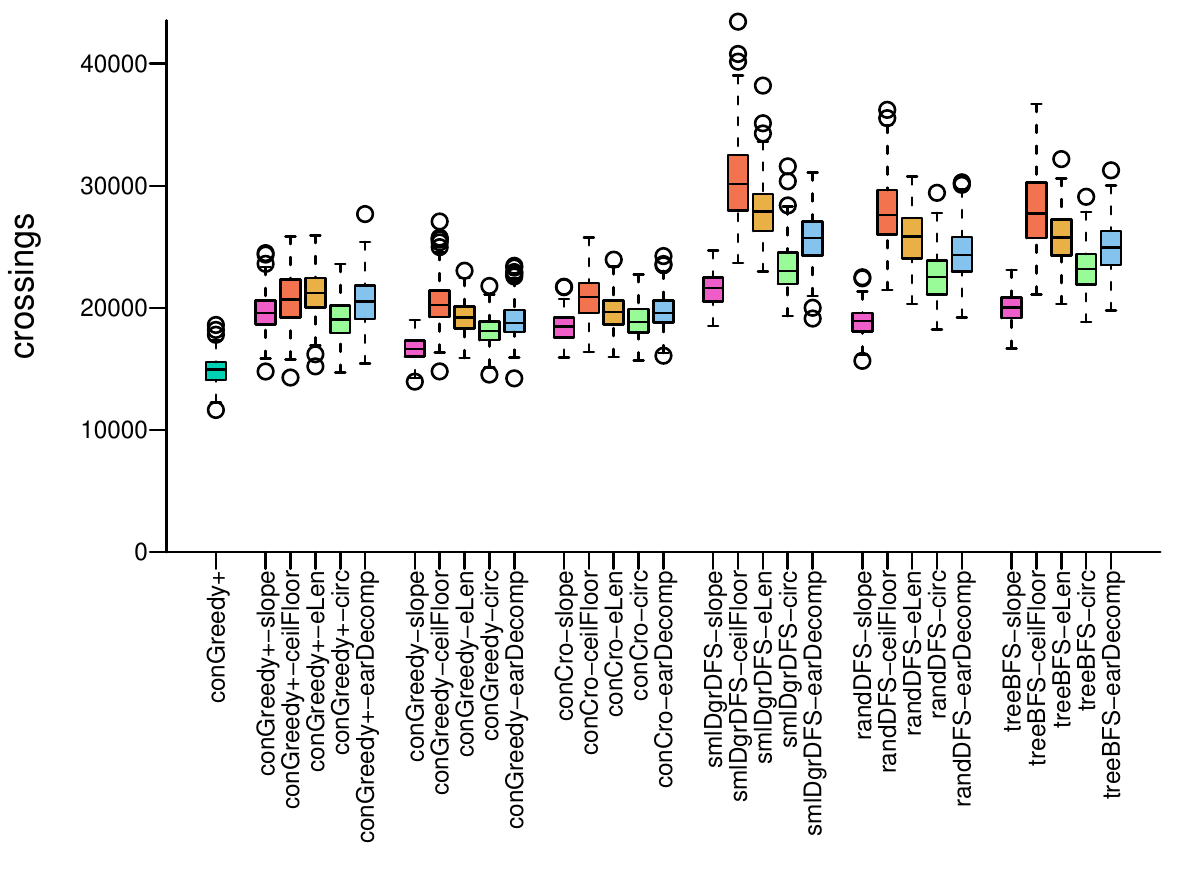}
 }\\
 \subfloat[Random (linear 3), $n = 250$, 10 pages.\label{fig:appendix:all:randomL:3k10}]{%
 \includegraphics[width=0.45\linewidth]{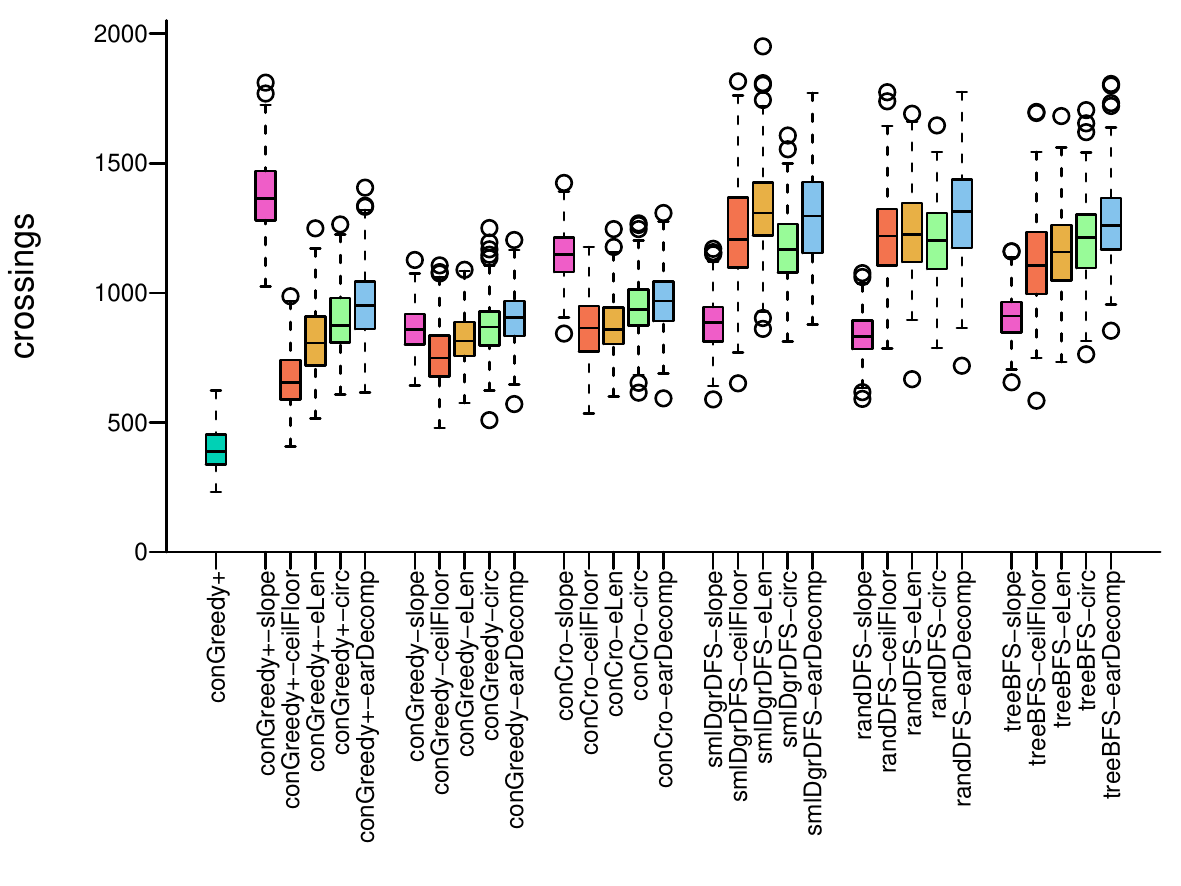}
 }\hfill
 \subfloat[Random (linear 6), $n = 250$, 10 pages.\label{fig:appendix:all:randomL:6k10}]{%
 \includegraphics[width=0.45\linewidth]{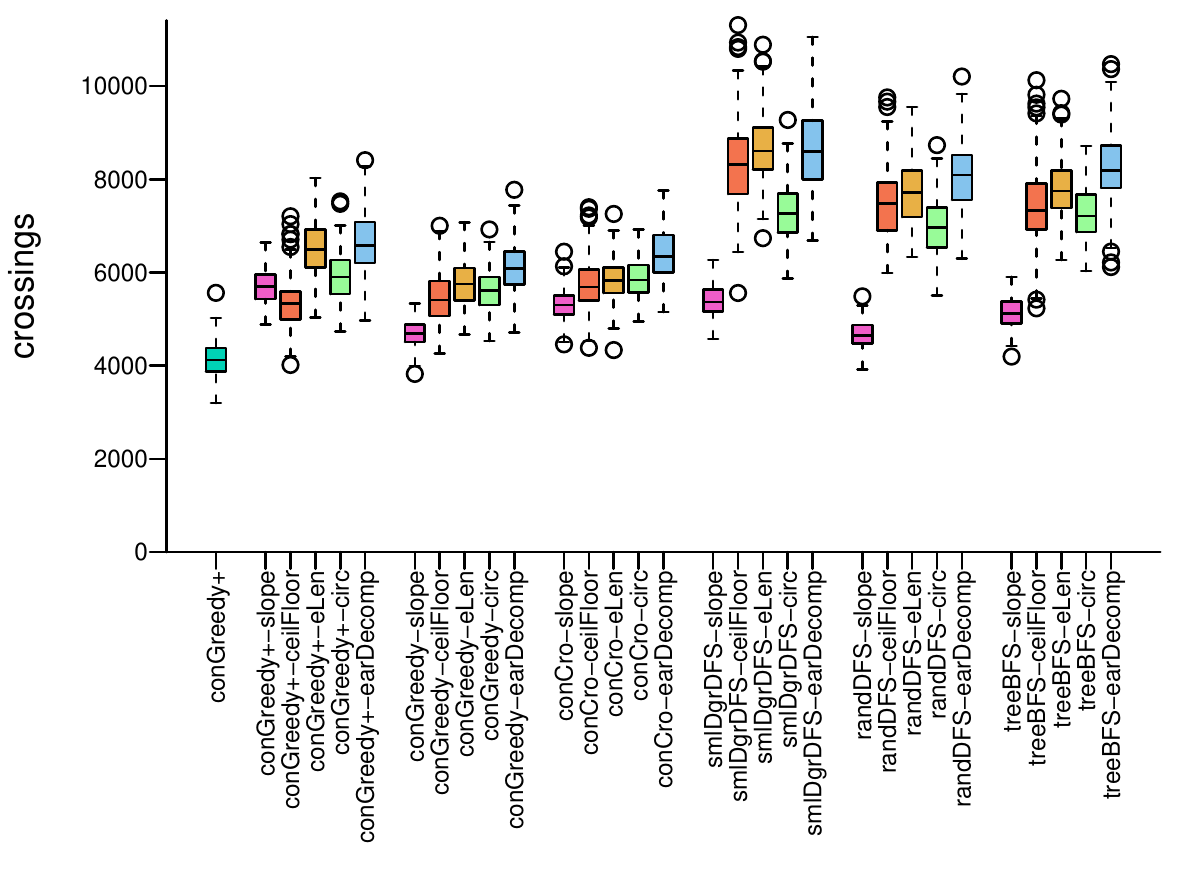}
 }\\
 \subfloat[Random (linear 3), $n = 250$, 15 pages.\label{fig:appendix:all:randomL:3k15}]{%
 \includegraphics[width=0.45\linewidth]{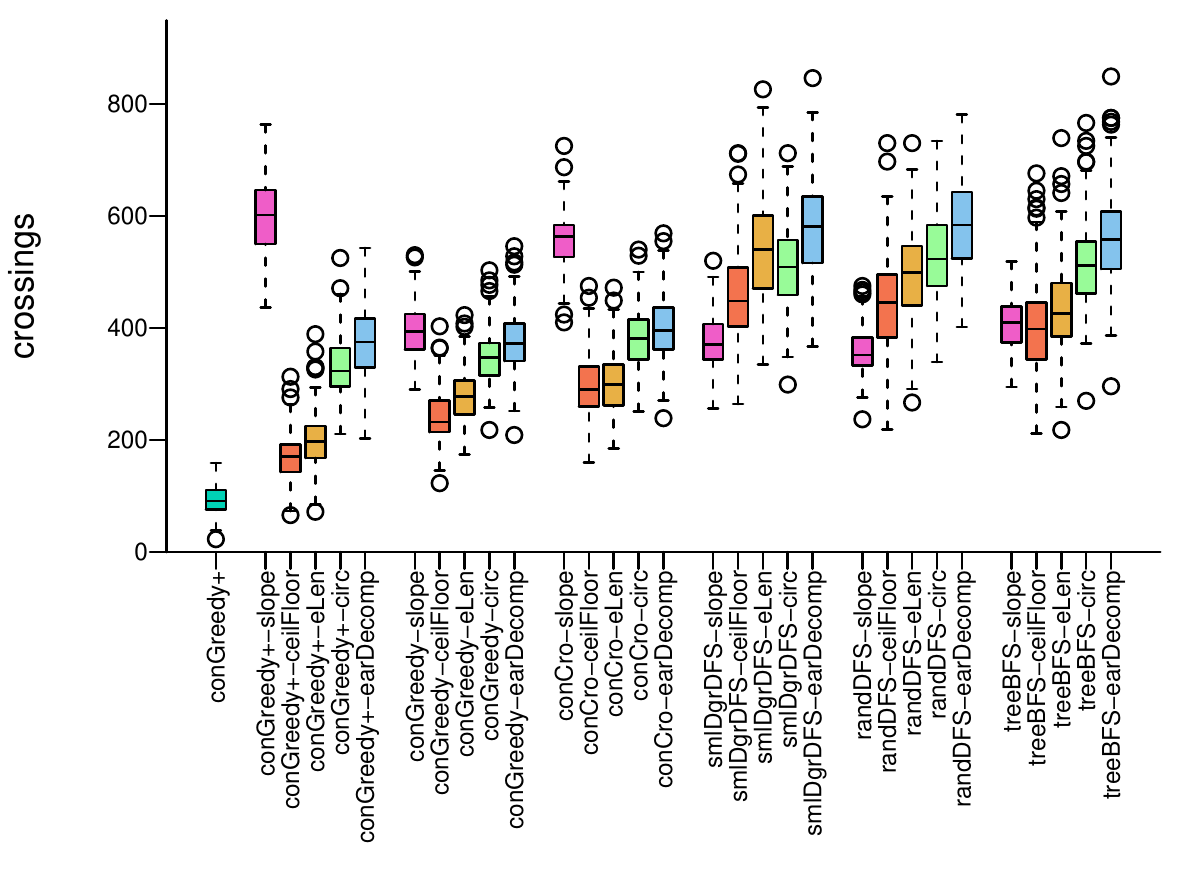}
 } \hfill
 \subfloat[Random (linear 6), $n = 250$, 15 pages.\label{fig:appendix:all:randomL:6k15}]{%
 \includegraphics[width=0.45\linewidth]{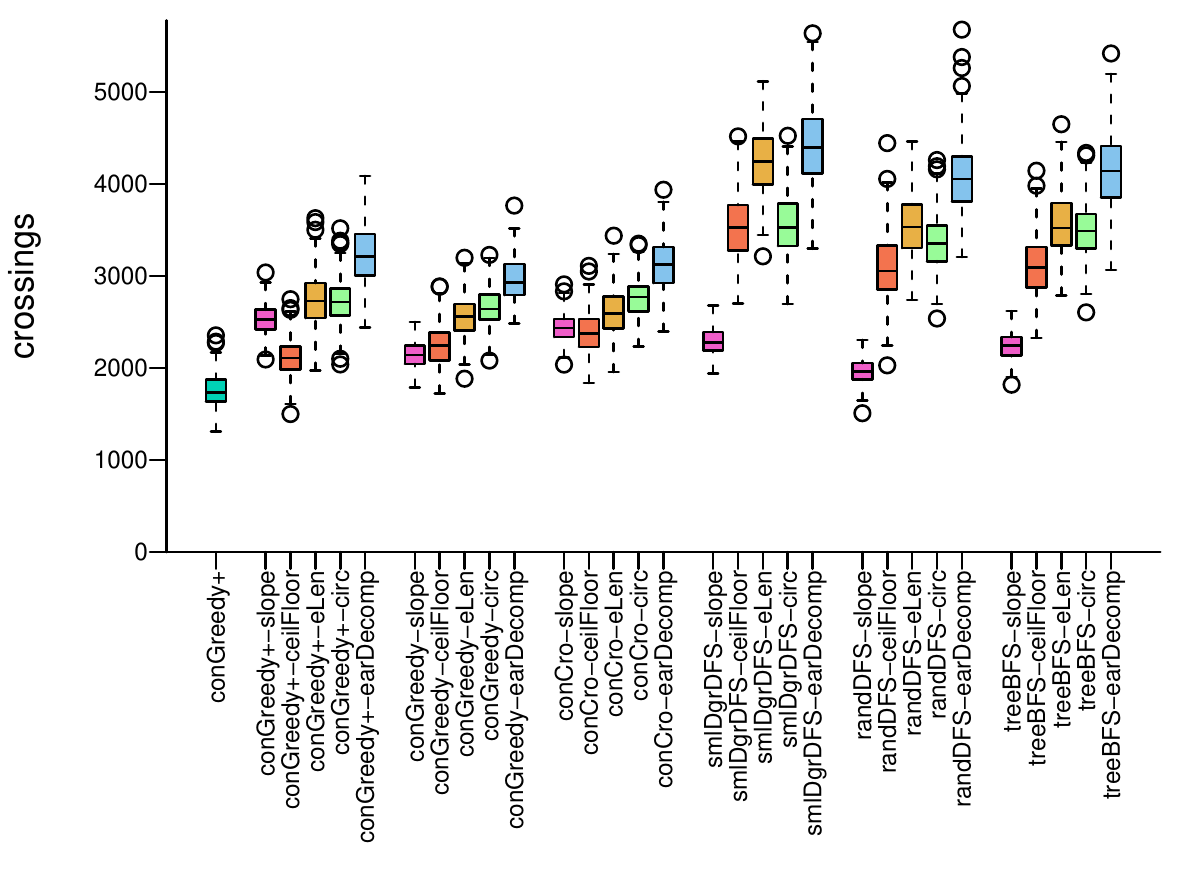}
 }
 \caption{All heuristics on random graphs with linear edge density and with 250 vertices. The
 differences between PA heuristics becomes more apparent for both higher density and more pages.
 The performance of \SLOPE gets significantly better with higher density. The search based
 VO heuristics perform nearly equally, as do the greedy VO heuristics with \CONGREEDY however
 slightly in the lead.}
 \label{fig:appendix:all:randomL}
\end{figure} 

% random quadratic 
\begin{figure}[htb]
 \centering
 \subfloat[3 pages.]{%
 \includegraphics[width=0.45\linewidth]{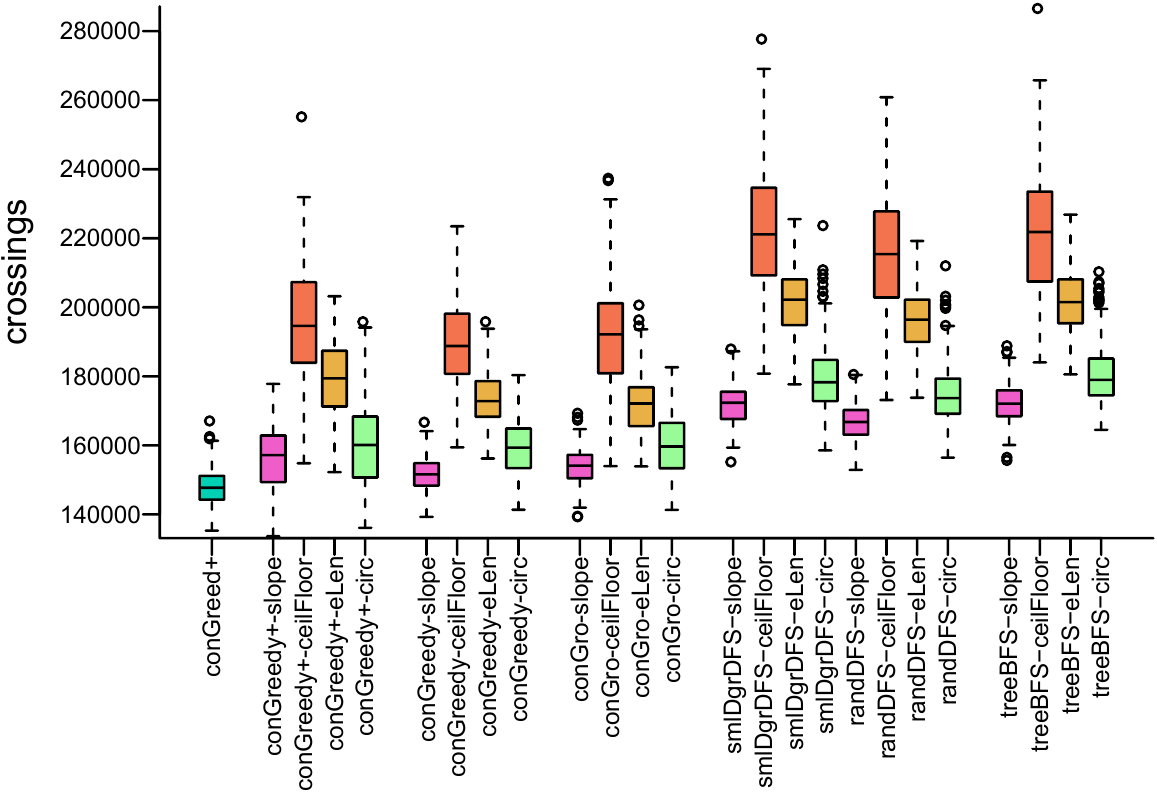}
 }\hfill
 \subfloat[6 pages.]{%
 \includegraphics[width=0.45\linewidth]{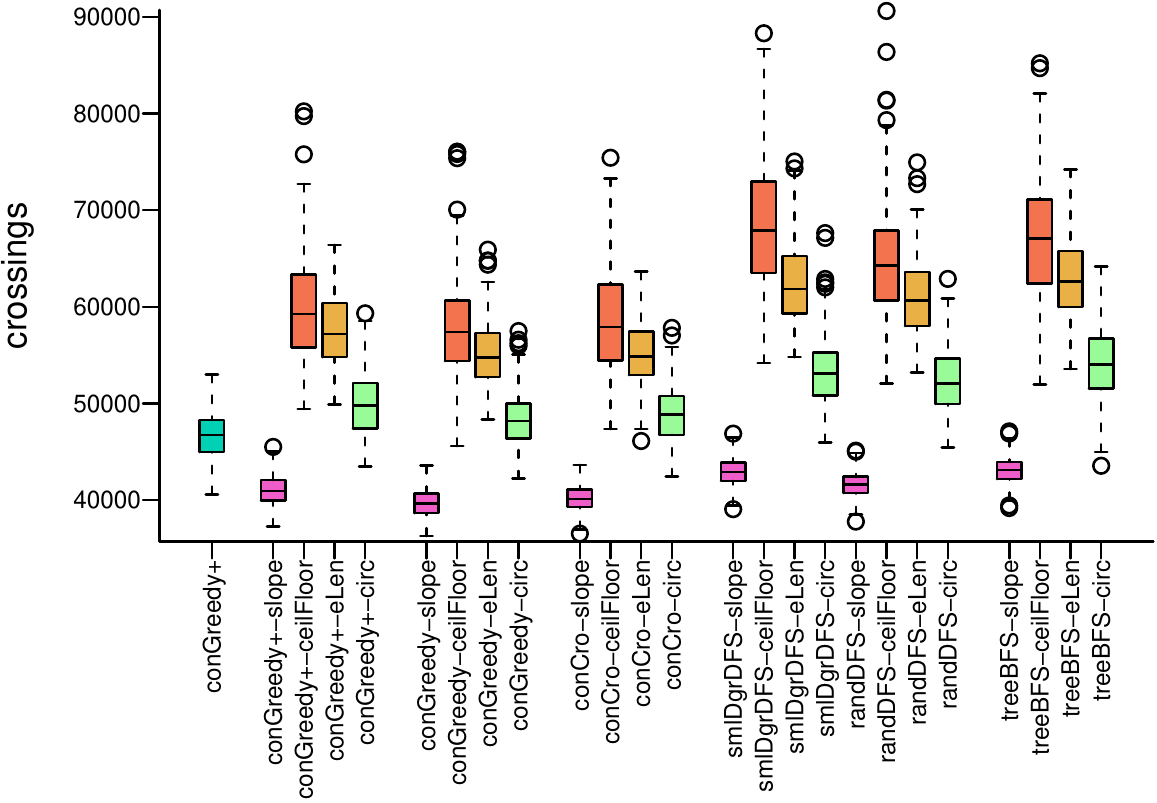}
 }\\
 \subfloat[9 pages.]{%
 \includegraphics[width=0.45\linewidth]{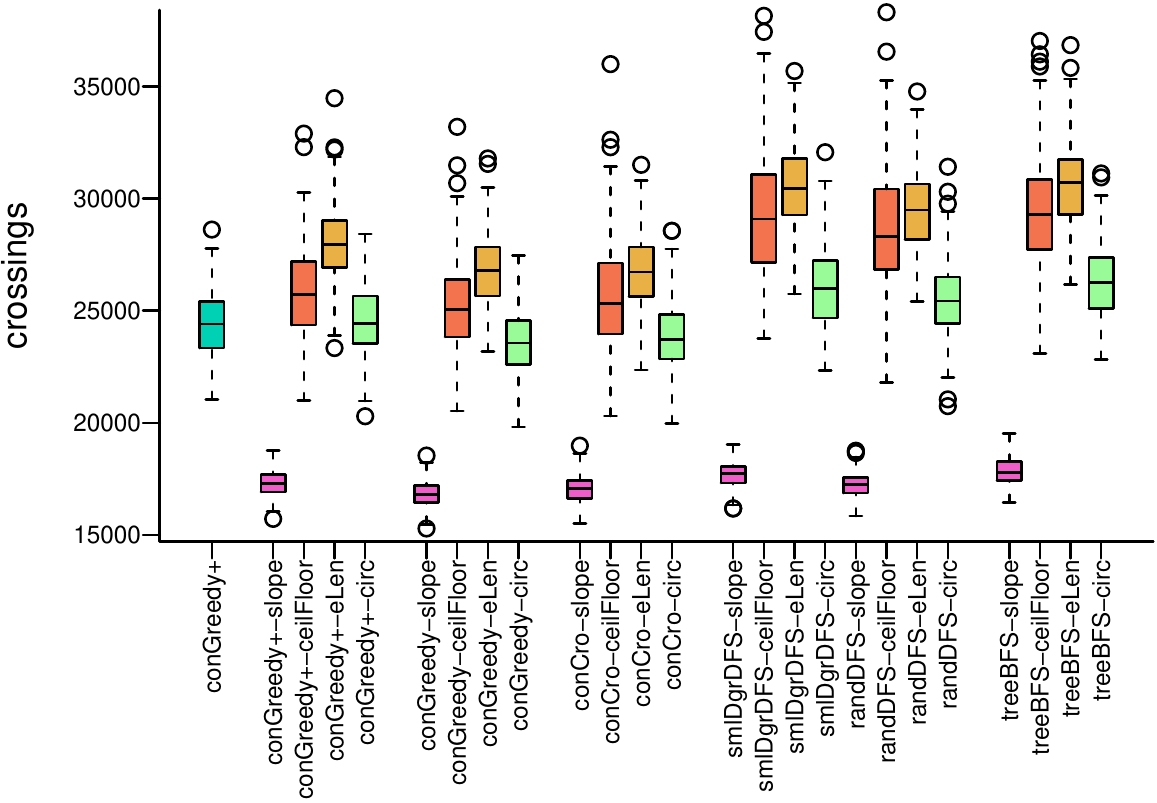}
 } $\quad$
 \subfloat[17 pages.]{%
 \includegraphics[width=0.3\linewidth]{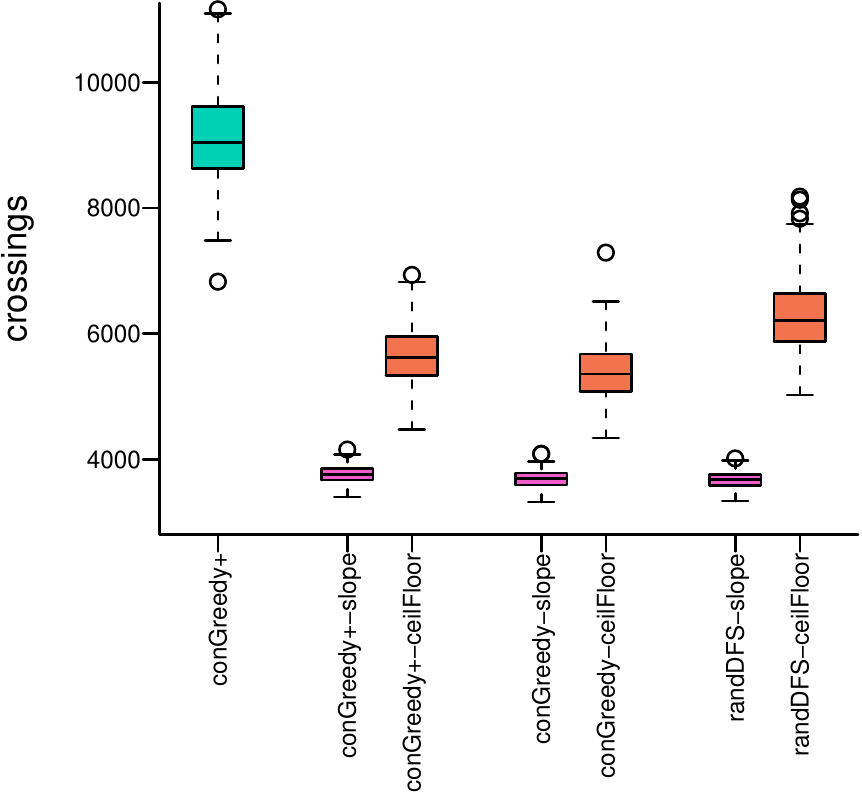}
 }
 \caption{Heuristics on random graph with 100 vertices and a density of 50\%.
 For 3 pages, \FCONGREEDY is the best heuristic. For all other number of pages, \SLOPE
 is the best PA heuristic and it becomes more important to use it than the choice of the VO
 heuristic. For dense graphs and higher number of pages we restricted some
 tests to only those heuristics that performed best in subsets of test graphs.}
 \label{fig:appendix:all:randomQuadratic}  
\end{figure}

% random quadratic (tiles)
\begin{figure}[htb]
\centering
\includegraphics[width=0.8\textwidth]{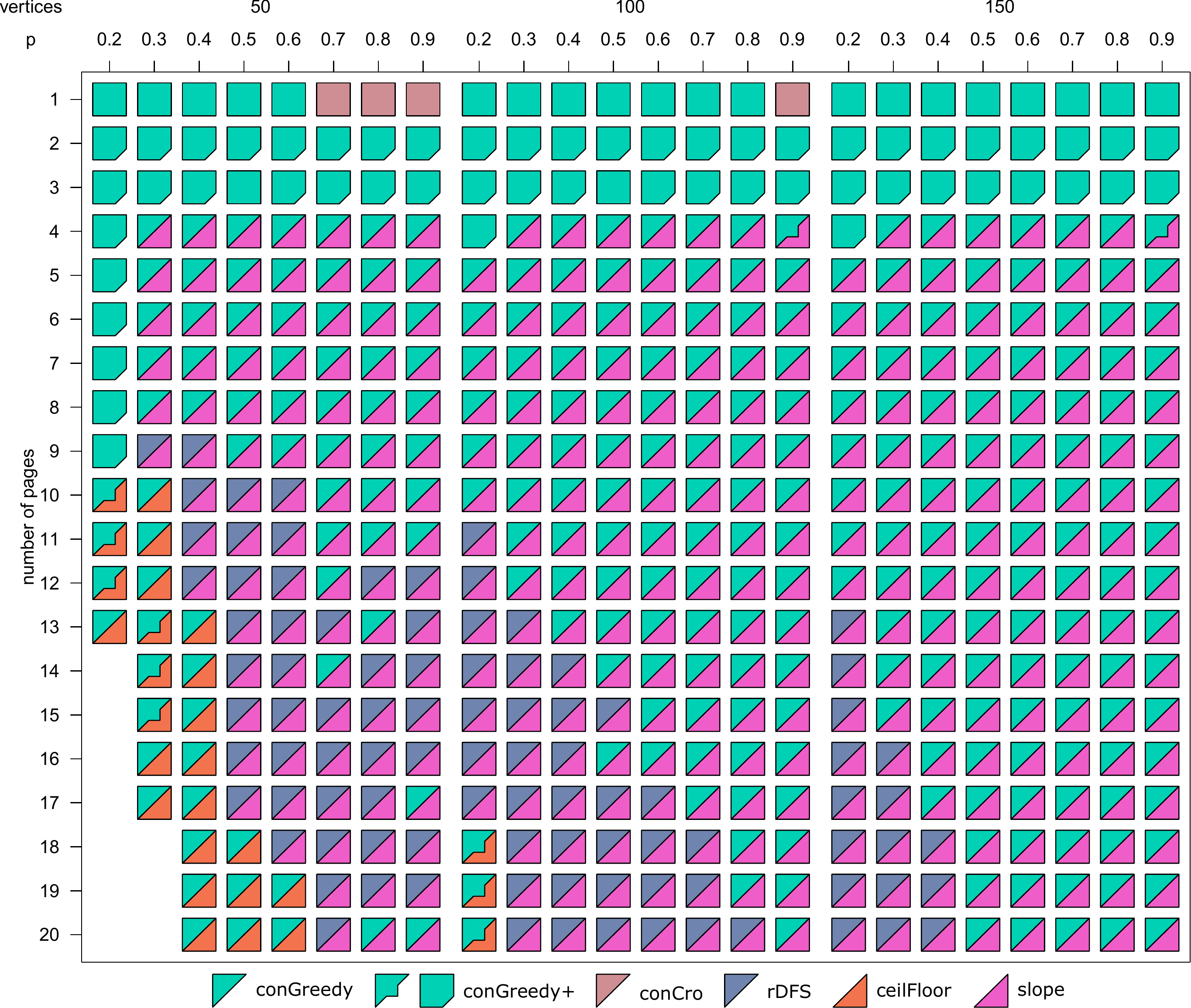}
\caption{Tile diagram for random graphs with different edge probability, i.e. quadratic number of
edges in terms of vertices. \SLOPE is the dominant heuristic for higher density and higher number
of vertices. For higher number of vertices \CONGREEDY gets better also for higher number of pages
than \RDFS.}
\label{fig:tiles:randomQuadratic}
\end{figure}

% running time
\begin{figure}
\centering
\subfloat[VO heuristics.\label{fig:time:heuristics:vo}]{%
      \includegraphics[width=0.45\textwidth]{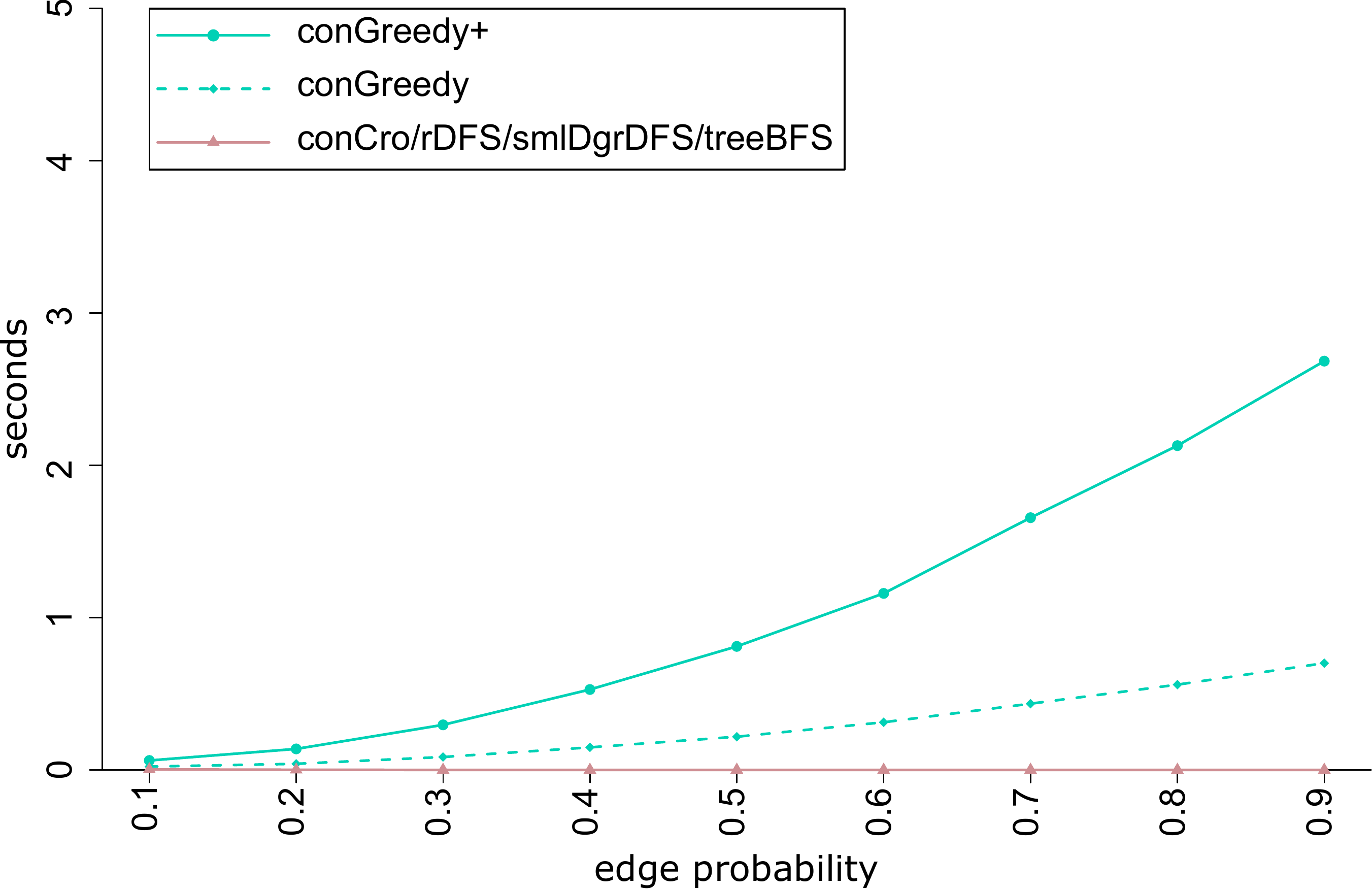}
}
$\quad$
\subfloat[PA heuristics.\label{fig:time:heuristics:pa}]{%
      \includegraphics[width=0.45\textwidth]{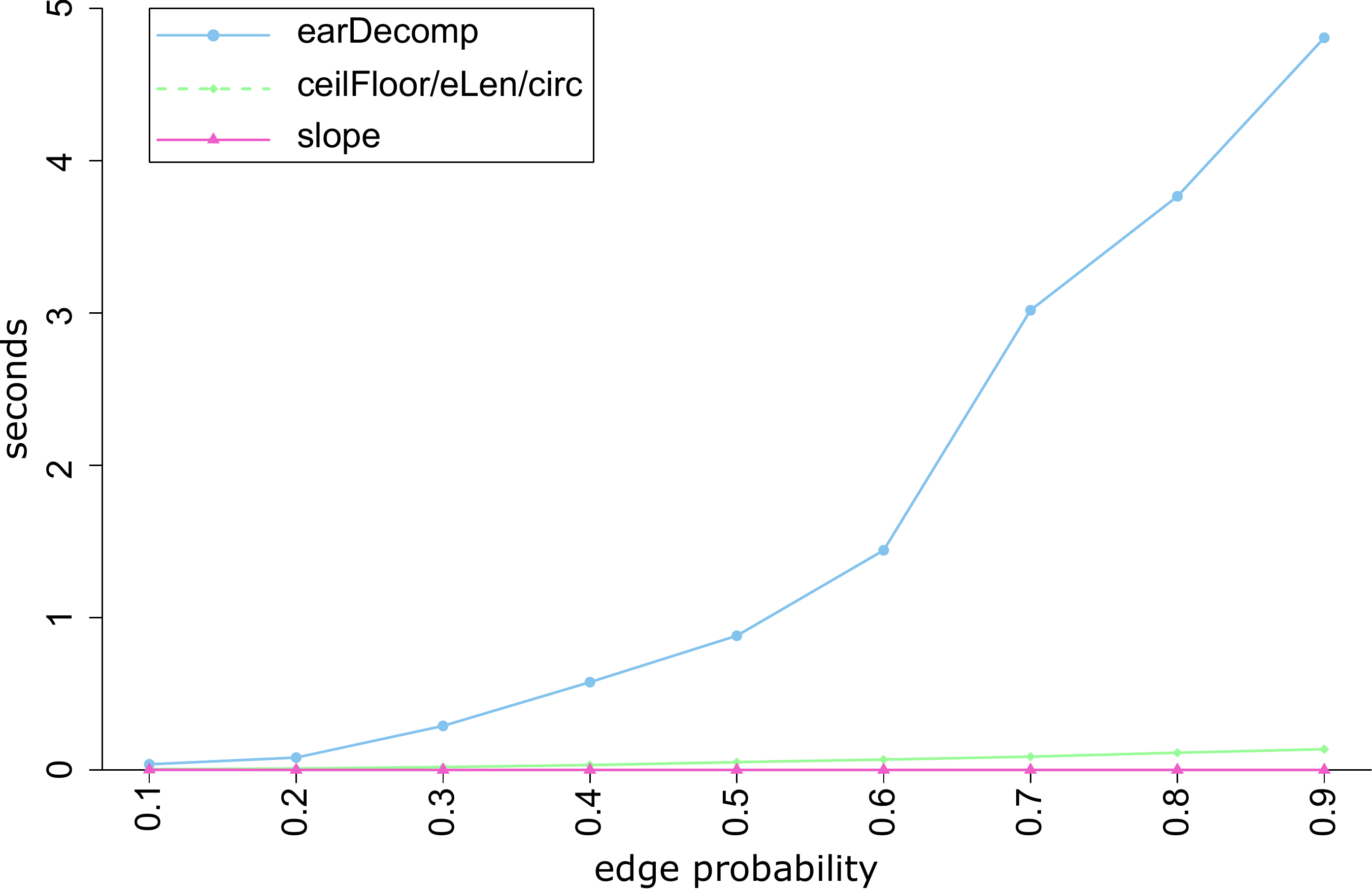}
 }
\caption{Running time of the construction heuristics for random graphs
with quadratic number of edges, 100 vertices and for four pages. The performance of VO heuristics
matches their asymptotic running times, while for the PA heuristics, \EAR performs significantly
worse than \ELEN, \CIRCULAR, with the same asymptotic running time.}
\label{fig:appendix:time:heuristics}
\end{figure}

% optimisation 
\begin{figure}[htb]
\centering
\subfloat[Topological planar, $n = 250$, 2 pages.\label{fig:opti:results:planar}]{%
      \includegraphics[width=0.36\textwidth]{opti/planar-n250-k2}
}
\hfill
\subfloat[Topological 1-planar, $n = 250$, 3 pages.\label{fig:opti:results:onePlanar}]{%
\includegraphics[width=0.36\linewidth]{opti/planar-n250-k2}
}\\
 \subfloat[4-tree, $n = 250$, 3 pages.\label{fig:opti:results:fourTree}]{%
 \includegraphics[width=0.36\linewidth]{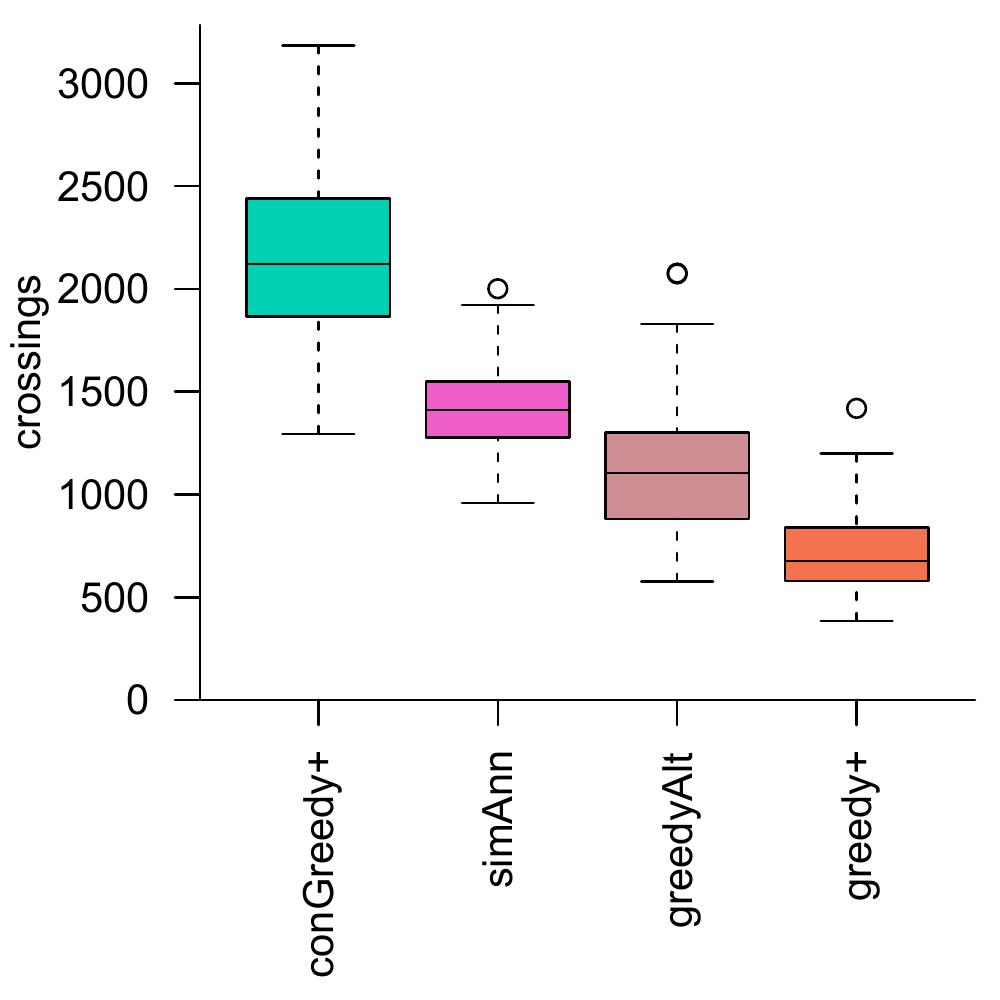}
 }
 \hfill
 \subfloat[Random (linear density 4), $n = 250$, 3
 pages.\label{fig:opti:results:planar}]{%
      \includegraphics[width=0.36\textwidth]{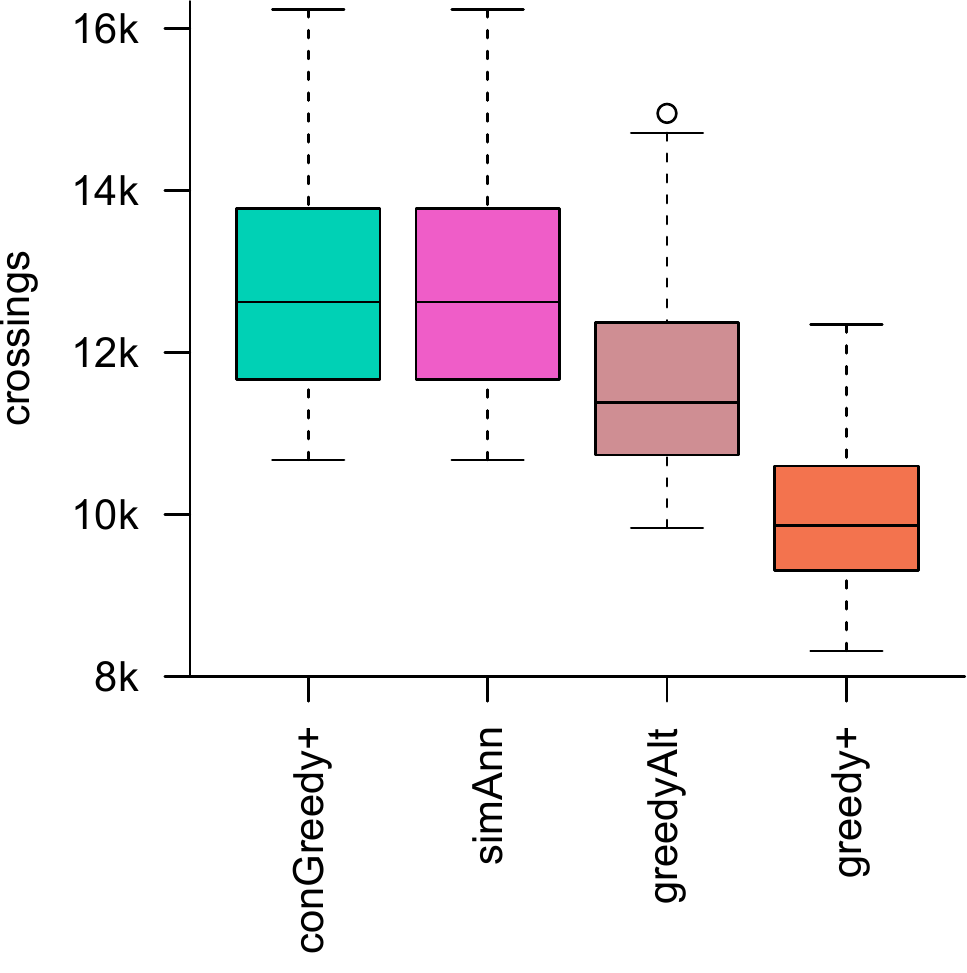}
 }\\
 \subfloat[Hypercube $Q_7$, 6 pages.\label{fig:opti:results:onePlanar}]{%
 \includegraphics[width=0.36\linewidth]{opti/Q7-k6}
 }
 \hfill
 \subfloat[Random (quadratic density 5), $n = 100$, 6 pages.\label{fig:opti:results:fourTree}]{%
 \includegraphics[width=0.36\linewidth]{opti/randomQ5-n100-k6}
 }
\caption{Local search heuristics performance in terms of crossings.}
\label{fig:appendix:opti1}
\end{figure}

\end{document}